\documentclass[12pt]{article}
\usepackage[utf8]{inputenc}
\usepackage[T1]{fontenc}
\usepackage{amsmath,amssymb,amsthm,commath,enumitem,mathtools}
\usepackage{bm}
\usepackage{kbordermatrix}
\usepackage[singlelinecheck=false]{caption}
\usepackage[margin=2.54cm]{geometry}
\usepackage[normalem]{ulem}
\usepackage{natbib}
\usepackage{graphicx}
\usepackage{booktabs}
\usepackage{color} 
\usepackage[hyphens]{url} 
\usepackage[onehalfspacing]{setspace}
\usepackage[small]{titlesec}
\usepackage{enumitem}
\usepackage[ruled,vlined]{algorithm2e}
\usepackage[tableposition=top,labelfont=bf]{caption}
\usepackage{layouts}
\usepackage{xcolor}
\usepackage{makecell}
\usepackage{tabularray}
\usepackage{microtype}
\graphicspath{ {./figures/} }
\usepackage{fancyhdr}

\allowdisplaybreaks
\let\originalleft\left
\let\originalright\right
\renewcommand{\left}{\mathopen{}\mathclose\bgroup\originalleft}
\renewcommand{\right}{\aftergroup\egroup\originalright}
\makeatletter
\newcommand*\bigcdot{\mathpalette\bigcdot@{.5}}
\newcommand*\bigcdot@[2]{\mathbin{\vcenter{\hbox{\scalebox{#2}{$\m@th#1\bullet$}}}}}
\makeatother

\title{Markov switching zero-inflated space-time multinomial models for comparing multiple infectious diseases}

\author{Dirk Douwes-Schultz$^{1}$\footnote{{{\it Corresponding author}: Dirk Douwes-Schultz, Department of Mathematics and Statistics, University of Calgary, 2500 University Drive NW, Calgary, AB, Canada, T2N 1N4. {\it E-mail}: {\tt
				dirk.douwesschultz@ucalgary.ca}.}}, Alexandra M. Schmidt$^{1}$, Laís Picinini Freitas$^{2}$ \\ and Marilia Sá Carvalho$^{3}$ \\
                	$^{1}$\textit{Department of Epidemiology, Biostatistics and Occupational Health,} \\ \textit{McGill University, Canada } \\
                    $^{2}$\textit{Department of Epidemiology and Quantitative Methods in Health,} \\
                    \textit{Sergio Arouca National School of Public Health,} \\
                    \textit{Oswaldo Cruz Foundation, Brazil } \\
				$^{3}$\textit{Scientific Computing Program,} \\ \textit{Oswaldo Cruz Foundation, Brazil }}
				
\date{\today}

\begin{document}

\maketitle

\begin{abstract}
Univariate zero-inflated models are increasingly being used to account for excess zeros in spatio-temporal infectious disease counts. However, the multivariate case is challenging due to the need to account for correlations across space, time and disease in both the count
 and zero-inflated components of the model. We are interested in comparing the transmission dynamics of several co-circulating infectious diseases across space and time, where some of the diseases can be absent for long periods. We first assume there is a baseline disease that is well-established and always present in the region. The other diseases switch between periods of presence and absence in each area through a series of coupled Markov chains, which account for long periods of disease absence, disease interactions and disease spread from neighboring areas. Since we are mainly interested in comparing the diseases, we assume the cases of the present diseases in an area jointly follow an autoregressive multinomial model. We use the multinomial model to investigate whether there are associations between certain factors, such as temperature, and differences in the transmission intensity of the diseases. Inference is performed using efficient Bayesian Markov chain Monte Carlo methods based on jointly sampling all unknown presence indicators. We apply the model to spatio-temporal counts of dengue, Zika, and chikungunya cases in Rio de Janeiro, during the first triple epidemic there.

\end{abstract}

\section{Introduction} \label{m3:intro}

{\color{black} In epidemiology, there has been a growing interest in analyzing multivariate counts of co-circulating infectious diseases across space and time to compare transmission dynamics \citep{freitas_spacetime_2019}.} {\color{black}In this paper}, we analyze bi-weekly reported case counts of Zika, chikungunya, and dengue in the 160 neighborhoods of Rio de Janeiro, Brazil, between 2015-2016, during the first triple epidemic there. {\color{black}The substantive aim of our analysis is to investigate how the transmission of the three diseases differs by factors such as temperature. For instance, if Zika transmits less effectively at lower temperatures compared to dengue, it would imply that its range of spread is more limited, which is helpful information for controlling the diseases \citep{tesla_temperature_2018}.}  Since we are mainly interested in comparing the diseases, we condition on the total number of cases in each area and bi-week and use a multinomial distribution to model their relative allocations \citep{dabney_issues_2005,dreassi_polytomous_2007,schmidt_poisson-multinomial_2022}. 

\begin{figure}[!t]
 	\centering
 	\includegraphics[width=.82\textwidth]{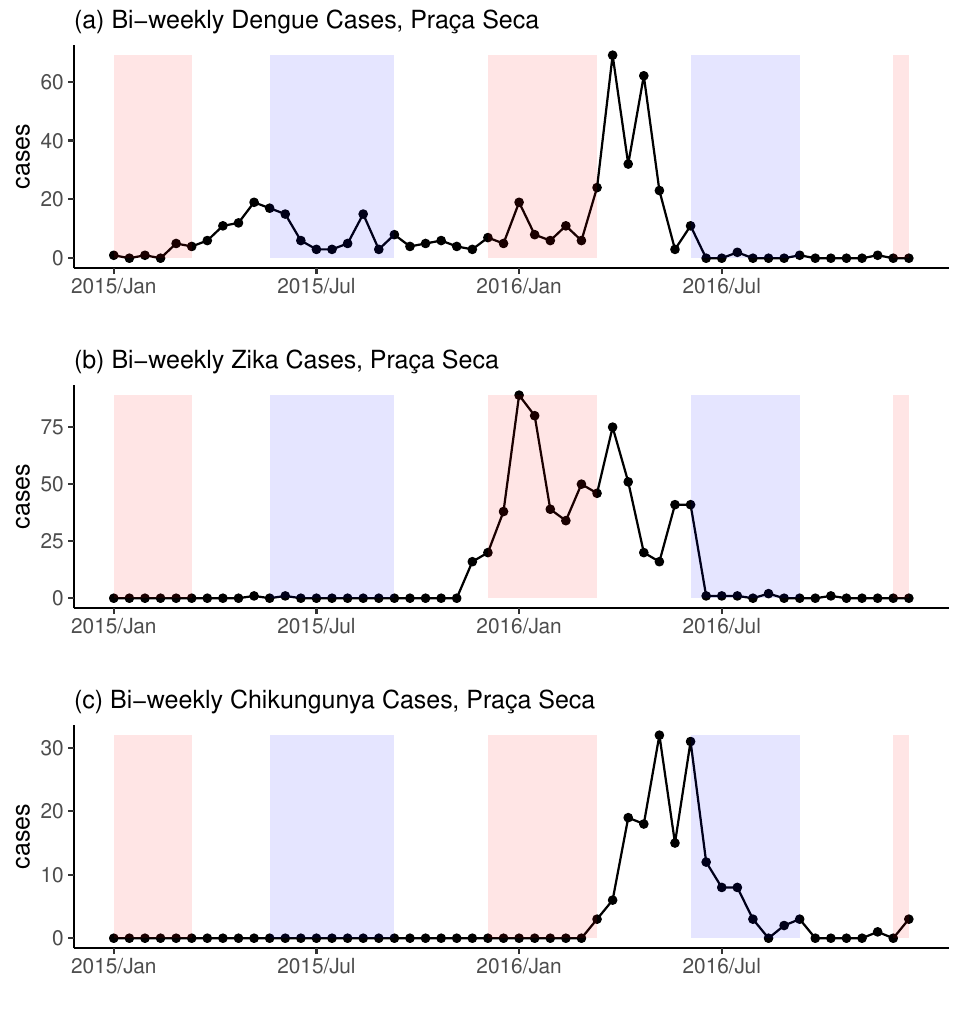}
	\caption{Bi-weekly reported case counts of dengue (a), Zika (b), and chikungunya (c) in Praça Seca, a west neighborhood of Rio de Janeiro, Brazil, between 2015-2016. Summer/winter seasons are highlighted in red/light blue. \label{fig3:fig_intro}} 
\end{figure}

A potential issue with applying a multinomial model to multivariate spatio-temporal infectious disease counts is that many infectious diseases, especially vector-borne diseases, can be
 absent for long periods of time in an area \citep{bartlettMeaslesPeriodicityCommunity1957,coutinhoThresholdConditionsNonAutonomous2006,adamsHowImportantVertical2010}. This behavior is illustrated in Figure \ref{fig3:fig_intro}, which shows bi-weekly reported case counts of dengue, Zika, and chikungunya in the western Rio neighborhood of Praça Seca, between 2015-2016. While dengue cases are reported consistently throughout the two years, very few Zika cases and no chikungunya cases are reported in the first year. These patterns are typical of most neighborhoods. If a disease is absent in an area, then there should be no chance of cases being reported. However, a multinomial model will always assign a positive probability to cases being reported. {\color{black}Therefore, there will likely be many more zeros in the counts than can be realistically produced by a fitted multinomial model (we investigate this in more detail in Section \ref{m3:application}) \citep{koslovsky_bayesian_2023}. Generally, when the fitted model cannot produce enough zeros, we say that the data contains excess zeros, which is very common when modeling infectious disease counts across space and time \citep{arabSpatialSpatioTemporalModels2015}.}  Also, we do not want to compare the transmission dynamics of a present disease and an absent one, as it would likely overestimate the favorability of the present disease.

{\color{black}When analyzing a single disease, zero-inflated count models \citep{lambertZeroInflatedPoissonRegression1992} are often used to deal with excess zeros in spatio-temporal infectious disease counts \citep{fernandesModellingZeroinflatedSpatiotemporal2009,arabSpatialSpatioTemporalModels2015,wangdiSpatialTemporalPatterns2018}. With these approaches, the presence/absence of the disease is first generated through a Bernoulli process  \citep{fernandesModellingZeroinflatedSpatiotemporal2009}. Then, if the disease is absent in an area, it is assumed that no cases will be reported. Conversely, if the disease is present, the number of reported cases is generated through a count process, typically negative binomial or Poisson \citep{aktekinAnalysisIncomeInequality2015}. A zero that arises from the count process is often interpreted as the result of the disease going undetected, referred to as a false zero \citep{vergne2014zero}. This is opposed to a true zero from the Bernoulli process, which corresponds to the actual absence of the disease. Random effects can be added to both the count and Bernoulli (zero-inflated) parts of the model to account for correlations across space and time \citep{torabiZeroinflatedSpatiotemporalModels2017, giorgiBivariateGeostatisticalModelling2018}. For example, \cite{hoefSpaceTimeZeroinflated2007} included random walks to account for correlations across time and conditional autoregressive effects \citep{besag1991bayesian} to account for correlations across space in both parts.

While the above models are popular in the univariate case, very little attention has been given to multivariate spatio-temporal zero-inflated disease modeling in general. To the best of our knowledge, no zero-inflated multinomial models have been proposed for space-time data. \cite{pavani_bayesian_2022} and \cite{rotejanaprasert_spatiotemporal_2021} both considered zero-inflated spatio-temporal multivariate Poisson models. They used spatial and temporal random effects to account for correlations in the data. There is a close relationship between multivariate Poisson and multinomial models for multivariate disease counts \citep{schmidt_poisson-multinomial_2022}. However, we prefer the multinomial approach, as it eliminates many nuisance parameters if the interest is only in comparing the diseases; see the supplementary materials (SM) Section 1. In addition, the above multivariate Poisson proposals have some important limitations that make them unsuitable for our application, including not allowing the probability of disease presence to change over time. Vector-borne diseases are highly seasonally driven and are more likely to be present in the summer and absent in the winter \citep{coutinhoThresholdConditionsNonAutonomous2006}. Also, if an infectious disease is present in an area, it should be more likely to be present in that area and neighboring areas in the next time \citep{douwes-schultzZerostateCoupledMarkov2022a}. The challenge with scaling up the univariate random effect models described in the previous paragraph to the multivariate case is the need to include random effects correlated across space, time and disease (not just space and time) in both the count and zero-inflated parts of the model, which would be computationally challenging.}

To account for long periods of disease absence, {\color{black}we will extend the zero-inflated multinomial modeling framework of \cite{zeng_zero-inflated_2022} to accommodate spatio-temporal infectious disease data}. {\color{black} Similar to the univariate models described above,} zero-inflated multinomial models allow categories (in our case diseases) to be absent from observations. If a category is absent, its multinomial probability is fixed at 0, {\color{black} meaning the count for that category will be 0}. The presence of the categories is usually treated as random and generated through a series of independent Bernoulli processes \citep{koslovsky_bayesian_2023}. While a zero-inflated multinomial model would allow some diseases to be absent in certain areas, existing approaches have made two independence assumptions that are inappropriate for spatio-temporal infectious disease counts. Firstly, they have assumed that the multivariate counts are independent conditional on the present categories of each observation. Infectious disease counts usually exhibit a high amount of correlation across space and time due to disease transmission \citep{bauerStratifiedSpaceTime2018}. Secondly, past approaches assumed that if a category is present in one observation, it does not affect whether that category or other categories are present in other observations (independence of categorical presence). If an infectious disease is present in an area, then it is more likely to be present in the next time period, as disease presence and absence are usually persistent \citep{coutinhoThresholdConditionsNonAutonomous2006}. Also, the presence of one disease may affect whether other diseases are present due to potential disease interactions \citep{paul_multivariate_2008}. Finally, due to between-area disease spread \citep{grenfellTravellingWavesSpatial2001}, the status of the disease in an area will affect its presence in neighboring areas.

To address the above issues, we first choose a baseline disease that is well-established in the region and assume it is always present. This avoids all diseases being absent in an area at once, which would be impossible if the total number of cases were not 0. {\color{black} As we discuss in Section \ref{m3:modelZI}, this problem was not considered by \cite{zeng_zero-inflated_2022}.} We then assume that all other diseases switch back and forth between periods of presence and absence in each area through a series of nonhomogeneous coupled Markov chains. {\color{black}That is, we allow the probabilities of disease presence to depend on whether the disease was present previously in the area; whether other diseases were present previously, to account for interactions; covariates, like temperature; and the prevalence of the disease in neighboring areas, to account for spread. Therefore, unlike \cite{pavani_bayesian_2022} and \cite{rotejanaprasert_spatiotemporal_2021}, we allow the risk of disease presence to change dynamically across space, time and disease. Also, unlike \cite{zeng_zero-inflated_2022}, the presence of the categories is not independent between observations. For instance, if a disease is present in an area, this disease, and potentially other diseases, can be more likely to be present again. For the count part of our zero-inflated model, we assume that if a disease is absent in an area, no cases will be reported. If a disease is present, we assume that the odds of a new case being from that disease relative to the baseline disease (relative odds) follows an autoregressive log-linear model. We allow the log relative odds to depend on past counts in the area and neighboring areas \citep{fokianos_regression_2003,tepe_spatio-temporal_2020}. Therefore, unlike \cite{zeng_zero-inflated_2022}, we do not assume the multivariate counts are independent conditional on the presence of the diseases, and we account for the transmission of the present diseases between neighboring areas \citep{bauerStratifiedSpaceTime2018}. Finally, our model can be written as an autoregressive Markov switching model, meaning Bayesian inference is very tractable through the forward filtering backward sampling algorithm \citep{fruhwirth-schnatterFiniteMixtureMarkov2006}.}

The rest of this paper is organized as follows. Section \ref{m3:model} introduces our proposed autoregressive multinomial model for comparing the transmission dynamics of multiple infectious diseases in a space-time setting. Section \ref{m3:modelZI} extends the model to incorporate zero inflation and deal with long periods of disease absence. Section \ref{m3:inferproc} details our Bayesian inferential procedure, which utilizes the forward filtering backward sampling (FFBS) algorithm of \cite{chibCalculatingPosteriorDistributions1996} for efficient inference. In Section \ref{m3:application}, we apply our model to cases of dengue, Zika and chikungunya in Rio de Janeiro during the first triple epidemic there. We close with a general discussion in Section \ref{m3:discussion}.

\section{An Autoregressive Multinomial Model for Comparing Transmission Dynamics} \label{m3:model}

Assume we have reported case counts for $K$ different infectious diseases across $i=1,\dots,N$ areas and $t=1,\dots,T$ time periods. Let $\bm{y}_{it}=(y_{1it},\dots,y_{Kit})^T$ represent the vector of case counts for all diseases in area $i$ during the time interval $(t-1,t]$, where $y_{kit}$ represents the reported case counts for disease $k$. We will first present our model without zero inflation and then extend to the zero-inflated case in Section \ref{m3:modelZI}. Since we are mainly interested in comparing the diseases, we condition on the total number of cases and model $\bm{y}_{it}$ using a multinomial distribution,
 \begin{align} \label{eqn:multinom}
\bm{y}_{it}|\text{total}_{it}, \bm{y}_{t-1} \sim \text{Multinom}(\bm{\pi}_{it},\text{total}_{it}), 
\end{align} where $\text{total}_{it}=\sum_{k=1}^{K}y_{kit}$ and $\bm{y}_{t-1}=(\bm{y}_{1(t-1)},\dots,\bm{y}_{N(t-1)})^{T}$ {\color{black}is the vector of all case counts observed at the previous time point}, giving the model an autoregressive structure. In (\ref{eqn:multinom}), $\bm{\pi}_{it}=(\pi_{1it},\dots,\pi_{Kit})^T$ represents the expected relative distribution of the disease counts at time $t$ in area $i$ given the previously observed counts $\bm{y}_{t-1}$. If conditions in $(t-1,t]$ favor some diseases over others, then the share of the favored diseases should grow relative to those less favored. Therefore, we model the {\color{black}odds of a new case being from disease $k$ relative to the baseline disease (relative odds) as,}
\begin{align} \label{eqn:relative_odds}
\frac{\pi_{kit}}{\pi_{1it}} &= \lambda_{kit}^* = \lambda_{kit}\frac{\left(y_{ki(t-1)}+1\right)^{\zeta_k}}{\left(y_{1i(t-1)}+1\right)^{\zeta_1}},
\end{align} for $k=2,\dots,K$, where we assume disease 1 is the baseline disease, and we add 1 to the top and bottom to avoid dividing by zero, adding such constants is common in autoregressive count models \citep{liboschikTscountPackageAnalysis2017,fritz_interplay_2022}. The parameters $\zeta_k$ and $\zeta_1$, assumed to be between 0 and 1, are meant to account for nonhomogeneous mixing \citep{minin_spatio-temporal_2019}. That is, as cases increase, there tends to be a dampening effect on transmission as individuals avoid infection and governments implement interventions. In (\ref{eqn:relative_odds}), $\lambda_{kit}>0$, which we link to covariates below, is a parameter that measures the favorability of conditions for the transmission of disease $k$ relative to the baseline disease during $(t-1,t]$. For instance, if $\lambda_{kit}>1$ ($\lambda_{kit}<1$) then we would expect the share of disease $k$ to grow (shrink) relative to the baseline disease in $(t-1,t]$, after adjusting for nonhomogeneous mixing with $\zeta_k$ and $\zeta_1$. If $\zeta_k \approx \zeta_1 << 1$, like for our motivating example in Section \ref{m3:application}, then this means that if $\lambda_{kit}>1$, the share of disease $k$ will grow initially from an equal share. However, as disease $k$ becomes more dominant, the share will be pulled back towards equality as individuals avoid infection from the dominant disease.

We link $\lambda_{kit}$ to covariates and random effects using a log-linear model,
\begin{align} \label{eqn:log_linear}
\log(\lambda_{kit})= \alpha_{0k}+\alpha_{0ki}+\bm{x}_{kit}^T \bm{\alpha}_k+\phi_{kit},
\end{align} where $\alpha_{0ki} \sim N(0,\sigma_{k}^2)$ is a normal random intercept meant to account for between area differences and $\bm{x}_{kit}$ is a vector of covariates that might affect the favorability of disease $k$ relative to the baseline disease. {\color{black}We also considered a spatially correlated intrinsic conditional auto-regressive (ICAR) model for the intercepts, $\alpha_{0ki} \sim \text{ICAR}(\sigma_{k,\text{ICAR}}^{2})$ \citep{besag1991bayesian}. However, for our motivating example in Section 4, while our proposed model converged, we were unable to obtain convergence for some of the comparison models when using ICAR intercepts. As the results of the normal random intercept and ICAR models were very similar, {\color{black}see SM Section 11,} we decided to use normal random intercepts in Section 4 to ensure that all comparisons were fair. We give the results of the ICAR model applied to our motivating example in SM Section 11, and the code is available on GitHub \url{https://github.com/Dirk-Douwes-Schultz/MS_ZIARMN_code}. We note that the number of cases in neighboring areas may affect the favorability of the diseases, as individuals will mix with those in neighboring areas \citep{grenfellTravellingWavesSpatial2001}. Therefore{\color{black}, following \cite{bauerStratifiedSpaceTime2018} and \cite{bracher_endemic-epidemic_2022},} we allow functions of disease cases in neighboring areas, such as the prevalence across neighboring areas, to be added to the covariate vector $\bm{x}_{kit}$ to account for the geographical spread of the disease. Also, we allow previous counts of other diseases, e.g., $\log(y_{ji(t-1)}+1)$ for $j\neq k$, to be added to $\bm{x}_{kit}$ to account for potential disease interactions \citep{freitas_spacetime_2019}, see Section \ref{m3:application} for an example.

We model the random effects $\phi_{kit}$ using a multivariate normal distribution \citep{xia_logistic_2013,zeng_zero-inflated_2022},
\begin{align} \label{eqn:logistic_normal}  \bm{\phi}_{it} &=
(\phi_{2it}, \dots, \phi_{Kit})^T \sim \text{MVN}_{K-1}(\bm{0},\bm{\Sigma}),
\end{align} where $\bm{\Sigma}$ is a $K-1$ by $K-1$ variance-covariance matrix. These random effects are mainly meant to account for overdispersion, which is a very common issue when modeling multinomial counts \citep{xia_logistic_2013}. {\color{black}Since $\phi_{kit}$ is correlated across diseases,} they also allow for a more flexible correlation structure between the individual disease counts {\color{black} compared to the standard multinomial distribution}; see SM Section 2.

There is an epidemiological interpretation of the parameters in Equations (\ref{eqn:multinom})-(\ref{eqn:logistic_normal}). As we show in SM Section 1, if all diseases follow a Reed-Frost chain binomial SIR model \citep{abbey_examination_1952,vynnycky_introduction_2010,bauerStratifiedSpaceTime2018} with roughly the same serial interval, the time between successive generations of cases, then we can derive Equations (\ref{eqn:multinom})-(\ref{eqn:logistic_normal}). Under this derivation, $\lambda_{kit}$ in (\ref{eqn:relative_odds}) represents the ratio of the effective reproduction number of disease $k$ and the baseline disease. The effective reproduction number is an important measure of disease transmission in epidemiology, representing the average number of new infections produced by a single infectious individual in the current population before they recover \citep{vynnycky_introduction_2010}. The covariate effect $\alpha_{kj}$ then represents the difference between the effect of covariate $x_{kitj}$ on the effective reproduction number of disease k and the baseline disease; see SM Section 1 for more details. 

While the Reed-Frost model is commonly used \citep{minin_spatio-temporal_2019}, it makes assumptions that are not appropriate for many diseases \citep{abbey_examination_1952}, and the model can be sensitive to underreporting and reporting delay when it comes to estimating the effective reproduction number \citep{bracher_marginal_2021,quick_regression_2021}. Regardless, the Reed-Frost derivation does reveal an important source of potential confounding due to differences in the susceptible populations, i.e., the population of individuals who are not immune to the disease. From the Reed-Frost derivation, see SM Equation (8), ideally, we would add $\log(\delta_{ki(t-1)}/\delta_{1i(t-1)})$ to $\log(\lambda_{kit})$ in Equation (\ref{eqn:log_linear}), where $\delta_{ki(t-1)}$ is the size of the susceptible population for disease $k$ in area $i$ at time $t-1$. This makes sense intuitively; if a disease has a larger susceptible population compared to another, then transmission for that disease will naturally be favored. The effect of any covariates correlated with the ratio of the susceptible populations could then be confounded. We conduct a sensitivity analysis to attempt to control for this for our motivating example in SM Section 9.

  We will call the model defined by (\ref{eqn:multinom})-(\ref{eqn:logistic_normal}) the autoregressive multinomial (ARMN) model. As detailed in the introduction, the ARMN model is not appropriate if some of the diseases are absent in an area, as it will always assign a positive probability to cases being reported. {\color{black} In a simulation study in SM Section 4, we show that ignoring long periods of disease absence leads to poor coverage of the 95\% credible intervals and posteriors centered far from the actual parameter values.} In the next subsection, we will keep the definitions from above and extend the ARMN model to deal with long periods of disease absence by adapting the proposal of \cite{zeng_zero-inflated_2022} to our space-time setting.

\subsection{Incorporating zero-inflation} \label{m3:modelZI}

Let $S_{kit}$, for $k=1,\dots,K$, be an indicator for the presence of disease $k$, so that $S_{kit}=1$ if disease $k$ was present in area $i$ during time $t$ and $S_{kit}=0$ if disease $k$ was absent. {\color{black} Following traditional zero-inflated epidemiological models \citep{vergne2014zero,douwes-schultzZerostateCoupledMarkov2022a}, we will assume that if a disease is absent ($S_{kit}=0$), no cases will be reported (a true or structural 0). Note that this means if all diseases are absent at once in an area ($S_{1it}=S_{2it}\dots=S_{Kit}=0$), the total number of cases would be 0 ($\text{total}_{it}=0$). However, a multinomial model conditions on the total count and does not generate it. Therefore, as it makes sense in many epidemiological applications, we will assume that the baseline disease is well established in the region and always present. That is, we assume $S_{1it}=1$ for all $i$ and $t$. We then assume that if a disease is present, the odds of a new case being from that disease relative to the baseline disease is given by Equation (\ref{eqn:relative_odds}) above, leading to the following zero-inflated multinomial model, \begin{align}
\begin{split} \label{eqn:zeng_douwes}
\bm{y}_{it}|\bm{S}_{it},\text{total}_{it}, \bm{y}_{t-1} &\sim \text{Multinom}(\bm{\pi}_{it},\text{total}_{it}),\text{where} \\[5pt]
\pi_{kit}&=0 \text{\hspace{.9cm} if $S_{kit}$=0 (absent),} \\[5pt]
\frac{\pi_{kit}}{\pi_{1it}} &= \lambda_{kit}^* \text{\hspace{.5cm} if $S_{kit}$=1 (present), for $k=2,\dots,K$},
\end{split}
\end{align} where $\bm{S}_{it}=(S_{2it},\dots,S_{Kit})^T$.
Note from (\ref{eqn:zeng_douwes}) that we can have zero-reported cases of a disease ($y_{kit}=0$) even if the disease is present ($S_{kit}=1$). Following \cite{vergne2014zero} and \cite{douwes-schultzZerostateCoupledMarkov2022a}, we interpret this as the disease being present but undetected by the surveillance system (a false or at-risk 0).
} 

{\color{black} We can compare (\ref{eqn:zeng_douwes}) with the zero-inflated multinomial framework of \cite{zeng_zero-inflated_2022}, who assumed}
\begin{align} \label{eqn:zeng}
\begin{split} 
\pi_{1it} = \frac{S_{1it}}{S_{1it}+\sum_{j=2}^{K}S_{jit}\lambda_{jit}^*}, \hspace{1cm}
\pi_{kit} = \frac{S_{kit}\lambda_{kit}^*}{S_{1it}+\sum_{j=2}^{K}S_{jit}\lambda_{jit}^*},
\end{split}
\end{align} for $k=2,\dots,K$. An immediate issue with (\ref{eqn:zeng}) is that we cannot have $S_{1it}=\dots=S_{Kit}=0$ as we would divide by 0. Even if we were to define $\bm{\pi}_{it}=\bm{0}$ if $S_{1it}=\dots=S_{Kit}=0$, we would have the same issue that this implies $\text{total}_{it}=0$. There is nothing in the model of \cite{zeng_zero-inflated_2022} to prevent $S_{1it}=\dots=S_{Kit}=0$.  They may have overlooked this as they dealt with the case of $K \geq 50$ and, therefore, it is very unlikely  $S_{1it}=\dots=S_{Kit}=0$. However, in our motivating example, we have $K=3$. {Note that Zeng's model (\ref{eqn:zeng}) assuming $S_{1it}=1$ for all $i$ and $t$ is equivalent to (\ref{eqn:zeng_douwes}).}

The presence of a disease in an area should depend on whether it was present previously, as disease presence and absence are often highly persistent \citep{coutinhoThresholdConditionsNonAutonomous2006,douwes-schultzZerostateCoupledMarkov2022a}, and may also depend on whether other diseases were present previously, as the diseases can interact with one another \citep{paul_multivariate_2008,sherlock_coupled_2013}. Additionally, the presence of a disease may depend on environmental or socioeconomic factors, such as temperature and water supply in the case of vector-borne diseases \citep{schmidtPopulationDensityWater2011,xuClimateVariationDrives2017}. Therefore, we model $S_{kit}$ as,
\begin{align} \label{eqn:presence}
\begin{split}
S_{kit}| \bm{S}_{i(t-1)}, \bm{y}_{t-1} &\sim \text{Bern}(p_{kit}) \\[5pt]
\text{logit}(p_{kit}) &= \eta_{0k} + \bm{z}_{kit}^T\bm{\eta}_k + \underbrace{\rho_k^{AR} S_{ki(t-1)}}_{\substack{\text{autoregressive}\\ \text{effect}}} + \underbrace{\sum_{j>1,\, j \neq k}\rho_{jk}^{DI} S_{ji(t-1)}}_{\substack{\text{disease interaction} \\ \text{effect}}} ,
\end{split} 
\end{align}  where $\bm{z}_{kit}$ is a vector of space-time covariates that might affect the presence of disease $k$. As the disease may spread from neighboring areas \citep{grenfellTravellingWavesSpatial2001}, we allow functions of disease cases in neighboring areas to be added to $\bm{z}_{kit}$, hence the dependence on $\bm{y}_{t-1}$ in (\ref{eqn:presence}). For instance, the prevalence of the disease in neighboring areas could be added, see Section \ref{m3:application}. Note, as $\rho_k^{AR}$ in Equation (\ref{eqn:presence}) increases and $\eta_{0k}$ decreases, the probability of getting a consecutive period of disease presence or absence approaches 1. Therefore, (\ref{eqn:presence}) can account for long consecutive periods of disease presence and absence, which are often observed for infectious diseases (see Figure \ref{fig3:fig_intro} and \cite{coutinhoThresholdConditionsNonAutonomous2006}).

We will call the ARMN model modified by Equations (\ref{eqn:zeng_douwes}) and (\ref{eqn:presence}) the Markov switching zero-inflated autoregressive multinomial (MS-ZIARMN) model. Note, from Equation (\ref{eqn:zeng_douwes}), $\pi_{1it}+\dots+
\pi_{Kit}=1$ for any possible values of the parameters and presence indicators. Also note that if diseases $k$ and $j$ are present in area $i$ during time $t$ we have $\pi_{kit}/\pi_{jit} = \lambda_{kit}^{*}/\lambda_{jit}^{*}$, which is the same as in the ARMN model described in Equations (\ref{eqn:relative_odds})-(\ref{eqn:logistic_normal}). That is, the MS-ZIARMN model preserves the relative distribution of the present diseases from the ARMN model. Finally, if $K=2$ and we assume Equation (\ref{eqn:zeng_douwes}), where $\log(\lambda_{2it}^*)=\text{logit}(\pi_{2it})=\alpha_{02i}+\bm{x}_{2it}^T \bm{\alpha}_2$ and $\text{logit}(p_{2it}) = \eta_{02} + \bm{z}_{2it}^T\bm{\eta}_2$, we get the well known and popular zero-inflated binomial (ZIB) model of \cite{hall_zero-inflated_2000}. Therefore, our model without space-time effects could be seen as a multinomial extension of \cite{hall_zero-inflated_2000}.

\section{Inferential Procedure} \label{m3:inferproc}

Here, we assume the number of diseases $K$ is small enough for matrix multiplication with a $2^K$ by $2^K$ matrix to be computationally feasible.  This should be the case in most epidemiological applications; for instance, in our motivating example, we compare $K=3$ diseases. Our inferential strategy revolves around reparametrizing the MS-ZIARMN model as a Markov switching model \citep{fruhwirth-schnatterFiniteMixtureMarkov2006}, which allows for very efficient Bayesian inference using the forward filtering backward sampling (FFBS) algorithm \citep{chibCalculatingPosteriorDistributions1996}. To simplify the explanations, we will assume, without loss of generality, that $K=3$.

In the case of $K=3$, we have $\bm{S}_{it}=(S_{2it},S_{3it})^T$. Let $S_{it}^*$ be an indicator for the possible values of $\bm{S}_{it}$, so that $S_{it}^*=1$ if $\bm{S}_{it}=(1,1)^T$, $S_{it}^*=2$ if $\bm{S}_{it}=(0,1)^T$, $S_{it}^*=3$ if $\bm{S}_{it}=(1,0)^T$ and $S_{it}^*=4$ if $\bm{S}_{it}=(0,0)^T$. Then, 
from Equation (\ref{eqn:zeng_douwes}),
\begin{align} \label{eqn:mixture}
\bm{y}_{it} | \bm{y}_{t-1}, S_{it}^*, \text{total}_{it} \sim 
\begin{cases}
\text{Multinom}\left(\left(\frac{1}{1+\lambda_{2it}^*+\lambda_{3it}^*},\frac{\lambda_{2it}^*}{1+\lambda_{2it}^*+\lambda_{3it}^*},\frac{\lambda_{3it}^*}{1+\lambda_{2it}^*+\lambda_{3it}^*}\right)^T,\text{total}_{it}\right), &\text{ if $S_{it}^*=1$}  \\[5pt]
\text{Multinom}\left(\left(\frac{1}{1+\lambda_{3it}^*},0,\frac{\lambda_{3it}^*}{1+\lambda_{3it}^*}\right)^T,\text{total}_{it}\right), &\text{ if $S_{it}^*=2$} \\[5pt]
\text{Multinom}\left(\left(\frac{1}{1+\lambda_{2it}^*},\frac{\lambda_{2it}^*}{1+\lambda_{2it}^*},0\right)^T,\text{total}_{it}\right), &\text{ if $S_{it}^*=3$} \\[5pt]
\text{Multinom}\left(\left(1,0,0\right)^T,\text{total}_{it}\right), &\text{ if $S_{it}^*=4$}, 
\end{cases}
\end{align} meaning, for $K=3$, the MS-ZIARMN model is a mixture of 4 different multinomial distributions where the multinomial probabilities are fixed at 0 for the diseases absent.

Note that, from Equation (\ref{eqn:presence}), as $\bm{S}_{it}$ only depends on $\bm{S}_{i(t-1)}$, $\bm{y}_{t-1}$ and covariates, $S_{it}^*$ follows a first order nonhomogeneous Markov chain. 
Let $\Gamma(S_{it}^*|\bm{y}_{t-1})$ be the transition matrix of $S_{it}^*$, where $\Gamma(S_{it}^*|\bm{y}_{t-1})_{jk}=P(S_{it}^*=k|S_{i(t-1)}^*=j,\bm{y}_{t-1})$ for $j,k=1,2,3,4$, $i=1,\dots,N$ and $t=2,\dots,T$. As $S_{it}^*$ is an indicator for $\bm{S}_{it}$, the transition matrix can be derived through 
\begin{align*}
P(\bm{S}_{it}=\bm{s}_{it}|\bm{S}_{i(t-1)}=\bm{s}_{i(t-1)},\bm{y}_{t-1})=\prod_{k=2}^{3}P(S_{kit}=s_{kit}|S_{2i(t-1)}=s_{2i(t-1)},S_{3i(t-1)}=s_{3i(t-1)},\bm{y}_{t-1}),
\end{align*} for $\bm{s}_{i(t-1)},\bm{s}_{it}=(1,1),(0,1),(1,0),(0,0)$, using Equation (\ref{eqn:presence}).  Finally, as $S_{i2}^*$ depends on $S_{i1}^*$, we require an initial distribution for the Markov chain, that is, $P(S_{i1}^*=j)$ for $j=1,2,3,4$ and $i=1,\dots, N$. The modeler can set the initial distribution based on how likely they believe the diseases to be present at the beginning of the study period.

Given $p(\bm{y}_{it} | \bm{y}_{t-1}, S_{it}^*, \text{total}_{it})$ and $\Gamma(S_{it}^*|\bm{y}_{t-1})$ for $i=1,\dots N$ and $t=2,\dots T$, and $p(S_{i1}^*)$ for $i=1,\dots,N$, we completely define a Markov switching model \citep{fruhwirth-schnatterFiniteMixtureMarkov2006}. A Markov switching model assumes a time series can be described by several submodels, often called states or regimes, where switching between submodels is governed by a first-order Markov chain \citep{hamiltonNewApproachEconomic1989}. In our case, the submodels are given by the 4 multinomial distributions in (\ref{eqn:mixture}), and we switch between these submodels through the transition matrix $\Gamma(S_{it}^*|\bm{y}_{t-1})$. Typically, $S_{it}^*$ is called the state indicator. 

Let  $\bm{S}^*=(\bm{S}_{1}^*,\dots,\bm{S}_{T}^*)^T$ be the vector of all state indicators, where $\bm{S}_t^*=(S_{1t}^*,\dots,S_{Nt}^*)^T$; $\bm{y}=(\bm{y}_1,\dots,\bm{y}_T)^T$ be the vector of all observations; and $\bm{\beta}$  the vector of all parameters in the multinomial part of the model 
(see (\ref{eqn:relative_odds})-(\ref{eqn:logistic_normal})). Further, let $\bm{\theta}$ be the vector of all parameters in the Markov chain part of the model  
(see (\ref{eqn:presence})), and, finally, let $\bm{v}=(\bm{\beta},\bm{\theta})^T$ be the vector of all model parameters. The joint distribution of $\bm{S}^*$ and $\bm{y}$ given $\bm{v}$ is given by,
\begin{align} \label{eqn:likelihood}
p(\bm{S}^*,\bm{y}|\bm{v}) = \prod_{i=1}^{N}\prod_{t=2}^{T} p(\bm{y}_{it} | \bm{y}_{t-1}, S_{it}^*, \text{total}_{it},\bm{\beta}) \prod_{i=1}^{N}p(S_{i1}^*)\prod_{t=2}^{T}p(S_{it}^*|S_{i(t-1)}^*,\bm{y}_{t-1},\bm{\theta}).
\end{align} Note, from Equation (\ref{eqn:mixture}), $S_{it}^*$ is only fully known when $y_{1it},y_{2it},y_{3it}>0$, in which case we must have $S_{it}^*=1$. It is possible, assuming $K$ is not too large, to use the forward filter \citep{hamiltonNewApproachEconomic1989} to completely marginalize $\bm{S}^*$ from (\ref{eqn:likelihood}) and calculate $p(\bm{y}|\bm{v})${\color{black}, which as a function of $\bm{v}$ is the marginal likelihood function,} see SM Section 3. However, we want to make inferences about $\bm{S}^*$ to investigate when the model believes the diseases were present or absent for model checking, as detailed in SM Section 6. Therefore, 
$\bm{S}^*$ is estimated along with $\bm{v}$ by sampling both from their joint posterior distribution $p(\bm{v},\bm{S}^*|\bm{y}) \propto p(\bm{S}^*,\bm{y}|\bm{v})p(\bm{v})$, where $p(\bm{v})$ is the prior distribution of $\bm{v}$. {\color{black}We will now discuss the prior specification in detail.}

We specified a conjugate inverse-Wishart prior for $\bm{\Sigma}$ with $K$ degrees of freedom and identity matrix, as it implies the correlations in $\bm{\Sigma}$ have marginal uniform prior distributions \citep{gelman_bayesian_2013}. {\color{black}For all non-bounded parameters, such as the intercepts and covariate effects, we used wide normal priors with a standard deviation of 5. For the precisions $1/\sigma^2_k$ we used conjugate $\text{gamma}(.1,.1)$ priors.}

As the joint posterior is not available in closed form, we resorted to Markov chain Monte Carlo (MCMC) methods; in particular, we used a hybrid Gibbs sampling algorithm with some steps of the Metropolis-Hastings algorithm to sample from it. The full details of the Gibbs sampler are given in SM Section 3. We sampled all of $\bm{S}^*$ jointly from $p(\bm{S}^*|\bm{v},\bm{y})$ using the FFBS algorithm for Markov switching models \citep{chibCalculatingPosteriorDistributions1996}. We found our MCMC algorithm mixed much faster when jointly sampling $\bm{S}^*$ compared to sampling each presence indicator individually. This is typically the case for Markov switching models \citep{fruhwirth-schnatterFiniteMixtureMarkov2006}, and was one of our main motivations for reparametrizing the MS-ZIARMN model as a Markov switching model.

Our MCMC algorithm was implemented using the R package Nimble \citep{valpineProgrammingModelsWriting2017}. We implemented the FFBS algorithm using Nimble's custom sampler feature. All other MCMC samplers mentioned in SM Section 3 are built into Nimble. All code and data are available on GitHub \url{https://github.com/Dirk-Douwes-Schultz/MS_ZIARMN_code}. {\color{black}In SM Section 4, we provide a simulation study, which shows that our proposed Gibbs
sampler can recover the true parameters of an MS-ZIARMN model that is specified like in
our motivating example in Section 4. }

\section{Application to  Counts of Dengue, Zika and Chikungunya}  \label{m3:application}

Since 2015, several countries in Latin America have experienced triple epidemics of dengue, Zika and chikungunya, placing an immense burden on the local populations \citep{rodriguez-morales_arboviral_2016,bisanzio_spatio-temporal_2018}. Between 2015 and 2016, Rio de Janeiro experienced its first triple epidemic of these diseases. Despite being transmitted by the same vector, the \emph{Aedes aegypti}, \cite{freitas_spacetime_2019}, using space-time cluster detection methods, found that Zika and dengue clusters were much more likely in the western region of Rio (there were no chikungunya clusters found in the west). In contrast, other regions contained many clusters of each disease, though often at different points in time. However, they did not investigate how these differences in clustering might depend on covariates or other factors. \cite{schmidt_poisson-multinomial_2022}, using a spatial multinomial regression model, found that cases of chikungunya were more likely in urban areas and Zika cases were less likely in population-dense areas, compared to dengue cases.  However, a spatial analysis cannot investigate how space-time factors might have affected differences in the transmission intensity of the diseases, which is of great interest in this example. For instance, laboratory studies have shown that the transmission of Zika may be more sensitive to temperature compared to dengue, which is important as it implies Zika will not be able to spread as effectively into North America and the highlands \citep{tesla_temperature_2018}. 

Therefore, we apply our MS-ZIARMN model to cases of the three diseases, $K=3$, across the 160 neighborhoods of Rio de Janeiro, $N=160$, for each bi-week between 2015-2016, $T=52$. Our model seems appropriate for this example, as there appear to be long periods of disease absence in both the time series of chikungunya and Zika cases, see Figure \ref{fig3:fig_intro} for example (overall 65\% of Zika cases are equal to 0 and 75\% of chikungunya cases are equal to 0).

\subsection{Model specification and fitting}
\label{m3:appfitting}

We modeled the Rio data with a bi-weekly time step as this roughly corresponds to the serial interval of the three diseases \citep{majumder_estimating_2016,riou_comparative_2017}. We took dengue as the baseline disease (k=1). {\color{black}Recall, from Section \ref{m3:modelZI}, that this means we are assuming that dengue was present in every neighborhood of Rio throughout the study period. We believe this is a reasonable assumption, as dengue has been circulating in Rio since 1986 and is well established there \citep{teixeira_dengue_2009}.} We then considered $k=2$ to represent Zika and $k=3$ to represent chikungunya. 

For $\bm{x}_{kit}$ in (\ref{eqn:log_linear}), covariates that may affect the favorability of Zika or chikungunya transmission relative to dengue transmission, we first took all spatial covariates considered in \cite{schmidt_poisson-multinomial_2022}: the percentage of neighborhood $i$ covered in green area, $\text{verde}_i$; the social development index of neighborhood $i$, $\text{SDI}_i$; and the population density of neighborhood $i$, $\text{popdens}_i$. We additionally considered the percentage of neighborhood $i$ that was occupied by favelas (slums), $\text{favela}_i$. These spatial covariates were chosen as the transmission of arboviral diseases can be significantly influenced by socioeconomic factors such as water supply and urbanicity \citep{schmidtPopulationDensityWater2011}, see \cite{schmidt_poisson-multinomial_2022} for more details. As for space-time covariates, we first took the average weekly maximum temperature in neighborhood $i$ during bi-week $t$, $\text{temp}_{it}$. Temperature is an important factor to consider, since if the transmission of Zika or chikungunya is more sensitive (less sensitive) to temperature compared to dengue, their range of spread will be smaller (bigger) \citep{tesla_temperature_2018,mercier_impact_2022}. Note that if there are many more cases of Zika or chikungunya in neighboring areas compared to dengue, then we would expect the share of those diseases to grow, all else being equal, due to between-area mixing \citep{stoddardHousetohouseHumanMovement2013}. Therefore, we considered the previous prevalence of disease $k$ and disease $1$ (dengue) across neighboring areas, $\log\left(\sum_{j\in NE(i)}y_{kj(t-1)}/\sum_{j\in NE(i)}\text{pop}_j+1\right)$ and 
$\log\left(\sum_{j\in NE(i)}y_{1j(t-1)}/\sum_{j\in NE(i)}\text{pop}_j+1\right)$, in $\bm{x}_{kit}$, where $NE(i)$ is the set of neighboring areas of neighborhood $i$ and $\text{pop}_j$ is the population of neighborhood $j$. We considered two areas neighbors if they shared a border. To simplify the model, we assumed the effect of neighboring dengue prevalence was the same on both $\log(\pi_{2it}/\pi_{1it})$ and $\log(\pi_{3it}/\pi_{1it})$, like with the within-area autoregression. Finally, it was speculated in \cite{freitas_spacetime_2019} that Zika circulation could have been inhibiting chikungunya transmission. To account for these kinds of disease interactions, we considered $\log(y_{3i(t-1)}+1)$ in $\bm{x}_{2it}$ and $\log(y_{2i(t-1)}+1)$ in $\bm{x}_{3it}$. That is, chikungunya cases were allowed to affect the favorability of Zika transmission relative to dengue transmission, and Zika cases were allowed to affect the favorability of chikungunya transmission relative to dengue transmission. 

As for covariates potentially associated with the presence of Zika or chikungunya, $\bm{z}_{kit}$ in (\ref{eqn:presence}), we considered mostly the same factors included in $\bm{x}_{kit}$. An exception is that we did not include cases of the other diseases, as the Markov chain defined in (\ref{eqn:presence}) already accounts for disease interactions through $\rho_{jk}^{DI}$. Also, since the presence is not defined relatively, we did not include neighboring dengue prevalence. 

We also need to specify the initial state distributions, see Section \ref{m3:inferproc}. As there were no chikungunya cases reported at all in the first year, we assumed chikungunya had only a 5 percent chance of being present at the start of 2015. For Zika, there were a few, but not many, cases reported through most of 2015; therefore, we assumed a 10 percent chance of initial presence. We fitted the MS-ZIARMN model specified above to the Rio data using our proposed Gibbs sampler from Section \ref{m3:inferproc}. We ran the Gibbs sampler for 250,000 iterations on three chains with an initial burn-in of 50,0000 iterations. All sampling was started from random values in the parameter space to avoid convergence to local modes. Convergence was checked using the Gelman-Rubin statistic ($<$1.05), the minimum effective sample size ($>$1000) and by visually examining the traceplots \citep{plummerCODAConvergenceDiagnosis2006}. {\color{black} The traceplots for all parameters (except the random effects) are shown in SM Section 10. It took around a day to run three chains of the Gibbs sampler in parallel on a laptop with an i9-12900HK, 2500 MHz, CPU.}

For comparison purposes, we also considered a model without zero inflation to ensure that there is justification for treating the many zeroes in the data, as hypothesized. That is, the ARMN model described in Equations (\ref{eqn:multinom})-(\ref{eqn:logistic_normal}) with the same covariates $\bm{x}_{2it}$ and $\bm{x}_{3it}$ as the MS-ZIARMN model. We also considered a zero-inflated model without the coupled Markov chains, that is, with $\rho_k^{AR}=0$ and $\rho_{jk}^{DI}=0$ for all $k$ and $j>1,\, j\neq k$ in Equation (\ref{eqn:presence}), which we will call the ZIARMN model. Without the coupled Markov chains, our model is a finite mixture and not a Markov mixture model, and the inference is greatly simplified, e.g., there is no need to use the FFBS algorithm. Finally, we compared with the zero-inflated multinomial model of \cite{zeng_zero-inflated_2022} as our approach is built on their framework. \cite{zeng_zero-inflated_2022} assumed that the probability a category was absent from an observation only depended on the category, so we let $\text{logit}(p_{kit})=\eta_{0k}$ for $k=2,3$ in Equation (\ref{eqn:presence}). We kept the rest of the model the same as the MS-ZIARMN model. Note that \cite{zeng_zero-inflated_2022} did not consider any covariates or space-time structure in the multinomial probabilities. However, this is inappropriate for our example and did not fit the data well (results not shown). Also, \cite{zeng_zero-inflated_2022} did not assume the baseline category was always present; however, this appears to be necessary when modeling a small number of categories, as discussed in Section \ref{m3:modelZI}.

Table \ref{tab3:comparemodels} shows the widely applicable information criterion (WAIC) \citep{gelmanUnderstandingPredictiveInformation2014} of the 4 considered models. We calculated the WAIC by marginalizing out the state indicators $\bm{S}^*$ as recommended by \cite{auger-methe_guide_2021}, see SM Section 5 for more details. Note that the model with the smallest WAIC is considered to have the best fit, and as a rule of thumb, a difference of 5 or more in the WAIC is considered significant. Table \ref{tab3:comparemodels} shows it is not only important to account for zero inflation, but also covariates and correlations across space, time and disease in the zero-inflated process (i.e., the process generating disease presence). Indeed, the model of \cite{zeng_zero-inflated_2022} had a worse fit compared to the ARMN model (which does not account for zero-inflation), despite the other zero-inflated models having a much better fit. 

\renewcommand{\arraystretch}{1.5}
\begin{table}[t]
\centering
\caption{\label{tab3:comparemodels} Shows the WAIC of the 4 considered models from Section \ref{m3:appfitting} fitted to the Rio data. The best fitting model, the one with the lowest WAIC, is bolded.}
\begin{tabular}{l|llll}
\hline
\textbf{Model}   & Zeng (2022) & ARMN & ZIARMN & \textbf{MS-ZIARMN} \\ 
\textbf{WAIC}     &  23,986 &   23,974 &  22,756 & \textbf{22,091}   \\
\hline
\end{tabular}
\end{table}

\subsection{Results} \label{m3:results}

Table \ref{tab3:multinomest} shows the estimated coefficients from the multinomial part of the fitted MS-ZIARMN model, that is, the intercepts and covariate effects from Equation (\ref{eqn:log_linear}). From the intercept row, at average values of the covariates, Zika transmission was favored over dengue and chikungunya transmission. That is, under average conditions, the share of Zika cases tended to grow over time in an area (at least from an equal share, see the nuances in Section \ref{m3:model}). It is interesting to compare these estimated intercepts with those from the ARMN and Zeng (2022) models in SM Table 1. Unlike the MS-ZIARMN model, the Zeng (2022) and ARMN models both estimated that, on average, Zika and chikungunya transmission was much less intense in Rio compared to dengue transmission. This does not correspond to our knowledge of the epidemiology of these diseases and is likely due to a failure to properly account for the many zeroes. Zika is generally considered more transmissible due to \emph{Aedes aegypti} transmitting Zika at a higher rate \citep{freitas_spacetime_2019}. This further illustrates it is important to account for zero inflation and to model the zero inflation properly (e.g., account for correlations).

\begin{table}[t]
\centering
\caption{\label{tab3:multinomest} Posterior means and 95\% posterior credible intervals (in parentheses) for the estimated coefficients from the multinomial part of the fitted MS-ZIARMN model. The intercept row shows $\lambda_{kit}$ for $k=2$ (Zika) and $k=3$ (chik.) in a typical area at average values of the covariates. Recall, if $\lambda_{kit}>1$ ($\lambda_{kit}<1$) then the share of disease $k$ relative to dengue will tend to grow (shrink) over time. Other rows show the ratio of $\pi_{kit}/\pi_{1it}$ (relative odds ratio) or the ratio of $\lambda_{kit}$ (rate ratio) (both are the same, see Equations (\ref{eqn:relative_odds})-(\ref{eqn:log_linear})) corresponding to a unit increase in the covariate. All covariates are standardized. Significant effects are bolded. See Section \ref{m3:appfitting} for an explanation of the covariates.}

\begin{tabular}{lccc} 
 \hline                             &       & \multicolumn{2}{c}{\textbf{Relative Odds Ratio or Rate Ratio}}            \\ \hline 
      \textbf{Covariates}  &    {\color{black}\textbf{Parameter}}                & \textbf{Zika-dengue} & \textbf{chik.-dengue}   \\  \hline
Intercept  & {\color{black}$\exp({\alpha_0})$}  &1.14 (1.03, 1.26)   & 1.02 (.91, 1.13)    \\
$\text{verde}_i$& {\color{black}$\exp({\alpha_{\text{verde}}})$}& 1.02 (.95, 1.09)  & .92 (.85, 1) \\
$\text{SDI}_i$& {\color{black}$\exp({\alpha_{\text{SDI}}})$}& 1.07 (.99, 1.15)    & 1.01 (.92, 1.11)\\
$\text{popdens}_i$&{\color{black}$\exp({\alpha_{\text{popdens}}})$} & 1.02 (.94, 1.11)  & 1.06 (.96, 1.16)  \\
$\text{favela}_i$&{\color{black}$\exp({\alpha_{\text{favela}}})$} & .98 (.91, 1.05) & .94 (.87, 1.03) \\ 
$\text{temp}_{it}$&{\color{black}$\exp({\alpha_{\text{temp}}})$} & \textbf{1.14 (1.09, 1.20)} & \textbf{.85 (.80, .90)} \\
Neighborhood dengue prevalence&{\color{black}$\exp({\alpha_{\text{NeiDeng}}})$} & \textbf{.70 (.67, .74)} & \textbf{.70 (.67, .74)} \\
 Neighborhood Zika prevalence&{\color{black}$\exp({\alpha_{\text{NeiZika}}})$} & \textbf{1.59 (1.48, 1.70)} & -- \\
Neighborhood chik. prevalence&{\color{black}$\exp({\alpha_{\text{NeiChik}}})$} & -- & \textbf{1.43 (1.36, 1.50)} \\
Previous Zika cases&{\color{black}$\exp({\alpha_{\text{Zika}}})$} & -- & \textbf{.90 (.85, .96)} \\
Previous chik. cases&{\color{black}$\exp({\alpha_{\text{Chik}}})$} & .98 (.95, 1.02) & -- \\
\hline
\end{tabular}
\end{table}

Interestingly, unlike in \cite{schmidt_poisson-multinomial_2022}, we did not find strong evidence that any of the spatial covariates were important for explaining differences in the transmission of the diseases. We found that Zika transmission was favored at warmer temperatures and chikungunya transmission was favored at colder temperatures, compared to dengue transmission. Both of these findings have been observed in laboratory studies \citep{tesla_temperature_2018,mercier_impact_2022}. Indeed, a concern with chikungunya is that it could spread into colder areas of Europe that are unreachable by dengue \citep{mercier_impact_2022}. To better quantify 
the temperature effects, we plot $\lambda_{kit}$, from Equation (\ref{eqn:relative_odds}), versus temperature for $k=2$ (Zika) and $k=3$ (chikungunya) in Figure \ref{fig3:fig2}. From the figure, we have strong evidence that, under otherwise average conditions, Zika transmission was favored above 33.2 degrees and chikungunya transmission was favored below 32 degrees, compared to dengue transmission (the average temperature in Rio during the study was 33.8 degrees, with the middle 95 percent being from 27.2 to 39.1 degrees).

 \begin{figure}[!t]
 	\centering
 	\includegraphics[width=\textwidth]{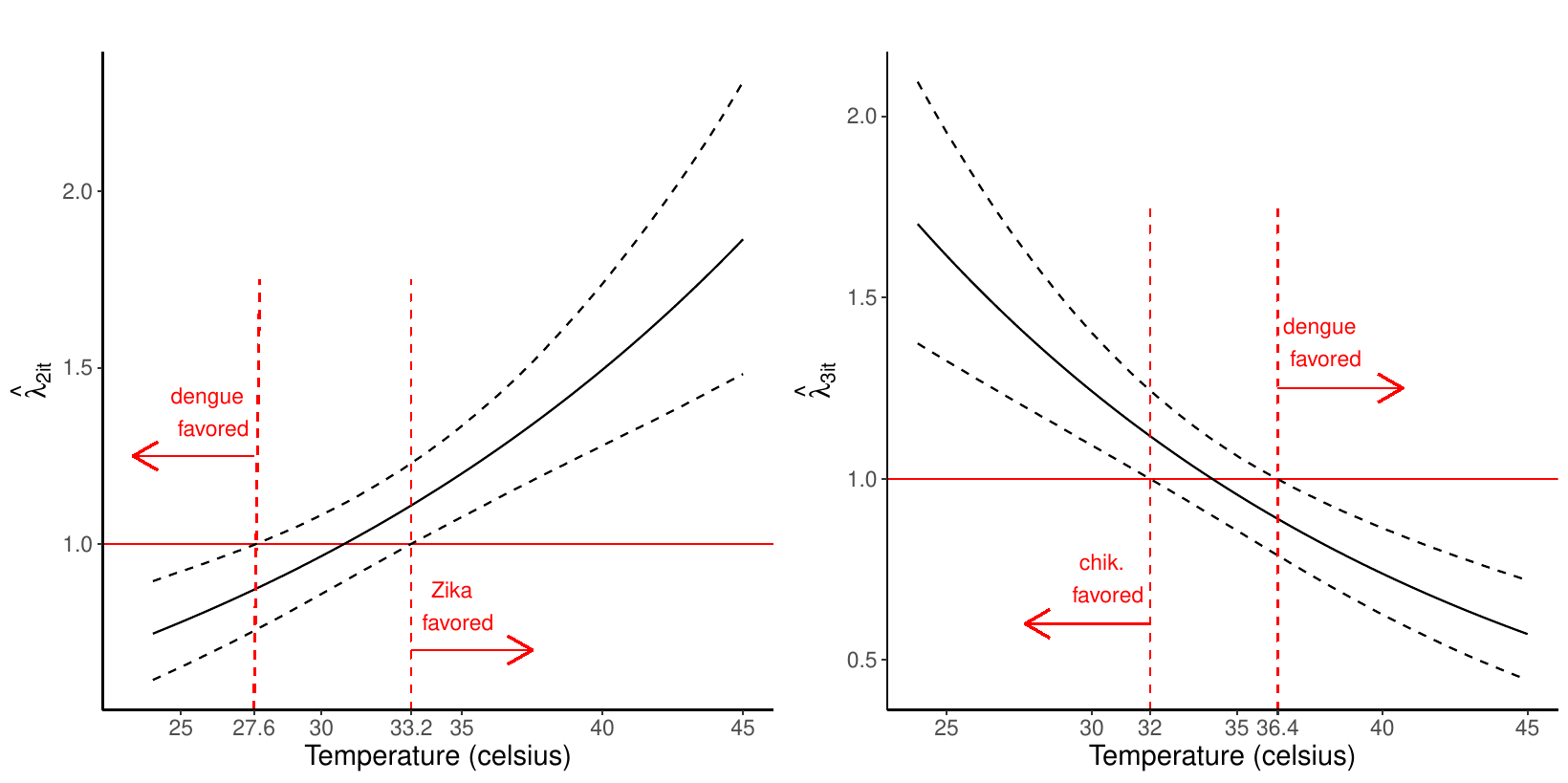}
	\caption{Shows the posterior means (black solid line) and 95\% posterior credible intervals (black dashed lines) of $\lambda_{kit}$ versus temperature for $k=2$ (Zika) (left) and $k=3$ (chikungunya) (right). We assumed an average area, other covariates fixed at their average values and a null space-time random effect. Recall, if $\lambda_{kit}>1$ ($\lambda_{kit}<1$) then the share of disease $k$ relative to dengue will tend to grow (shrink) over time. The red horizontal dashed lines are drawn at temperatures below and above which  $\lambda_{kit}$ is significantly different from 1.\label{fig3:fig2}} 
\end{figure} 

The neighborhood prevalence effects are by far the strongest in Table \ref{tab3:multinomest} for explaining differences in the relative distributions of the diseases. For instance, we estimated that a standard deviation increase in the log of the prevalence of Zika across neighboring areas during the previous week was associated with a 59 (48, 70) percent increase in the odds of a new case being Zika relative to dengue. As \emph{Aedes aegypti} primarily bite during the day, its arboviruses can spread rapidly between areas \citep{stoddardHousetohouseHumanMovement2013}. We found that areas with more Zika cases were less favorable towards the transmission of chikungunya relative to dengue. That is, we estimated that a standard deviation increase in the log of the previous number of Zika cases in an area was associated with a 10 (4, 15) percent decrease in the odds of a new case being chikungunya relative to dengue. This does not necessarily mean that Zika transmission inhibited chikungunya transmission in a non-relative sense. It could be that Zika transmission encouraged both dengue and chikungunya transmission, but the effect on dengue transmission was larger. Therefore, this result needs to be interpreted with some caution.

We estimated $\zeta_1$, $\zeta_2$ and $\zeta_3$ from Equation (\ref{eqn:relative_odds}) as .44 (.38, .49), .36 (.31, .32) and .43 (.36, .51), respectively, indicating a high degree of nonhomogeneous mixing. That is, the model predicts that if one of the diseases becomes very dominant in a neighborhood, the relative distribution of the diseases will move towards a uniform distribution, perhaps because individuals avoid infection from the dominant disease. However, as avoiding one of the diseases should help avoid the others, as they are transmitted by the same vector, $\zeta_k$ could also be mainly reflecting differences in disease immunity, which are difficult to account for explicitly (see SM Section 9). From Equations (\ref{eqn:log_linear})-(\ref{eqn:logistic_normal}), we estimated the standard deviation of $\phi_{2it}$ as .75 (.71, .79), of $\phi_{3it}$ as .80 (.75, .86) and the correlation between $\phi_{2it}$ and $\phi_{3it}$ as .59 (.52, .65). This implies that the correlations between the disease cases, marginalizing the random effects and conditional on their total, is about what we would expect without the random effects, see the simulation study in SM Section 2. That is, the random effects do not induce any excess correlations between the disease counts. However, they do induce a large amount of overdispersion, as seen in Figure \ref{fig3:fig4}.

\begin{table}[t]
\centering
\caption{\label{tab3:markovest} Posterior means and 95\% posterior credible intervals (in parentheses) for the estimated parameters from the Markov chain part of the fitted MS-ZIARMN model. The intercept row shows the probability of Zika presence and the logit of the probability of chikungunya presence, assuming no previous disease presence in the area, no cases in neighboring areas and at average values of the other covariates. The Zika column shows the ratio of $p_{2it}/(1-p_{2it})$ (odds of Zika presence) corresponding to a unit increase in the covariate. The chik. column shows the untransformed estimates (not exponentiated, see text). All covariates are standardized except the neighborhood prevalences, which are given in units of 1 case per 25,000 (average combined size of neighboring areas) at small prevalences.}

\begin{tabular}{lccc} 
 \hline                           &         & {\textbf{Odds Ratio}}   &  {\textbf{Untransformed}}        \\ \hline 
             \textbf{Covariate}  & {\color{black}\textbf{Parameter}}              & \textbf{Zika presence} & \textbf{chik. presence}  \\[-3pt]
             &{\color{black}$\text{Zika col.}\,\,\,|\,\,\,\text{chik. col.}$}&& \\\hline
Intercept  &{\color{black}$\exp(\eta_0)/(1+\exp(\eta_0))\,\,\,|\,\,\,\eta_{0}$} & .1 (.08, .12)  & -9.06 (-11.97, -7)    \\
$\text{verde}_i$&{\color{black}$\exp(\eta_{\text{verde}})\,\,\,|\,\,\,\eta_{\text{verde}}$} & 1.14 (.98, 1.31)  & 08 (-.20 .37)  \\
$\text{SDI}_i$& {\color{black}$\exp(\eta_{\text{SDI}})\,\,\,|\,\,\,\eta_{\text{SDI}}$}& 1.11 (.96, 1.29)    & .12 (-.18, .42)\\
$\text{popdens}_i$&{\color{black}$\exp(\eta_{\text{popdens}})\,\,\,|\,\,\,\eta_{\text{popdens}}$} & 1 (.85, 1.19)  & .02 (-.33, .37)  \\
$\text{favela}_i$& {\color{black}$\exp(\eta_{\text{favela}})\,\,\,|\,\,\,\eta_{\text{favela}}$}& 1.03 (.89, 1.19) & 0 (-.3, .29) \\ 
$\text{temp}_{it}$&{\color{black}$\exp(\eta_{\text{temp}})\,\,\,|\,\,\,\eta_{\text{temp}}$} & \textbf{1.44 (1.24, 1.66)} & \textbf{1.11 (.6, 1.7)} \\ 
Neighborhood Zika prevalence&{\color{black}$\exp(\eta_{\text{NeiZika}})\,\,\,|\,\,\,\eta_{\text{NeiZika}}$} & \textbf{1.70 (1.55, 1.87)} & -- \\
Neighborhood chik. prevalence&{\color{black}$\exp(\eta_{\text{NeiChik}})\,\,\,|\,\,\,\eta_{\text{NeiChik}}$} & -- & \textbf{1.09 (.85, 1.35)} \\
Previous Zika presence&{\color{black}$\exp(\rho_2^{AR})\,\,\,|\,\,\,\rho_{23}^{DI}$} & \textbf{6.41 (3.95, 9.94)} & \textbf{5.12 (3.2, 8)} \\
Previous chik. presence& {\color{black}$\exp(\rho_{32}^{DI})\,\,\,|\,\,\,\rho_{3}^{AR}$} & \textbf{3.46 (2.42, 4.98)} & \textbf{11.55 (8.13, 16.12)} \\
\hline
\end{tabular}
\end{table}

Table \ref{tab3:markovest} shows the estimated parameters from the Markov chain component of the model, that is, from Equation (\ref{eqn:presence}). The Zika column shows the ratio of the odds of Zika presence corresponding to a unit increase in the covariate. However, the intercept for chikungunya is highly negative, making the odds ratios difficult to interpret. Therefore, we did not transform the chikungunya estimates. From the intercept row, it was very unlikely for chikungunya to emerge in an area under average conditions. Indeed, chikungunya emergence was only likely after Zika had emerged and temperatures rose in the next summer. In contrast, the emergence of Zika was much more random, occurring with a 10 (8, 12) percent chance bi-weekly under average conditions. Once chikungunya did emerge, it was likely to stay in the area (probability of .88 (.63, .99) assuming no neighboring cases, no Zika presence in the previous bi-week and otherwise average conditions). {\color{black} In contrast, Zika persistence was more reliant on the presence of cases in neighboring areas (there was only a 41 (32, 49) percent chance of Zika being present if it was present previously and there were no neighboring cases).} Both Zika and chikungunya were more likely to be present at high temperatures and when there were cases in neighboring areas. The effects of neighborhood prevalence are especially large, again illustrating the diseases' high rate of geographical spread. The presence of Zika was an indicator of chikungunya presence, and vice versa. This could be unrelated to disease interactions and due to unmeasured confounding, as the diseases share the same vector. 

{\color{black}Some of the chikungunya estimates in Table \ref{tab3:markovest} are extreme. Chikungunya cases were first reported in most neighborhoods between bi-weeks 28 and 34 (measured from the beginning of 2015). This implies a significant change in the probability of chikungunya presence around bi-week 28 across all neighborhoods. Therefore, the extreme estimates in Table \ref{tab3:markovest} are likely due to the
model attempting to fit this large and fast shift in the probability of chikungunya presence. In light of the above, the chikungunya estimates in Table \ref{tab3:markovest} should be interpreted with caution, as there may be unmeasured confounders driving the change in the risk of chikungunya presence around bi-week 28. For example, spread from travel outside the city \citep{nunes2015emergence}, which is difficult to measure and may be related to factors such as temperature. In contrast, Zika cases emerged much more randomly across the city, as reflected in Table \ref{tab3:markovest}.}

\begin{figure}[!t]
 	\centering
 	\includegraphics[width=.8\textwidth]{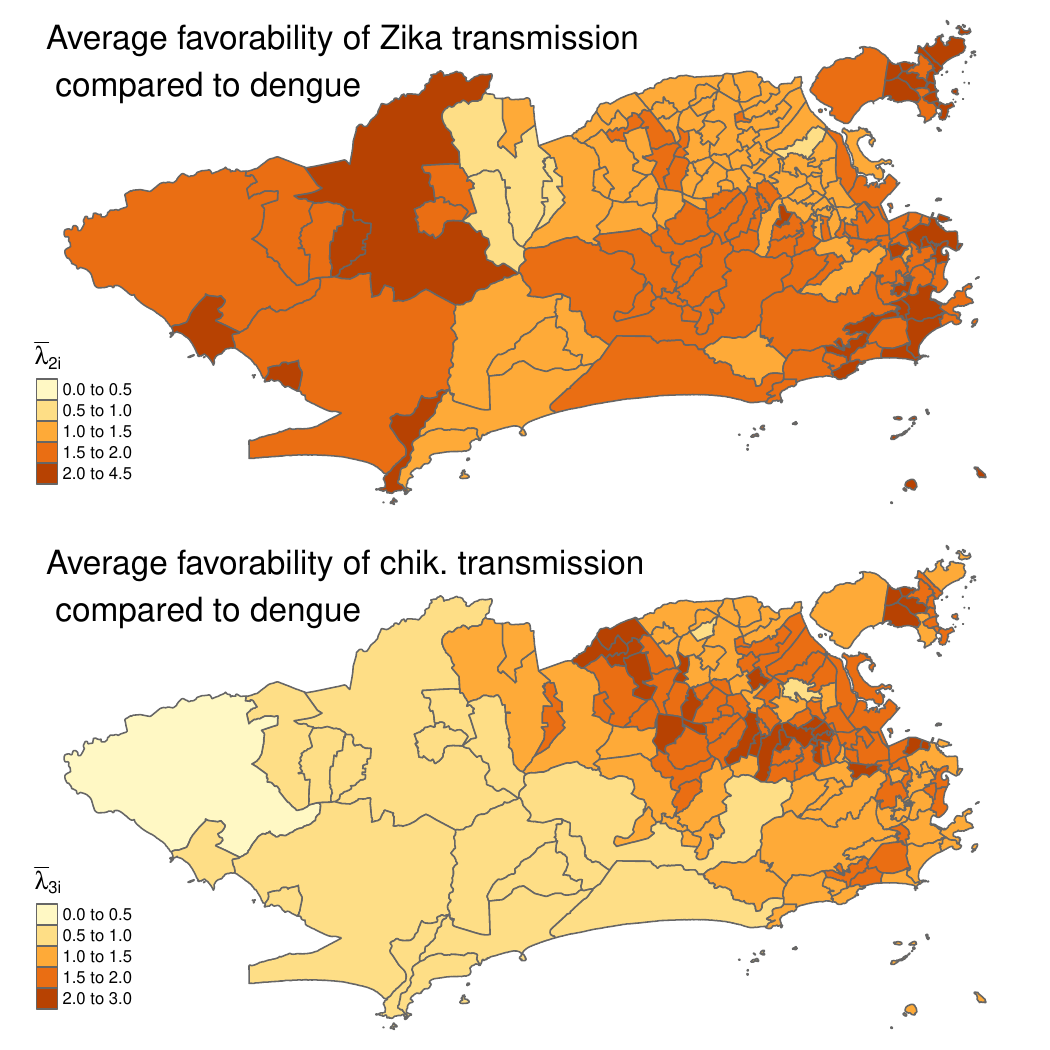}
	\caption{Shows the posterior mean of ${\bar{\lambda}_{ki}=\left(1/\sum_{t=2}^{T}S_{kit}\right)\sum_{t=2}^{T}\lambda_{kit}S_{kit}}$ for $k=2$ (Zika) (top map) and $k=3$ (chik.) (bottom map). If $\bar{\lambda}_{ki}>1$ ($\bar{\lambda}_{ki}<1$) then the transmission of disease $k$ was favored (disfavored) on average over dengue when it was present in the neighborhood, and the share of disease $k$ tended to grow (shrink) relative to dengue in the neighborhood.\label{fig3:fig3}} 
\end{figure}

The maps in Figure \ref{fig3:fig3} show the posterior mean of ${\bar{\lambda}_{ki}=\left(1/\sum_{t=2}^{T}S_{kit}\right)\sum_{t=2}^{T}\lambda_{kit}S_{kit}}$ (average value of $\lambda_{kit}$ when disease $k$ was present in neighborhood $i$) for $k=2$ (Zika) (top map) and $k=3$ (chik.) (bottom map). From the bottom map, chikungunya transmission was disfavored in the west of the city when it was present, and the share of chikungunya cases tended to decline there. This was partly due to temperatures being 2-3 degrees higher on average in the west (see SM Figure 2) and to chikungunya transmission being disfavored at warmer temperatures relative to the other two diseases (see Table \ref{tab3:multinomest}). Also, there were many Zika cases in the west of the city, and chikungunya transmission was disfavored in areas with many Zika cases. Zika transmission was favored throughout most of the city when it was present. This can be partly explained by the inherent favorability of Zika, i.e., the intercept in Table \ref{tab3:multinomest}; by temperatures typically being higher in Rio when Zika was present (the temperature covariate was centered using the overall average temperature); and by Zika transmission being favored at higher temperatures.

\begin{figure}[!t]
 	\centering
 	\includegraphics[width=\textwidth]{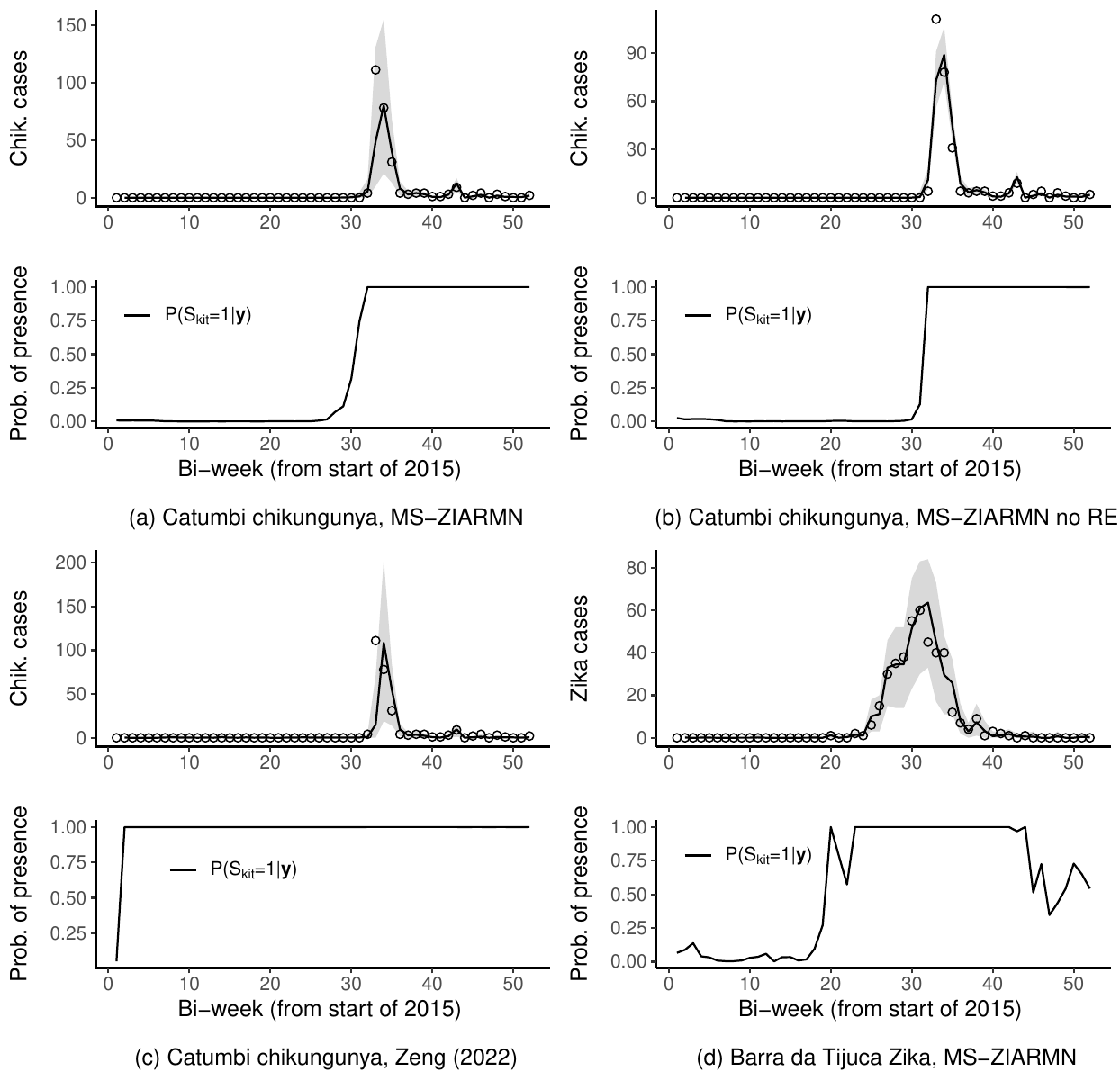}
	\caption{Top graphs show the posterior means (solid lines) and 95\% posterior credible intervals (shaded areas) of the fitted values $y_{kit}'$, see SM Section 6. The bottom graphs show the posterior probability that the disease was present in each bi-week, $P(S_{kit}=1|\bm{y})$ versus $t$. "MS-ZIARMN no random effect" refers to the MS-ZIARMN model fit without including $\phi_{kit}$ from Equations (\ref{eqn:log_linear})-(\ref{eqn:logistic_normal}). \label{fig3:fig4}} 
\end{figure}

Figure \ref{fig3:fig4} shows fitted values {\color{black}in the top graphs of each panel} from three models. SM Section 6 explains how we constructed the fitted values by drawing new space-time random effects and simulating from the fitted model. {\color{black}The bottom graphs of each panel in Figure \ref{fig3:fig4} show the posterior probability the disease was present in the area, $P(S_{kit}=1|\bm{y})$; see SM Section 6.}  {\color{black}If an area reports cases of disease $k$, then we know that it is present there, that is, $P(S_{kit}=1|\bm{y})=1$ if $y_{kit}>0$. If an area does not report cases, the disease could be absent or present but undetected, the probability of the latter is given by $P(S_{kit}=1|\bm{y})$.}  By comparing panels (a) and (b), the random effects $\phi_{kit}$, from Equations (\ref{eqn:log_linear})-(\ref{eqn:logistic_normal}), are an important component of the model and induce a large amount of necessary overdispersion. From panel (c), the model of \cite{zeng_zero-inflated_2022}, which does not account for correlations in disease presence, estimates chikungunya as always present in Catumbi, except for the initial state (this is also true for Zika and in all other areas). Therefore, the model presented in \cite{zeng_zero-inflated_2022} is too restrictive to capture the likely complex space-time distribution of disease presence in the city. {\color{black}Panel (d) shows Zika likely circulated undetected at the end of 2016}. One way to think of $P(S_{kit}=1|\bm{y})$ (from the bottom graphs) is as a weight for the observation's contribution to the likelihood of the multinomial parameters of the model. For example, from panel (a), the MS-ZIARMN model does not use any chikungunya observations before around week 28 in Catumbi to inform the multinomial estimates in Table \ref{tab3:multinomest}. In contrast, the model of \cite{zeng_zero-inflated_2022}, from panel (c), will use all observations to inform the multinomial model, which likely causes bias due to the large number of 0s, see the discussion at the beginning of Section \ref{m3:results} {\color{black}and our simulation study in SM Section 4.}

The model takes disease absence to mean there was no chance of cases being reported. However, if the diseases are present in very small amounts, the chance of cases being reported could be close enough to 0 that it is indistinguishable. Indeed, if we replaced the 0s in Equation (\ref{eqn:mixture}) by say .005, and normalized the probabilities, we would likely get similar parameter estimates. Therefore, disease presence should be interpreted closer to "a feasible chance of cases being reported" than "the presence of at least one infection".

\section{Discussion} \label{m3:discussion}

We have proposed a zero-inflated spatio-temporal multinomial model for comparing the transmission dynamics of multiple co-circulating infectious diseases. Our approach can account for long periods of disease absence while investigating how certain factors are related to differences in the transmission intensity of the diseases. Past zero-inflated multinomial models \citep{tang_zero-inflated_2019,zeng_zero-inflated_2022,koslovsky_bayesian_2023} have made
independence assumptions that are inappropriate for spatio-temporal infectious disease counts, such as assuming categorical presence was independent between observations. We accounted for correlations across space, time and disease in both the multinomial and zero-inflated components through a combination of autoregression, regressing on past observations, and by assuming the presence of the diseases followed a series of coupled Markov chains. This approach allowed for efficient and computationally feasible Bayesian inference using the FFBS algorithm. {\color{black} We also showed, in a simulation study in SM Section 4, that failing to account for zero-inflation often leads to many important parameters not being covered by the credible intervals, which further lends support to our approach.}

We assume that the cases of the diseases present in an area jointly follow a multinomial model, conditioning on their total. It is more standard in the disease mapping literature to jointly model multiple disease counts with some form of a multivariate Poisson distribution \citep{jack_estimating_2019}. There are advantages to both approaches and we give a more detailed comparison of the two (without model fitting) in SM Section 1.2. If one is only interested in comparing the transmission of the diseases, like here, then a multinomial model eliminates many nuisance parameters and any shared space-time factors. A multivariate Poisson approach could estimate the overall effects of a covariate on transmission, not just the relative effects, and does not need to assume one of the diseases is always present. As mentioned in the introduction, only \cite{pavani_bayesian_2022} and \cite{rotejanaprasert_spatiotemporal_2021} have considered zero-inflated spatio-temporal multivariate Poisson models. Both relied on random effects to capture correlations in the data. However, these proposals have some important limitations, including not allowing the probabilities of disease presence to change over time. These modeling assumptions are inappropriate for our motivating example and likely for others as well. For instance, Figure \ref{fig3:fig4} clearly shows the probabilities of disease presence changing over time. An approach based on random effects could become intractable as one would need to consider random effects correlated across three dimensions, space, time and disease, in both the count and zero-inflated components of the model. Our approach seems much more computationally feasible and could be easily extended to the multivariate Poisson case, see SM Section 1.2. We will focus on these extensions and comparisons with the multinomial model in future work.

In our application, we found that Zika generally had more intense transmission compared to dengue and chikungunya, but was also not able to transmit as well at a lower temperature. This was not a major factor in tropical Rio, see Figure \ref{fig3:fig3}, but could be important in North America and Europe. In contrast, chikungunya transmission was relatively higher at lower temperatures, meaning it might fare better in colder regions compared to Zika and dengue. Laboratory studies have come to similar conclusions \citep{tesla_temperature_2018,mercier_impact_2022}. Alternative models that did not account for zero inflation or did not model correlations in disease presence \citep{zeng_zero-inflated_2022} did not fit as well and produced estimates inconsistent with our knowledge about the epidemiology of the diseases.

There are also some important limitations of our work. Being an observational study, unmeasured confounding could be an important issue, especially concerning differences in the susceptible populations, which are unobservable. We tried using differences in cumulative incidence as a proxy for differences in the susceptible populations, in a sensitivity analysis in SM Section 9, and our results did not change. However, differences in cumulative incidence could be an imperfect proxy due to, for example, changes in reporting rates across space and time. Another limitation is that weather covariates, such as temperature, can have lagged and non-linear effects on the transmission of arboviral diseases \citep{lowe_nonlinear_2018}. As our multinomial model estimates differences in effects on transmission, our approach might be somewhat robust to non-linearity if the non-linear patterns are similar between diseases. It would be challenging to incorporate non-linear and lagged effects into our already complex modeling scheme. Shrinkage methods could help deal with the many parameters that would need to be added to the model by eliminating unimportant effects \citep{wangBayesianNonHomogeneousHidden2022}. {\color{black} Also, the effect of weather covariates on the incidence of arboviral diseases may change across space or time \citep{nobre2005spatio}. To accommodate this, the covariate effects could be modeled as random walks \citep{nobre2005spatio} or other spatio-temporal random effects in future work. However, this would be computationally challenging since covariates can enter the model in each of the relative odds and each of the probabilities of presence.} {\color{black} Finally, another limitation of our approach is that a multinomial model is not appropriate for predicting future disease counts. To predict future cases of a disease, a multinomial model would need to know the total number of cases across all diseases in the future. A possible solution is to build a separate model for the total count \citep{schmidt_poisson-multinomial_2022}, or a multivariate Poisson model could be used, as discussed above.}

\section*{Acknowledgements}
This work is part of the PhD thesis of D. Douwes-Schultz under the supervision of A. M. Schmidt in the Graduate Program of Biostatistics at McGill University, Canada. Douwes-Schultz is grateful for financial support from IVADO and the Canada First Research Excellence Fund/Apogée (PhD Excellence Scholarship 2021-9070375349). Schmidt is grateful for financial support from the Natural Sciences and Engineering Research Council (NSERC) of Canada (Discovery Grant RGPIN-2017-04999) and from Institut de valorisation des données (IVADO) (PRF-2019-6839748021). L. P. Freitas is supported by the Visiting Researcher Program of the Sergio Arouca National School of Public Health (ENSP), Oswaldo Cruz Foundation (Fiocruz). M. S. Carvalho acknowledges support from the National Council for Scientific and Technological Development (CNPq) - Brazil [307450/2021-0] and Fundação Carlos Chagas Filho de Amparo à Pesquisa do Estado do Rio de Janeiro (FAPERJ) - Brazil [E-26/203.967/2024]. This research was enabled in part by support provided by Calcul Québec (www.calculquebec.ca) and Compute Canada (www.computecanada.ca).

\bibliography{ms}

\begin{thebibliography}{62}
\expandafter\ifx\csname natexlab\endcsname\relax\def\natexlab#1{#1}\fi
\expandafter\ifx\csname url\endcsname\relax
  \def\url#1{\texttt{#1}}\fi
\expandafter\ifx\csname urlprefix\endcsname\relax\def\urlprefix{URL: }\fi

\bibitem[{Abbey(1952)}]{abbey_examination_1952}
Abbey, H. (1952) An {examination} of the {Reed}-{Frost} {theory} of {epidemics}.
\newblock \textit{Human Biology}, \textbf{24}, 201--233.

\bibitem[{Adams and Boots(2010)}]{adamsHowImportantVertical2010}
Adams, B. and Boots, M. (2010) How important is vertical transmission in mosquitoes for the persistence of dengue? {Insights} from a mathematical model.
\newblock \textit{Epidemics}, \textbf{2}, 1--10.

\bibitem[{Aktekin and Musal(2015)}]{aktekinAnalysisIncomeInequality2015}
Aktekin, T. and Musal, M. (2015) Analysis of income inequality measures on human immunodeficiency virus mortality: A spatiotemporal {{Bayesian}} perspective.
\newblock \textit{Journal of the Royal Statistical Society: Series A (Statistics in Society)}, \textbf{178}, 383--403.

\bibitem[{Arab(2015)}]{arabSpatialSpatioTemporalModels2015}
Arab, A. (2015) Spatial and spatio-temporal models for modeling epidemiological data with excess zeros.
\newblock \textit{International Journal of Environmental Research and Public Health}, \textbf{12}, 10536--10548.

\bibitem[{Auger-Méthé et~al.(2021)Auger-Méthé, Newman, Cole, Empacher, Gryba, King et~al.}]{auger-methe_guide_2021}
Auger-Méthé, M., Newman, K., Cole, D., Empacher, F., Gryba, R., King, A.~A. et~al. (2021) A guide to state–space modeling of ecological time series.
\newblock \textit{Ecological Monographs}, \textbf{91}, e01470.

\bibitem[{Bartlett(1957)}]{bartlettMeaslesPeriodicityCommunity1957}
Bartlett, M.~S. (1957) Measles periodicity and community size.
\newblock \textit{Journal of the Royal Statistical Society: Series A (General)}, \textbf{120}, 48--60.

\bibitem[{Bauer and Wakefield(2018)}]{bauerStratifiedSpaceTime2018}
Bauer, C. and Wakefield, J. (2018) Stratified space\textendash time infectious disease modelling, with an application to hand, foot and mouth disease in {{China}}.
\newblock \textit{Journal of the Royal Statistical Society Series C (Applied Statistics)}, \textbf{67}, 1379--1398.

\bibitem[{Besag et~al.(1991)Besag, York and Molli{\'e}}]{besag1991bayesian}
Besag, J., York, J. and Molli{\'e}, A. (1991) Bayesian image restoration, with two applications in spatial statistics.
\newblock \textit{Annals of the Institute of Statistical Mathematics}, \textbf{43}, 1--20.

\bibitem[{Bisanzio et~al.(2018)Bisanzio, Dzul-Manzanilla, Gomez-Dantés, Pavia-Ruz, Hladish, Lenhart et~al.}]{bisanzio_spatio-temporal_2018}
Bisanzio, D., Dzul-Manzanilla, F., Gomez-Dantés, H., Pavia-Ruz, N., Hladish, T.~J., Lenhart, A. et~al. (2018) Spatio-temporal coherence of dengue, chikungunya and {Zika} outbreaks in {Merida}, {Mexico}.
\newblock \textit{PLOS Neglected Tropical Diseases}, \textbf{12}, e0006298.

\bibitem[{Bracher and Held(2021)}]{bracher_marginal_2021}
Bracher, J. and Held, L. (2021) A marginal moment matching approach for fitting endemic-epidemic models to underreported disease surveillance counts.
\newblock \textit{Biometrics}, \textbf{77}, 1202--1214.

\bibitem[{Bracher and Held(2022)}]{bracher_endemic-epidemic_2022}
--- (2022) Endemic-epidemic models with discrete-time serial interval distributions for infectious disease prediction.
\newblock \textit{International Journal of Forecasting}, \textbf{38}, 1221--1233.

\bibitem[{Chib(1996)}]{chibCalculatingPosteriorDistributions1996}
Chib, S. (1996) Calculating posterior distributions and modal estimates in {{Markov}} mixture models.
\newblock \textit{Journal of Econometrics}, \textbf{75}, 79--97.

\bibitem[{Coutinho et~al.(2006)Coutinho, Burattinia, Lopeza and Massada}]{coutinhoThresholdConditionsNonAutonomous2006}
Coutinho, F. A.~B., Burattinia, M.~N., Lopeza, L.~F. and Massada, E. (2006) Threshold conditions for a non-autonomous epidemic system describing the population dynamics of dengue.
\newblock \textit{Bulletin of Mathematical Biology}, \textbf{68}, 2263--2282.

\bibitem[{Dabney and Wakefield(2005)}]{dabney_issues_2005}
Dabney, A.~R. and Wakefield, J.~C. (2005) Issues in the mapping of two diseases.
\newblock \textit{Statistical Methods in Medical Research}, \textbf{14}, 83--112.

\bibitem[{{Douwes-Schultz} and Schmidt(2022)}]{douwes-schultzZerostateCoupledMarkov2022a}
{Douwes-Schultz}, D. and Schmidt, A.~M. (2022) Zero-state coupled {Markov} switching count models for spatio-temporal infectious disease spread.
\newblock \textit{Journal of the Royal Statistical Society: Series C (Applied Statistics)}, \textbf{71}, 589--612.

\bibitem[{Dreassi(2007)}]{dreassi_polytomous_2007}
Dreassi, E. (2007) Polytomous {disease} {mapping} to {detect} {uncommon} {risk} {factors} for {related} {diseases}.
\newblock \textit{Biometrical Journal}, \textbf{49}, 520--529.

\bibitem[{Fernandes et~al.(2009)Fernandes, Schmidt and Migon}]{fernandesModellingZeroinflatedSpatiotemporal2009}
Fernandes, M. V.~M., Schmidt, A.~M. and Migon, H.~S. (2009) Modelling zero-inflated spatio-temporal processes.
\newblock \textit{Statistical Modelling}, \textbf{9}, 3--25.

\bibitem[{Fokianos and Kedem(2003)}]{fokianos_regression_2003}
Fokianos, K. and Kedem, B. (2003) Regression {theory} for {categorical} {time} {series}.
\newblock \textit{Statistical Science}, \textbf{18}, 357--376.

\bibitem[{Freitas et~al.(2019)Freitas, Cruz, Lowe and Sá~Carvalho}]{freitas_spacetime_2019}
Freitas, L.~P., Cruz, O.~G., Lowe, R. and Sá~Carvalho, M. (2019) Space–time dynamics of a triple epidemic: dengue, chikungunya and {Zika} clusters in the city of {Rio} de {Janeiro}.
\newblock \textit{Proceedings of the Royal Society B: Biological Sciences}, \textbf{286}, 20191867.

\bibitem[{Fritz and Kauermann(2022)}]{fritz_interplay_2022}
Fritz, C. and Kauermann, G. (2022) On the interplay of regional mobility, social connectedness and the spread of {COVID}-19 in {Germany}.
\newblock \textit{Journal of the Royal Statistical Society: Series A (Statistics in Society)}, \textbf{185}, 400--424.

\bibitem[{{Fr{\"u}hwirth-Schnatter}(2006)}]{fruhwirth-schnatterFiniteMixtureMarkov2006}
{Fr{\"u}hwirth-Schnatter}, S. (2006) \textit{Finite {{Mixture}} and {{Markov Switching Models}}}.
\newblock Springer {{Series}} in {{Statistics}}. {New York}: {Springer-Verlag}.

\bibitem[{Gelman et~al.(2013)Gelman, Carlin, Stern, Dunson, Vehtari and Rubin}]{gelman_bayesian_2013}
Gelman, A., Carlin, J.~B., Stern, H.~S., Dunson, D.~B., Vehtari, A. and Rubin, D.~B. (2013) \textit{Bayesian {Data} {Analysis}}.
\newblock Boca Raton: Chapman and Hall/CRC, 3rd edn.

\bibitem[{Gelman et~al.(2014)Gelman, Hwang and Vehtari}]{gelmanUnderstandingPredictiveInformation2014}
Gelman, A., Hwang, J. and Vehtari, A. (2014) Understanding predictive information criteria for {{Bayesian}} models.
\newblock \textit{Statistics and Computing}, \textbf{24}, 997--1016.

\bibitem[{Giorgi et~al.(2018)Giorgi, Schl{\"u}ter and Diggle}]{giorgiBivariateGeostatisticalModelling2018}
Giorgi, E., Schl{\"u}ter, D.~K. and Diggle, P.~J. (2018) Bivariate geostatistical modelling of the relationship between {{Loa}} loa prevalence and intensity of infection.
\newblock \textit{Environmetrics}, \textbf{29}, e2447.

\bibitem[{Grenfell et~al.(2001)Grenfell, Bj{\o}rnstad and Kappey}]{grenfellTravellingWavesSpatial2001}
Grenfell, B.~T., Bj{\o}rnstad, O.~N. and Kappey, J. (2001) Travelling waves and spatial hierarchies in measles epidemics.
\newblock \textit{Nature}, \textbf{414}, 716--723.

\bibitem[{Hall(2000)}]{hall_zero-inflated_2000}
Hall, D.~B. (2000) Zero-{inflated} {Poisson} and {binomial} {regression} with {random} {effects}: {A} {case} {study}.
\newblock \textit{Biometrics}, \textbf{56}, 1030--1039.

\bibitem[{Hamilton(1989)}]{hamiltonNewApproachEconomic1989}
Hamilton, J.~D. (1989) A new approach to the economic analysis of nonstationary time series and the business cycle.
\newblock \textit{Econometrica}, \textbf{57}, 357--384.

\bibitem[{Hoef and Jansen(2007)}]{hoefSpaceTimeZeroinflated2007}
Hoef, J. M.~V. and Jansen, J.~K. (2007) Space\textemdash time zero-inflated count models of harbor seals.
\newblock \textit{Environmetrics}, \textbf{18}, 697--712.

\bibitem[{Jack et~al.(2019)Jack, Lee and Dean}]{jack_estimating_2019}
Jack, E., Lee, D. and Dean, N. (2019) Estimating the {changing} {nature} of {Scotland}'s {health} {inequalities} by using a {multivariate} {spatiotemporal} {model}.
\newblock \textit{Journal of the Royal Statistical Society Series A (Statistics in Society)}, \textbf{182}, 1061--1080.

\bibitem[{Koslovsky(2023)}]{koslovsky_bayesian_2023}
Koslovsky, M.~D. (2023) A {Bayesian} zero-inflated {Dirichlet}-multinomial regression model for multivariate compositional count data.
\newblock \textit{Biometrics}, \textbf{79}, 3239--3251.

\bibitem[{Lambert(1992)}]{lambertZeroInflatedPoissonRegression1992}
Lambert, D. (1992) Zero-inflated {Poisson} regression, with an application to defects in manufacturing.
\newblock \textit{Technometrics}, \textbf{34}, 1--14.

\bibitem[{Liboschik et~al.(2017)Liboschik, Fokianos and Fried}]{liboschikTscountPackageAnalysis2017}
Liboschik, T., Fokianos, K. and Fried, R. (2017) Tscount: {{An}} {{R}} package for analysis of count time series following generalized linear models.
\newblock \textit{Journal of Statistical Software}, \textbf{82}, 1--51.

\bibitem[{Lowe et~al.(2018)Lowe, Gasparrini, Meerbeeck, Lippi, Mahon, Trotman et~al.}]{lowe_nonlinear_2018}
Lowe, R., Gasparrini, A., Meerbeeck, C. J.~V., Lippi, C.~A., Mahon, R., Trotman, A.~R. et~al. (2018) Nonlinear and delayed impacts of climate on dengue risk in {Barbados}: {A} modelling study.
\newblock \textit{PLoS Medicine}, \textbf{15}, e1002613.

\bibitem[{Majumder et~al.(2016)Majumder, Cohn, Fish and Brownstein}]{majumder_estimating_2016}
Majumder, M., Cohn, E., Fish, D. and Brownstein, J. (2016) Estimating a feasible serial interval range for {Zika} fever.
\newblock \textit{Bulletin of the World Health Organization}.

\bibitem[{Mercier et~al.(2022)Mercier, Obadia, Carraretto, Velo, Gabiane, Bino et~al.}]{mercier_impact_2022}
Mercier, A., Obadia, T., Carraretto, D., Velo, E., Gabiane, G., Bino, S. et~al. (2022) Impact of temperature on dengue and chikungunya transmission by the mosquito {Aedes} albopictus.
\newblock \textit{Scientific Reports}, \textbf{12}, 6973.

\bibitem[{Nobre et~al.(2005)Nobre, Schmidt and Lopes}]{nobre2005spatio}
Nobre, A.~A., Schmidt, A.~M. and Lopes, H.~F. (2005) Spatio-temporal models for mapping the incidence of malaria in {Par{\'a}}.
\newblock \textit{Environmetrics: The official journal of the International Environmetrics Society}, \textbf{16}, 291--304.

\bibitem[{Nunes et~al.(2015)Nunes, Faria, de~Vasconcelos, Golding, Kraemer, de~Oliveira, Azevedo, da~Silva, da~Silva, da~Silva et~al.}]{nunes2015emergence}
Nunes, M. R.~T., Faria, N.~R., de~Vasconcelos, J.~M., Golding, N., Kraemer, M.~U., de~Oliveira, L.~F., Azevedo, R. d. S. d.~S., da~Silva, D. E.~A., da~Silva, E. V.~P., da~Silva, S.~P. et~al. (2015) Emergence and potential for spread of chikungunya virus in {Brazil}.
\newblock \textit{BMC medicine}, \textbf{13}, 102.

\bibitem[{Paul et~al.(2008)Paul, Held and Toschke}]{paul_multivariate_2008}
Paul, M., Held, L. and Toschke, A.~M. (2008) Multivariate modelling of infectious disease surveillance data.
\newblock \textit{Statistics in Medicine}, \textbf{27}, 6250--6267.

\bibitem[{Pavani and Moraga(2022)}]{pavani_bayesian_2022}
Pavani, J. and Moraga, P. (2022) A {Bayesian} {joint} {spatio}-temporal {model} for {multiple} {mosquito}-{borne} {diseases}.
\newblock In \textit{New {Frontiers} in {Bayesian} {Statistics}} (eds. R.~Argiento, F.~Camerlenghi and S.~Paganin), Springer {Proceedings} in {Mathematics} \& {Statistics}, 69--77. Cham: Springer International Publishing.

\bibitem[{Plummer et~al.(2006)Plummer, Best, Cowles and Vines}]{plummerCODAConvergenceDiagnosis2006}
Plummer, M., Best, N., Cowles, K. and Vines, K. (2006) {{CODA}}: Convergence diagnosis and output analysis for {{MCMC}}.
\newblock \textit{R News}, \textbf{6}, 7--11.

\bibitem[{Quick et~al.(2021)Quick, Dey and Lin}]{quick_regression_2021}
Quick, C., Dey, R. and Lin, X. (2021) Regression {models} for {understanding} {COVID}-19 {epidemic} {dynamics} {with} {incomplete} {data}.
\newblock \textit{Journal of the American Statistical Association}, \textbf{116}, 1561--1577.

\bibitem[{Riou et~al.(2017)Riou, Poletto and Boëlle}]{riou_comparative_2017}
Riou, J., Poletto, C. and Boëlle, P.-Y. (2017) A comparative analysis of {chikungunya} and {Zika} transmission.
\newblock \textit{Epidemics}, \textbf{19}, 43--52.

\bibitem[{Rodriguez-Morales et~al.(2016)Rodriguez-Morales, Villamil-Gómez and Franco-Paredes}]{rodriguez-morales_arboviral_2016}
Rodriguez-Morales, A.~J., Villamil-Gómez, W.~E. and Franco-Paredes, C. (2016) The arboviral burden of disease caused by co-circulation and co-infection of dengue, chikungunya and {Zika} in the {Americas}.
\newblock \textit{Travel Medicine and Infectious Disease}, \textbf{14}, 177--179.

\bibitem[{Rotejanaprasert et~al.(2021)Rotejanaprasert, Lee, Ekapirat, Sudathip and Maude}]{rotejanaprasert_spatiotemporal_2021}
Rotejanaprasert, C., Lee, D., Ekapirat, N., Sudathip, P. and Maude, R.~J. (2021) Spatiotemporal distributed lag modelling of multiple {Plasmodium} species in a malaria elimination setting.
\newblock \textit{Statistical Methods in Medical Research}, \textbf{30}, 22--34.

\bibitem[{Schmidt et~al.(2022)Schmidt, Freitas, Cruz and Carvalho}]{schmidt_poisson-multinomial_2022}
Schmidt, A.~M., Freitas, L.~P., Cruz, O.~G. and Carvalho, M.~S. (2022) A {Poisson}-multinomial spatial model for simultaneous outbreaks with application to arboviral diseases.
\newblock \textit{Statistical Methods in Medical Research}, \textbf{31}, 1590--1602.

\bibitem[{Schmidt et~al.(2011)Schmidt, Suzuki, Thiem, White, Tsuzuki, Yoshida et~al.}]{schmidtPopulationDensityWater2011}
Schmidt, W.-P., Suzuki, M., Thiem, V.~D., White, R.~G., Tsuzuki, A., Yoshida, L.-M. et~al. (2011) Population density, water supply, and the risk of dengue fever in vietnam: Cohort study and spatial analysis.
\newblock \textit{PLOS Medicine}, \textbf{8}, e1001082.

\bibitem[{Sherlock et~al.(2013)Sherlock, Xifara, Telfer and Begon}]{sherlock_coupled_2013}
Sherlock, C., Xifara, T., Telfer, S. and Begon, M. (2013) A coupled hidden {Markov} model for disease interactions.
\newblock \textit{Journal of the Royal Statistical Society Series C (Applied Statistics)}, \textbf{62}, 609--627.

\bibitem[{Stoddard et~al.(2013)Stoddard, Forshey, Morrison, {Paz-Soldan}, {Vazquez-Prokopec} et~al.}]{stoddardHousetohouseHumanMovement2013}
Stoddard, S.~T., Forshey, B.~M., Morrison, A.~C., {Paz-Soldan}, V.~A., {Vazquez-Prokopec}, G.~M. et~al. (2013) House-to-house human movement drives dengue virus transmission.
\newblock \textit{Proceedings of the National Academy of Sciences of the United States of America}, \textbf{110}, 994--999.

\bibitem[{Tang and Chen(2019)}]{tang_zero-inflated_2019}
Tang, Z.-Z. and Chen, G. (2019) Zero-inflated generalized {Dirichlet} multinomial regression model for microbiome compositional data analysis.
\newblock \textit{Biostatistics}, \textbf{20}, 698--713.

\bibitem[{Teixeira et~al.(2009)Teixeira, Costa, Barreto and Barreto}]{teixeira_dengue_2009}
Teixeira, M.~G., Costa, M. d. C.~N., Barreto, F. and Barreto, M.~L. (2009) Dengue: twenty-five years since reemergence in {Brazil}.
\newblock \textit{Cadernos de Saúde Pública}, \textbf{25}, S7--S18.

\bibitem[{Tepe and Guldmann(2020)}]{tepe_spatio-temporal_2020}
Tepe, E. and Guldmann, J.-M. (2020) Spatio-temporal multinomial autologistic modeling of land-use change: {A} parcel-level approach.
\newblock \textit{Environment and Planning B: Urban Analytics and City Science}, \textbf{47}, 473--488.

\bibitem[{Tesla et~al.(2018)Tesla, Demakovsky, Mordecai, Ryan, Bonds, Ngonghala et~al.}]{tesla_temperature_2018}
Tesla, B., Demakovsky, L.~R., Mordecai, E.~A., Ryan, S.~J., Bonds, M.~H., Ngonghala, C.~N. et~al. (2018) Temperature drives {Zika} virus transmission: evidence from empirical and mathematical models.
\newblock \textit{Proceedings of the Royal Society B: Biological Sciences}, \textbf{285}, 20180795.

\bibitem[{Torabi(2017)}]{torabiZeroinflatedSpatiotemporalModels2017}
Torabi, M. (2017) Zero-inflated spatio-temporal models for disease mapping.
\newblock \textit{Biometrical Journal}, \textbf{59}, 430--444.

\bibitem[{de~Valpine et~al.(2017)de~Valpine, Turek, Paciorek, {Anderson-Bergman}, Lang and Bodik}]{valpineProgrammingModelsWriting2017}
de~Valpine, P., Turek, D., Paciorek, C.~J., {Anderson-Bergman}, C., Lang, D.~T. and Bodik, R. (2017) Programming with models: Writing statistical algorithms for general model structures with {NIMBLE}.
\newblock \textit{Journal of Computational and Graphical Statistics}, \textbf{26}, 403--413.

\bibitem[{Vergne et~al.(2014)Vergne, Paul, Chaengprachak, Durand, Gilbert, Dufour, Roger, Kasemsuwan and Grosbois}]{vergne2014zero}
Vergne, T., Paul, M.~C., Chaengprachak, W., Durand, B., Gilbert, M., Dufour, B., Roger, F., Kasemsuwan, S. and Grosbois, V. (2014) Zero-inflated models for identifying disease risk factors when case detection is imperfect: {Application} to highly pathogenic avian influenza {H5N1} in {Thailand}.
\newblock \textit{Preventive Veterinary Medicine}, \textbf{114}, 28--36.

\bibitem[{Vynnycky(2010)}]{vynnycky_introduction_2010}
Vynnycky, E. (2010) \textit{An {Introduction} to {Infectious} {Disease} {Modelling}}.
\newblock New York: Oxford University Press, USA, 1st edn.

\bibitem[{Wakefield et~al.(2019)Wakefield, Dong and Minin}]{minin_spatio-temporal_2019}
Wakefield, J., Dong, T.~Q. and Minin, V.~N. (2019) Spatio-{temporal} {analysis} of {surveillance} {data}.
\newblock In \textit{Handbook of {Infectious} {Disease} {Data} {Analysis}} (eds. L.~Held, N.~Hens, P.~O'Neil and J.~Wallinga), 455--475. Boca Raton: Chapman and Hall/CRC.

\bibitem[{Wang et~al.(2023)Wang, Chiang, Haneef, Rao, Moss and Vannucci}]{wangBayesianNonHomogeneousHidden2022}
Wang, E., Chiang, S., Haneef, Z., Rao, V., Moss, R. and Vannucci, M. (2023) Bayesian non-homogeneous hidden {Markov} model with variable selection for investigating drivers of seizure risk cycling.
\newblock \textit{The Annals of Applied Statistics}, \textbf{17}, 333--356.

\bibitem[{Wangdi et~al.(2018)Wangdi, Clements, Du and Nery}]{wangdiSpatialTemporalPatterns2018}
Wangdi, K., Clements, A. C.~A., Du, T. and Nery, S.~V. (2018) Spatial and temporal patterns of dengue infections in {{Timor}}-{{Leste}}, 2005\textendash 2013.
\newblock \textit{Parasites \& Vectors}, \textbf{11}, 1--9.

\bibitem[{Xia et~al.(2013)Xia, Chen, Fung and Li}]{xia_logistic_2013}
Xia, F., Chen, J., Fung, W.~K. and Li, H. (2013) A {logistic} {normal} {multinomial} {regression} {model} for {microbiome} {compositional} {data} {analysis}.
\newblock \textit{Biometrics}, \textbf{69}, 1053--1063.

\bibitem[{Xu et~al.(2017)Xu, Stige, Chan, Zhou, Yang, Sang et~al.}]{xuClimateVariationDrives2017}
Xu, L., Stige, L.~C., Chan, K.-S., Zhou, J., Yang, J., Sang, S. et~al. (2017) Climate variation drives dengue dynamics.
\newblock \textit{Proceedings of the National Academy of Sciences}, \textbf{114}, 113--118.

\bibitem[{Zeng et~al.(2022)Zeng, Pang, Zhao and Wang}]{zeng_zero-inflated_2022}
Zeng, Y., Pang, D., Zhao, H. and Wang, T. (2022) A {zero}-{inflated} {logistic} {normal} {multinomial} {model} for {extracting} {microbial} {compositions}.
\newblock \textit{Journal of the American Statistical Association}, \textbf{118}, 2356–--2369.

\end{thebibliography}


\begin{thebibliography}{24}
\expandafter\ifx\csname natexlab\endcsname\relax\def\natexlab#1{#1}\fi
\expandafter\ifx\csname url\endcsname\relax
  \def\url#1{\texttt{#1}}\fi
\expandafter\ifx\csname urlprefix\endcsname\relax\def\urlprefix{URL: }\fi

\bibitem[{Abbey(1952)}]{abbey_examination_1952}
Abbey, H. (1952) An {examination} of the {Reed}-{Frost} {theory} of {epidemics}.
\newblock \textit{Human Biology}, \textbf{24}, 201--233.

\bibitem[{Aogo et~al.(2023)Aogo, Zambrana, Sanchez, Ojeda, Kuan, Balmaseda, Gordon, Harris and Katzelnick}]{aogo_effects_2023}
Aogo, R.~A., Zambrana, J.~V., Sanchez, N., Ojeda, S., Kuan, G., Balmaseda, A., Gordon, A., Harris, E. and Katzelnick, L.~C. (2023) Effects of boosting and waning in highly exposed populations on dengue epidemic dynamics.
\newblock \textit{Science Translational Medicine}, \textbf{15}, eadi1734.

\bibitem[{Auger-Méthé et~al.(2021)Auger-Méthé, Newman, Cole, Empacher, Gryba, King et~al.}]{auger-methe_guide_2021}
Auger-Méthé, M., Newman, K., Cole, D., Empacher, F., Gryba, R., King, A.~A. et~al. (2021) A guide to state–space modeling of ecological time series.
\newblock \textit{Ecological Monographs}, \textbf{91}, e01470.

\bibitem[{Bauer and Wakefield(2018)}]{bauerStratifiedSpaceTime2018}
Bauer, C. and Wakefield, J. (2018) Stratified space\textendash time infectious disease modelling, with an application to hand, foot and mouth disease in {{China}}.
\newblock \textit{Journal of the Royal Statistical Society Series C (Applied Statistics)}, \textbf{67}, 1379--1398.

\bibitem[{Besag et~al.(1991)Besag, York and Molli{\'e}}]{besag1991bayesian}
Besag, J., York, J. and Molli{\'e}, A. (1991) Bayesian image restoration, with two applications in spatial statistics.
\newblock \textit{Annals of the Institute of Statistical Mathematics}, \textbf{43}, 1--20.

\bibitem[{Chen et~al.(2019)Chen, Khamthong and Lee}]{chenMarkovSwitchingIntegervalued2019}
Chen, C. W.~S., Khamthong, K. and Lee, S. (2019) Markov switching integer-valued generalized auto-regressive conditional heteroscedastic models for dengue counts.
\newblock \textit{Journal of the Royal Statistical Society: Series C (Applied Statistics)}, \textbf{68}, 963--983.

\bibitem[{Chib(1996)}]{chibCalculatingPosteriorDistributions1996}
Chib, S. (1996) Calculating posterior distributions and modal estimates in {{Markov}} mixture models.
\newblock \textit{Journal of Econometrics}, \textbf{75}, 79--97.

\bibitem[{Finkenstadt and Grenfell(2000)}]{finkenstadt_time_2000}
Finkenstadt, B.~F. and Grenfell, B.~T. (2000) Time {series} {modelling} of {childhood} {diseases}: {A} {dynamical} {systems} {approach}.
\newblock \textit{Journal of the Royal Statistical Society. Series C (Applied Statistics)}, \textbf{49}, 187--205.

\bibitem[{Fokianos and Tjøstheim(2011)}]{fokianos_log-linear_2011}
Fokianos, K. and Tjøstheim, D. (2011) Log-linear {Poisson} autoregression.
\newblock \textit{Journal of Multivariate Analysis}, \textbf{102}, 563--578.

\bibitem[{Freitas et~al.(2019)Freitas, Cruz, Lowe and Sá~Carvalho}]{freitas_spacetime_2019}
Freitas, L.~P., Cruz, O.~G., Lowe, R. and Sá~Carvalho, M. (2019) Space–time dynamics of a triple epidemic: dengue, chikungunya and {Zika} clusters in the city of {Rio} de {Janeiro}.
\newblock \textit{Proceedings of the Royal Society B: Biological Sciences}, \textbf{286}, 20191867.

\bibitem[{Gelman et~al.(2014)Gelman, Hwang and Vehtari}]{gelmanUnderstandingPredictiveInformation2014}
Gelman, A., Hwang, J. and Vehtari, A. (2014) Understanding predictive information criteria for {{Bayesian}} models.
\newblock \textit{Statistics and Computing}, \textbf{24}, 997--1016.

\bibitem[{Göertz et~al.(2017)Göertz, Vogels, Geertsema, Koenraadt and Pijlman}]{goertz_mosquito_2017}
Göertz, G.~P., Vogels, C. B.~F., Geertsema, C., Koenraadt, C. J.~M. and Pijlman, G.~P. (2017) Mosquito co-infection with {Zika} and chikungunya virus allows simultaneous transmission without affecting vector competence of {Aedes} aegypti.
\newblock \textit{PLoS Neglected Tropical Diseases}, \textbf{11}, e0005654.

\bibitem[{Held et~al.(2005)Held, Höhle and Hofmann}]{held_statistical_2004}
Held, L., Höhle, M. and Hofmann, M. (2005) A statistical framework for the analysis of multivariate infectious disease surveillance counts.
\newblock \textit{Statistical Modelling}, \textbf{5}, 187--199.

\bibitem[{{Knorr-Held} and Richardson(2003)}]{knorr-heldHierarchicalModelSpace2003b}
{Knorr-Held}, L. and Richardson, S. (2003) A hierarchical model for space\textendash time surveillance data on meningococcal disease incidence.
\newblock \textit{Journal of the Royal Statistical Society: Series C (Applied Statistics)}, \textbf{52}, 169--183.

\bibitem[{Lawson and Kim(2022)}]{lawson_bayesian_2022}
Lawson, A.~B. and Kim, J. (2022) Bayesian space-time {SIR} modeling of {Covid}-19 in two {US} states during the 2020–2021 pandemic.
\newblock \textit{PLoS One}, \textbf{17}, e0278515.

\bibitem[{Liboschik et~al.(2017)Liboschik, Fokianos and Fried}]{liboschikTscountPackageAnalysis2017}
Liboschik, T., Fokianos, K. and Fried, R. (2017) Tscount: {{An}} {{R}} package for analysis of count time series following generalized linear models.
\newblock \textit{Journal of Statistical Software}, \textbf{82}, 1--51.

\bibitem[{Paul et~al.(2008)Paul, Held and Toschke}]{paul_multivariate_2008}
Paul, M., Held, L. and Toschke, A.~M. (2008) Multivariate modelling of infectious disease surveillance data.
\newblock \textit{Statistics in Medicine}, \textbf{27}, 6250--6267.

\bibitem[{Plummer et~al.(2006)Plummer, Best, Cowles and Vines}]{plummerCODAConvergenceDiagnosis2006}
Plummer, M., Best, N., Cowles, K. and Vines, K. (2006) {{CODA}}: Convergence diagnosis and output analysis for {{MCMC}}.
\newblock \textit{R News}, \textbf{6}, 7--11.

\bibitem[{Reich et~al.(2013)Reich, Shrestha, King, Rohani, Lessler, Kalayanarooj, Yoon, Gibbons, Burke and Cummings}]{reich_interactions_2013}
Reich, N.~G., Shrestha, S., King, A.~A., Rohani, P., Lessler, J., Kalayanarooj, S., Yoon, I.-K., Gibbons, R.~V., Burke, D.~S. and Cummings, D. A.~T. (2013) Interactions between serotypes of dengue highlight epidemiological impact of cross-immunity.
\newblock \textit{Journal of The Royal Society Interface}, \textbf{10}, 20130414.

\bibitem[{Roberts and Sahu(1997)}]{roberts_updating_1997}
Roberts, G.~O. and Sahu, S.~K. (1997) Updating {schemes}, {correlation} {structure}, {blocking} and {parameterization} for the {Gibbs} {sampler}.
\newblock \textit{Journal of the Royal Statistical Society. Series B (Methodological)}, \textbf{59}, 291--317.

\bibitem[{Shaby and Wells(2010)}]{shabyExploringAdaptiveMetropolis2010}
Shaby, B.~A. and Wells, M.~T. (2010) Exploring an {{adaptive Metropolis algorithm}}.

\bibitem[{Tibbits et~al.(2014)Tibbits, Groendyke, Haran and Liechty}]{tibbitsAutomatedFactorSlice2014}
Tibbits, M.~M., Groendyke, C., Haran, M. and Liechty, J.~C. (2014) Automated factor slice sampling.
\newblock \textit{Journal of Computational and Graphical Statistics}, \textbf{23}, 543--563.

\bibitem[{Vynnycky(2010)}]{vynnycky_introduction_2010}
Vynnycky, E. (2010) \textit{An {Introduction} to {Infectious} {Disease} {Modelling}}.
\newblock New York: Oxford University Press, USA, 1st edn.

\bibitem[{Wakefield et~al.(2019)Wakefield, Dong and Minin}]{minin_spatio-temporal_2019}
Wakefield, J., Dong, T.~Q. and Minin, V.~N. (2019) Spatio-{temporal} {analysis} of {surveillance} {data}.
\newblock In \textit{Handbook of {Infectious} {Disease} {Data} {Analysis}} (eds. L.~Held, N.~Hens, P.~O'Neil and J.~Wallinga), 455--475. Boca Raton: Chapman and Hall/CRC.

\end{thebibliography}

\end{document}


\maketitle

\tableofcontents

\section{Reed-Frost Derivation of the ARMN Model}\label{app3:RF}

 In this section, we will derive the ARMN model, see Section 2 of the main manuscript, from a multivariate Reed-Frost model. The ARMN model stands independently based on the justifications given in Section 2. However, the Reed-Frost derivation gives a more epidemiological interpretation for some of the parameters and reveals important sources of potential confounding.

 The Reed-Frost model is a discrete-time susceptible-infectious-recovered (SIR) model originally developed by Lowell J. Reed and Wade Hampton Frost in a series of lectures at Johns Hopkins University in the late 1920s \citep{abbey_examination_1952}. The model takes a time step of one serial interval, the time between successive cases, and assumes that an infected individual does not become infectious till the next time step and then afterward is immune. See \cite{abbey_examination_1952} and Chapter 6 of \cite{vynnycky_introduction_2010} for more details. Following the autoregressive Poisson derivation of \cite{bauerStratifiedSpaceTime2018}, if disease $k$ follows a Reed-Frost model in area $i$ then the number of new infections in the interval $(t-1,t]$ is given by,
 \begin{align}
\begin{split} \label{eqn:reed_frost}
y_{kit} | y_{ki(t-1)} &\sim \text{Pois}(\Phi_{kit}) \\ 
\Phi_{kit} &= \frac{\delta_{ki(t-1)}}{\text{pop}_i} R_{kit} y_{ki(t-1)},
\end{split}
\end{align} where $\delta_{ki(t-1)}$ is the size of the susceptible population, the population not immune to the disease, for disease $k$ at time $t-1$ in area $i$, and $\text{pop}_{i}$ is the total population of area $i$. The parameter $R_{kit}$ represents the time-varying basic reproduction number of disease $k$. That is, the average number of new infections in $(t-1,t]$ resulting from a single infectious individual if the population were fully susceptible, i.e., if $\delta_{ki(t-1)}=\text{pop}_i$. The parameter $R_{kit}^*=\frac{\delta_{ki(t-1)}}{\text{pop}_i} R_{kit}$ represents the effective reproduction number, sometimes called the net reproduction number. That is, the average number of new infections resulting from a single infectious individual in the current, not fully susceptible, population. Both the basic reproduction and effective reproduction numbers are considered important measures of disease transmission in epidemiology \citep{vynnycky_introduction_2010}.

We make small adjustments to (\ref{eqn:reed_frost}). Following \cite{finkenstadt_time_2000}, we raise $y_{ki(t-1)}$ to the power of $0<\zeta_k<1$ to account for nonhomogeneous mixing. We also add 1 to $y_{ki(t-1)}$ to avoid the absorbing 0 state of the Poisson autoregressive process in (\ref{eqn:reed_frost}). It was shown by \cite{fokianos_log-linear_2011} that the parameter estimates are not sensitive to the constant added to $y_{ki(t-1)}$. Therefore, we considered
\begin{align} \label{eqn:reed_frost_mod}
\Phi_{kit} = \frac{\delta_{ki(t-1)}}{\text{pop}_i} R_{kit} \left(y_{ki(t-1)}+1\right)^{\zeta_k},
\end{align} in place of $\Phi_{kit}$ in Equation (\ref{eqn:reed_frost}).

We use correlated log-linear models for the basic reproduction numbers,
\begin{align} \label{eqn:rf_log_linear}
\log( R_{kit}) = \beta_{0ki}+\bm{x}_{it}^T\bm{\beta}_k+\psi_{kit}+b_{it},
\end{align} where $\beta_{0ki}\sim N(\beta_{0k},\sigma_{\beta,k}^2)$ is a normal random intercept meant to account for between area differences and $\bm{x}_{it}$ is a vector of space-time covariates that may affect the transmission of the diseases. If we know a covariate $x_{itl}$ does not affect the transmission of disease $k$, then we can fix $\beta_{kl}=0$. To account for overdispersion and correlation between the reproduction numbers,  we model the random effects $\psi_{kit}$ using a multivariate normal distribution,
\begin{align} \label{eqn:logistic_normal_RF}  \bm{\psi}_{it} &=
( \psi_{1it}, \dots, \psi_{Kit})^T \sim \text{MVN}_{K}(\bm{0},\bm{\Sigma}_{RF}),
\end{align} where $\bm{\Sigma}_{RF}$ is a $K$ by $K$ variance-covariance matrix. In (\ref{eqn:rf_log_linear}), $b_{it}$ represents shared space-time factors. This may be particularly relevant for arboviruses that share the same vector as in our motivating example. We do not specify any model for $b_{it}$ as the joint distribution of the disease counts conditional on their total does not depend on it, see below.

Equations (\ref{eqn:reed_frost})-(\ref{eqn:logistic_normal_RF}) define a multivariate Reed-Frost model for the diseases. Following the well-known relationship between the multinomial and Poisson distributions, we have that,
\begin{align} \label{eqn:multinom_rf}
\bm{y}_{it}|\text{total}_{it}, \bm{y}_{t-1} \sim \text{Multinom}(\bm{\pi}_{it},\text{total}_{it}), 
\end{align} where,
\begin{align}
\pi_{kit} = \frac{\Phi_{kit}}{\sum_{j=1}^{K}\Phi_{jit}}.
\end{align} The relative odds, relative to disease 1, are then given by,
\begin{align} \label{eqn:rf_relative_odds}
\frac{\pi_{kit}}{\pi_{1it}} = \frac{\Phi_{kit}}{\Phi_{1it}} =\frac{R_{kit}^*}{R_{1it}^*}\frac{\left(y_{ki(t-1)}+1\right)^{\zeta_k}}{\left(y_{1i(t-1)}+1\right)^{\zeta_1}},
\end{align} where,
\begin{align} \label{eqn:rf_ratio}
\log\left(\frac{R_{kit}^*}{R_{1it}^*}\right) = (\beta_{0ki}-\beta_{01i})+\bm{x}_{it}^T(\bm{\beta}_k-\bm{\beta}_1)+(\psi_{kit}-\psi_{1it})+(b_{it}-b_{it})+\log\left(\frac{\delta_{ki(t-1)}}{\delta_{1i(t-1)}}\right).
\end{align} Equations (\ref{eqn:multinom_rf})-(\ref{eqn:rf_ratio}) define an ARMN model where,
\begin{itemize}
 \item $\displaystyle \lambda_{kit}=\frac{R_{kit}^*}{R_{1it}^*}$; the ratio of the effective reproduction number of disease $k$ and disease 1.
  \item $\displaystyle \bm{x}_{kit}=\{x_{itl} \in \bm{x}_{it}\,:\,\beta_{kl} \neq 0\text{ or }\beta_{1l} \neq 0\}$; the set of covariates that affect either the transmission of disease $k$ or the transmission of disease 1.
  \item If $x_{itl} \in \bm{x}_{kit}$ let $l_k$ represent the index of $x_{itl}$ in $\bm{x}_{kit}$, then $\displaystyle \alpha_{kl_k}=\beta_{kl}-\beta_{1l}$; the difference between the effect of covariate $x_{itl}$ on the effective reproduction number of disease $k$ and disease $1$.
  \item $\alpha_{0ki}=\beta_{0ki}-\beta_{01i}$, $\alpha_{0k}=\beta_{0k}-\beta_{1k}$ and $\sigma_k^2=\sigma_{\beta,k}^2+\sigma_{\beta,1}^2$.
  \item $\phi_{kit}=\psi_{kit}-\psi_{1it}$ and $\Sigma_{kj}=\Sigma_{RF,kj}-\Sigma_{RF,k1}-\Sigma_{RF,j1}+\Sigma_{RF,11}$.
  \item $\displaystyle \log\left(\frac{\delta_{ki(t-1)}}{\delta_{1i(t-1)}}\right)$ is added as an offset to $\log(\lambda_{kit})$.
\end{itemize}

In the ARMN model, the shared factors $b_{it}$ are eliminated. Since we do not typically observe $\log\left(\frac{\delta_{ki(t-1)}}{\delta_{1i(t-1)}}\right)$ it is an important source of potential confounding. We discuss this issue extensively in Section 2 of the main manuscript and Section 9 below. Also, as we are modeling reported cases, there could be confounding due to changes in relative reporting rates. For example, consider a spatial covariate associated with areas more likely to report disease $k$ compared to disease $1$. In that case, it would seem like there was an increase in $R_{kit}^*/R_{1it}^*$ associated with the covariate even if there was no change in the true ratio.

\subsection{Extensions to account for disease interactions and geographical spread} \label{app3:RF_extensions}

The multivariate Reed-Frost model defined by Equations (\ref{eqn:reed_frost})-(\ref{eqn:logistic_normal_RF}) does not account for disease interactions, beyond the random effects $\psi_{kit}$, and disease spread between areas. To further account for disease interactions we can add $\log(y_{ji(t-1)}+1)$ to $\bm{x}_{it}$ and assume its corresponding coefficient in $\bm{\beta}_j$ is equal to 0. That is, we can assume previous cases of disease $j$ affect the basic reproduction number of the other diseases. This is justified in many instances, for example, it has been shown that mosquitoes infected by both Zika and chikungunya transmit Zika at a higher rate \citep{goertz_mosquito_2017}. In the multinomial form of the model, the above is equivalent to adding $\log(y_{ji(t-1)}+1)$ to $\bm{x}_{kit}$ for $j\neq k$ and $j\neq 1$, which we do for Zika and chikungunya in Section 4 of the main manuscript. Note that any interactions with the baseline disease would be absorbed by $\zeta_1$ or $\zeta_k$ in the multinomial representation of the model. A similar strategy for accounting for disease interactions in a multivariate Poisson model was used by \cite{paul_multivariate_2008}.

Several approaches have been proposed for extending the Reed-Frost model to deal with disease spread between areas \citep{minin_spatio-temporal_2019}. \cite{bauerStratifiedSpaceTime2018}, following \cite{held_statistical_2004}, added a sum of weighted previous cases in neighboring areas to the conditional means of the Reed-Frost model. That is, in the context of our model, ${\Phi_{kit}'=\Phi_{kit}+\lambda^{NE}_{k}\sum_{j \in NE(i)}\omega_{ji}y_{kj(t-1)}}$, where $\Phi_{kit}'$ is the conditional mean adjusted for geographical disease spread,  $\lambda^{NE}_{k}>0$ is an unknown spatial effect and $\omega_{ji}$ are fixed weights representing the influence area $j$ has on area $i$. However, an additive adjustment would be awkward in the multinomial form of the model (\ref{eqn:rf_relative_odds}). We can also consider a multiplicative adjustment, similar to \cite{lawson_bayesian_2022},
\begin{align*}
\Phi_{kit}' = \Phi_{kit}\left(\sum_{j \in NE(i)}\omega_{ji}y_{kj(t-1)}+1\right)^{\beta_k^{NE}},
\end{align*} where $\beta_k^{NE}$ is an unknown spatial effect, which can be negative. This is equivalent to adding $\log(\sum_{j \in NE(i)}\omega_{ji}y_{kj(t-1)}+1)$ to $\bm{x}_{it}$ and setting its coefficient in $\bm{\beta}_m$ for $m \neq k $ to 0. Unlike the additive adjustment, the multiplicative adjustment is convenient in the multinomial form of the model since it is the same as adding $\log(\sum_{j \in NE(i)}\omega_{ji}y_{kj(t-1)}+1)$ and $\log(\sum_{j \in NE(i)}\omega_{ji}y_{1j(t-1)}+1)$ to $\bm{x}_{kit}$. We do this in Section 4 of the main manuscript, where we consider the weights $\omega_{ji}=1/\sum_{m \in NE(i)} pop_m$, which gives the prevalence of the diseases across neighboring areas.

\subsection{The difference between multinomial and multivariate Poisson approaches} \label{app3:multinom_vs_multipoiss}

We can use the above derivation of the ARMN model from the multivariate Reed-Frost model to better understand the advantages of both approaches. Firstly, the ARMN model has $(K-1)/K$ times as many parameters, not counting the shared factors $b_{it}$, compared to the multivariate Reed-Frost model. For instance, in the case of $K=3$ diseases, the ARMN model has 66\% as many parameters, which is a substantial reduction in model complexity. The ARMN model also eliminates all shared factors $b_{it}$. This may be especially relevant for arboviruses that share the same vector, like in our motivating example.

An immediate advantage of the multivariate Reed-Frost approach is that it allows for estimating the effect of a covariate on the transmission of any of the diseases. The ARMN model can only estimate the differences in the effects, which the multivariate Reed-Frost model could also provide. The multivariate Reed-Frost model could be extended to deal with zero inflation and long periods of disease absence in a similar way to the MS-ZIARMN model. Following Section 2.1 of the main manuscript, we could replace $\Phi_{kit}$ by $\Phi_{kit}S_{kit}$ and model $\bm{S}_{it}$ as following a coupled Markov chain. In this case, we would not have to assume $S_{1it}$ is always equal to one. That is, we would not have to assume one of the diseases was always present like with the MS-ZIARMN model.

In conclusion, there seem to be advantages to both approaches. The multinomial approach is likely more appropriate when one is mainly interested in differences between the transmission of the diseases and we can assume one of the diseases is always present, like in our motivating example. It goes beyond the scope of this paper to do a comparison based on model fitting, given the complexity of the multivariate Reed-Frost model and any zero-inflated extensions. This will be the subject of further work.

\section{Simulation Study to Investigate Correlations Induced by the Random Effects}
\label{app3:sim}

We designed a simulation study to better understand the correlations between the disease counts, conditional on their total, induced by the random effects $\phi_{kit}$ in Equations (3)-(4) of the main manuscript. We simulated from the following multinomial distribution with multivariate normal random effects added to the log relative odds,
\begin{align} \label{eqn:multi_sim}
\begin{split}
(y_1,y_2,y_3)^T \,|\, \text{total},  \phi_2, \phi_3  &\sim \text{Multinom}(\bm{\pi}=(\pi_1,\pi_2,\pi_3)^T,\text{total}=y_1+y_2+y_3) \\[5pt]
\log\left(\frac{\pi_3}{\pi_1}\right) &=  \alpha_{03} + \phi_3 \\[5pt] 
\log\left(\frac{\pi_2}{\pi_1}\right) &=  \alpha_{02} + \phi_2 \\[5pt] 
(\phi_2,\phi_3)^T &\sim MVN_{2}\left((0,0)^T,  \left( {\begin{array}{cc}
   \sigma_2^2 & \rho \sigma_2 \sigma_3 \\
   \rho \sigma_2 \sigma_3 & \sigma_3^2 \\
  \end{array} } \right)\right).
\end{split}
\end{align} We fixed $\alpha_{02}=\log(1.14)$, $\alpha_{03}=0$, $\sigma_2=.75$ and $\sigma_3=.8$ based on the estimates from our motivating example in Section 4.2 of the main manuscript. We assumed $\text{total}=10$ as this was the average sum of dengue, Zika and chikungunya cases in a neighborhood.

\begin{figure}[!t]
 	\centering
 	\includegraphics[width=\textwidth]{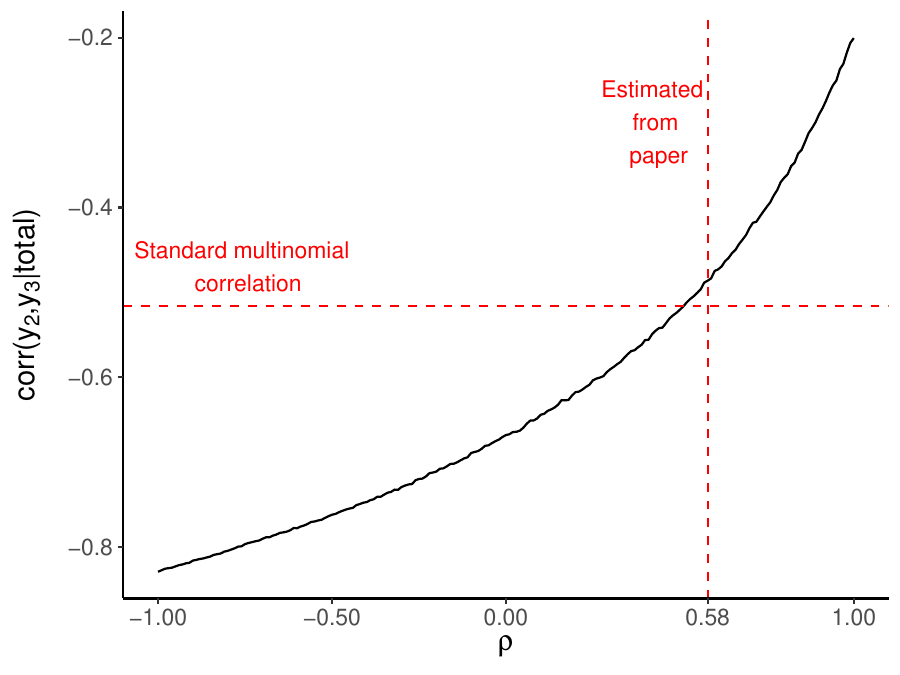}
	\caption{The black solid line shows $\text{corr}(y_2,y_3|\text{total})$ versus $\rho$, marginalizing the random effects, based on simulations of (\ref{eqn:multi_sim}). The red horizontal dashed line is drawn at $\text{corr}(y_2,y_3|\text{total})$ assuming the random effects are fixed at 0, i.e., $\phi_2=\phi_3=0$ in (\ref{eqn:multi_sim}). The red vertical dashed line is drawn at the value of $\rho$ estimated in Section 4 of the main text. \label{fig3:figSM1}} 
\end{figure}

Figure \ref{fig3:figSM1} shows the correlation between $y_2$ and $y_3$, conditional on the total count and marginalizing the random effects, versus $\rho$ based on simulations of (\ref{eqn:multi_sim}) (to the best of our knowledge there is no closed form solution). If there were no random effects, then
\begin{align*}
\text{corr}(y_2,y_3|\text{total})=\frac{-\text{total}\left(\frac{e^{\alpha_{02}}}{1+e^{\alpha_{02}}+e^{\alpha_{03}}}\right)\left(\frac{e^{\alpha_{03}}}{1+e^{\alpha_{02}}+e^{\alpha_{03}}}\right)}{\sqrt{\text{total}\left(\frac{e^{\alpha_{02}}}{1+e^{\alpha_{02}}+e^{\alpha_{03}}}\right)\left(\frac{1+e^{\alpha_{03}}}{1+e^{\alpha_{02}}+e^{\alpha_{03}}}\right)}\sqrt{\text{total}\left(\frac{e^{\alpha_{03}}}{1+e^{\alpha_{02}}+e^{\alpha_{03}}}\right)\left(\frac{1+e^{\alpha_{02}}}{1+e^{\alpha_{02}}+e^{\alpha_{03}}}\right)}},
\end{align*}
which is given by the horizontal red dashed line in Figure \ref{fig3:figSM1}. The correlation between the random effects $\rho$ needs to be greater than around .5 for the marginalized correlation between the counts to be greater than that produced by the standard multinomial distribution (i.e., without the random effects). This is an interesting and somewhat unintuitive result, as we would expect $\rho=0$ to be a more likely threshold for this. In Section 4 of the main manuscript, we estimated $\rho$ as .58 (.52, .65), and, therefore, the random effects did not likely induce any significant correlations between the counts in our motivating example.

From Figure \ref{fig3:figSM1}, even if $\rho$ is close to 1 the marginal correlation between the counts, conditional on the total, is negative. We do not see negative correlations as a limitation of the model. The correlation is conditional on the total count and, therefore, if one of the counts increases another must necessarily decrease. Regardless, based on Figure \ref{fig3:figSM1}, adding multivariate normal random effects to the log relative odds leads to a more flexible correlation structure among the disease counts compared to the standard multinomial distribution. We tried different values of $\alpha_{02}$ and $\alpha_{03}$ in (\ref{eqn:multi_sim}) and got similar results. Also, plotting $\text{corr}(y_1,y_3|\text{total})$ and $\text{corr}(y_1,y_2|\text{total})$ versus $\rho$ (not shown) led to the same conclusions.

\section{Hybrid Gibbs Sampling Algorithm}

In this Section, we describe the hybrid Gibbs sampling algorithm we use to sample from the joint posterior of $\bm{S}^*$ and $\bm{v}$, $p(\bm{S}^*,\bm{v}|\bm{y})$. We first separate the parameter vector into blocks $\bm{v}=(\bm{v}_1,\dots,\bm{v}_D)^T$, see below.  Secondly, we generate random initial values for the Gibbs sampler, $\bm{v}^{[0]}=(\bm{v}_1^{[0]},\dots,\bm{v}_D^{[0]})^T$ and $\bm{S}^{[0]}$. Then the following steps are repeated for $m=1,\dots,Q$, where $Q$ is large enough to ensure convergence,
\begin{enumerate}[\quad Step 1:]
\item Sample $\bm{v}_d^{[m]}$ from $p(\bm{v}_d|\bm{v}_1^{[m]},\dots,\bm{v}_{d-1}^{[m]},\bm{v}_{d+1}^{[m-1]},\dots,\bm{v}_{D}^{[m-1]},\bm{S}^{[m-1]},\bm{y})$ for $d=1,\dots,D$.
\item Sample $\bm{S}^{[m]}$ from $p(\bm{S}|\bm{v}^{[m]},\bm{y})$ using the forward filtering backward sampling (FFBS) algorithm \citep{chibCalculatingPosteriorDistributions1996}, see Section \ref{app3:FFBS} below.
\end{enumerate}

We sampled each $\bm{\phi}_{it}$ in Equations (3)-(4) of the main manuscript jointly using an adaptive blocked random walk metropolis step \citep{roberts_updating_1997,shabyExploringAdaptiveMetropolis2010}. For our motivating example, we sampled some of the regression coefficients in $\bm{v}$ jointly using an automated factor slice sampler \citep{tibbitsAutomatedFactorSlice2014}, as they mixed slowly and exhibited high posterior correlations. All other elements of $\bm{v}$ without conjugate priors were sampled individually using an adaptive random walk Metropolis step \citep{shabyExploringAdaptiveMetropolis2010}.

\subsection{Forward Filtering Backward Sampling (FFBS) Algorithm} \label{app3:FFBS}

In this Section, we provide the FFBS algorithm \citep{chibCalculatingPosteriorDistributions1996} for sampling from $p(\bm{S}^*|\bm{y},\bm{v})$ needed for our hybrid Gibbs sampler described above. Like in Section 3 of the main manuscript, we will assume $K=3$, which leads to 4 possible states for $S_{it}^*$. We will use the subscript $t_1:t_2$ to denote a temporally indexed vector subsetted to the interval $t_1$ to $t_2$, e.g., $\bm{S}^{*}_{i(t_1:t_2)}=(S_{it_1}^*,\dots,S_{it_2}^*)^T$. We will also use the subscript $(-i)$ to denote the vector with the ith element removed.

First note that $p(\bm{S}^*|\bm{y},\bm{v}) \propto p(\bm{S}^*,\bm{y}|\bm{v})$ which, from Equation (9) of the main text, factors into functions involving only $\bm{S}^{*}_{i(1:T)}$ for each $i$. Therefore, $\bm{S}^*_{1(1:T)},\dots,\bm{S}^*_{N(1:T)}$ are mutually independent given $\bm{y}$ and $\bm{v}$. This means to sample from $p(\bm{S}^*|\bm{y},\bm{v})$ we can sample from $p(\bm{S}^{*}_{i(1:T)}|\bm{y},\bm{v})$ for each $i$. A similar argument can be used to show that $\bm{S}^*_{1(1:t)},\dots,\bm{S}^*_{N(1:t)}$ are mutually independent conditional on either $\bm{y}_{1:t}$ and $\bm{v}$ or $\bm{y}_{1:(t-1)}$ and $\bm{v}$.

The initial forward filtering part of the FFBS algorithm starts at $t=2$ and goes recursively to $t=T$. For each $t$, we first calculate the predictive probabilities and then the filtered probabilities. The predictive probabilities are given by,
\begin{align} \label{eqn:pp}
\begin{split}
P(S_{it}^*=l | \bm{y}_{1:(t-1)},\bm{v}) &= \sum_{k=1}^{4} P(S_{it}^*=l|S_{i(t-1)}^*=k,\bm{y}_{t-1},\bm{v})P(S_{i(t-1)}^*=k|\bm{y}_{1:(t-1)},\bm{v}),
\end{split}
\end{align} for $l=1,\dots, 4$. For $t=2$, $P(S_{i(t-1)}^*=k|\bm{y}_{1:(t-1)},\bm{v})=P(S_{i1}^*=k|\bm{v})$, which is the initial state distribution. Recall that the modeler sets the initial state distribution based on how likely they believe the diseases to be present at the beginning of the study. The probability $P(S_{it}^*=l|S_{i(t-1)}^*=k,\bm{y}_{t-1},\bm{v})$ is the probability of the Markov chain in area $i$ transitioning from state $k$ at time $t-1$ to state $l$ at time $t$, i.e., $\Gamma(S_{it}^*|\bm{y}_{t-1})_{kl}$, see Section 3 of the main text.
The filtered probabilities are given by,
\begin{align} \label{eqn:FF}
\begin{split}
p(S_{it}^*|\bm{y}_{1:t},\bm{v})&= p(S_{it}^*|\bm{S}_{(-i)t}^*,\bm{y}_{1:t},\bm{v}) \text{\hspace{1cm}(by mutual independence)} \\
&\propto p(\bm{y}_t|\bm{S}^{*}_t,\bm{y}_{1:(t-1)},\bm{v})p(\bm{S}^{*}_t|\bm{y}_{1:(t-1)},\bm{v}) \\
&= \prod_{j=1}^{N} p(y_{jt}|S^{*}_{jt},\bm{y}_{1:(t-1)}, \text{total}_{jt},\bm{\beta})p(S^{*}_{jt}|\bm{y}_{1:(t-1)},\bm{v}) \text{\hspace{1cm}(by mutual independence)}  \\
& \propto p(y_{it}|S^{*}_{it},\bm{y}_{1:(t-1)}, \text{total}_{it},\bm{\beta})p(S^{*}_{it}|\bm{y}_{1:(t-1)},\bm{v}), \\[5pt]
\text{implying,}& \\[5pt]
P(S_{it}^*=l|\bm{y}_{1:t},\bm{v}) &= \frac{p(\bm{y}_{it} | \bm{y}_{t-1}, S_{it}^*=l, \text{total}_{it},\bm{\beta})P(S_{it}^*=l | \bm{y}_{1:(t-1)},\bm{v})}{p(\bm{y}_{it}|\bm{y}_{1:(t-1)},\text{total}_{it},\bm{v})}, \\[5pt] 
\text{where,}&  \\[5pt]
p(\bm{y}_{it}|\bm{y}_{1:(t-1)},\text{total}_{it},\bm{v}) &= \sum_{k=1}^{4} p(\bm{y}_{it} | \bm{y}_{t-1}, S_{it}^*=k, \text{total}_{it},\bm{\beta})P(S_{it}^*=k | \bm{y}_{1:(t-1)},\bm{v}).
\end{split}
\end{align} The distribution $p(\bm{y}_{it} | \bm{y}_{t-1}, S_{it}^*=l, \text{total}_{it},\bm{\beta})$ is given by Equation (8) of the main text. The probability $P(S_{it}^*=l | \bm{y}_{1:(t-1)},\bm{v})$ is the predictive probability for time $t$ calculated in the previous step.

Once the filtered probabilities have been calculated for $t=2,\dots, T$, the backward sampling part of the FFBS algorithm is performed. We have that,
\begin{align*}
p(\bm{S}^*_{i(1:T)}|\bm{y},\bm{v}) &= \left[\prod_{t=1}^{T-1}p(S^{*}_{it}|S^{*}_{i(t+1)},\bm{y}_{1:t},\bm{v})\right] p(S^{*}_{iT}|\bm{y},\bm{v}).
\end{align*} Also, note that,
\begin{align} \label{eqn:bs}
\begin{split}
p(S^{*}_{it}|S^{*}_{i(t+1)},\bm{y}_{1:t},\bm{v}) &\propto p(S^{*}_{i(t+1)}|S^{*}_{it},\bm{y}_{1:t},\bm{v})p(S^{*}_{it}|,\bm{y}_{1:t},\bm{v}), \\[5pt]
\text{implying},& \\[5pt]
P(S^{*}_{it}=l|S^{*}_{i(t+1)}=j,\bm{y}_{1:t},\bm{v}) &= \frac{P(S^{*}_{i(t+1)}=j|S^{*}_{it}=l,\bm{y}_{1:t},\bm{v})P(S^{*}_{it}=l|,\bm{y}_{1:t},\bm{v})}{\sum_{k=1}^{4}P(S^{*}_{i(t+1)}=j|S^{*}_{it}=k,\bm{y}_{1:t},\bm{v})P(S^{*}_{it}=k|,\bm{y}_{1:t},\bm{v})},
\end{split}
\end{align} for $l=1,\dots,4$. The probability $P(S^{*}_{i(t+1)}=j|S^{*}_{it}=l,\bm{y}_{1:t},\bm{v})$ is given by $\Gamma(S_{i(t+1)}^*|\bm{y}_{t})_{lj}$. The probability $P(S^{*}_{it}=l|,\bm{y}_{1:t},\bm{v})$ is the filtered probability for time $t$. 

Therefore, to sample $\bm{S}^{*[m]}_{i(1:T)} \sim p(\bm{S}^*_{i(1:T)}|\bm{y},\bm{v})$ we first sample $S^{*[m]}_{iT} \sim p(S^{*}_{iT}|\bm{y},\bm{v})$ using the final filtered probabilities. Then we work backwards sampling ${S^{*[m]}_{it} \sim p(S^{*}_{it}|S^{*}_{i(t+1)}=S^{*[m]}_{i(t+1)},\bm{y}_{1:t},\bm{v})}$ using the probabilities $P(S^{*}_{it}=l|S^{*}_{i(t+1)}=S^{*[m]}_{i(t+1)},\bm{y}_{1:t},\bm{v})$ for $l=1,\dots,4$, from (\ref{eqn:bs}), for $t=T-1, \dots, 1$.

Finally, we note that the likelihood of $\bm{v}$ given $\bm{y}$, conditioning on the totals, is provided by,
\begin{align*}
p(\bm{y}|\bm{v}) = \prod_{i=1}^{N}\prod_{t=2}^{T} p(\bm{y}_{it}|\bm{y}_{1:(t-1)},\text{total}_{it},\bm{v}),
\end{align*} which can be calculated using the forward filter, see Equation (\ref{eqn:FF}). Therefore, we could base inference about $\bm{v}$ on the marginal, concerning the states, likelihood. However, as explained in the main text, we are interested in making inferences about the unknown states as they represent when the model believes the diseases were present. The marginal likelihood contribution $p(\bm{y}_{it}|\bm{y}_{1:(t-1)},\text{total}_{it},\bm{v})$ is used, however, to calculate the WAIC as we explain in {\color{black}Section \ref{app3:WAIC} below}.

\newpage

{\color{black}
\section{{\color{black}Simulation Study to Assess Parameter Recovery}}

We designed a simulation study to ensure our hybrid Gibbs sampling algorithm could recover the true parameters of the MS-ZIARMN model. We simulated data from a slightly simplified version of the MS-ZIARMN model used in our motivating example in Section 4 of the main text. More specifically, we removed some of the covariate effects to reduce the computational cost of running multiple simulations. We simulated data from the following MS-ZIARMN model,
\begin{align}
\begin{split}
\log(\lambda_{kit}^{*}) &= \alpha_{0ki}+\alpha_{1k}\text{temp}_{it}+\alpha_{2k}\text{verde}_i+\zeta_{k}\log(y_{k,i,t-1}+1)-\zeta_{1}\log(y_{1,i,t-1}+1)+ \\ &\alpha_{6k}\log\left(\frac{\sum_{j \in NE(i)}y_{k,j,t-1}}{\sum_{j \in NE(i)}pop_j}+1\right)+\alpha_7 \log\left(\frac{\sum_{j \in NE(i)}y_{1,j,t-1}}{\sum_{j \in NE(i)}pop_j}+1\right)+\phi_{kit} \\[5pt]
\text{logit}(p_{kit}) &= \eta_{0k}+\eta_{1k}\text{temp}_{it}+\eta_{2k}\text{verde}_i+\eta_{8k}\log\left(\frac{\sum_{j \in NE(i)}y_{k,j,t-1}}{\sum_{j \in NE(i)}pop_j}+1\right)+ \\
& \rho_k^{AR} S_{k,i,t-1}+\rho_{jk}^{DI}S_{j,i,t-1} \\[5pt]
\alpha_{0ki} &\sim N(\alpha_{0k},\sigma^2_{k}) \\
\left(\phi_{2it},\phi_{3it}\right)^T & \sim \text{MVN}_2\left(\bm{0},\bm{\Sigma}\right) \\
S_{2,i,1} &\sim \text{Bernoulli}(.1), \hspace{.25cm} S_{3,i,1} \sim \text{Bernoulli}(.05),
\end{split} \label{eqn:sim_model}
\end{align} for $i=1,\dots,160$, $t=2,\dots,52$, $k=2,3=K$ and $j\neq k = 2,3$. We took $\text{temp}_{it}$, $\text{verde}_i$, $pop_j$, $NE(i)$, $\text{total}_{it}$ in Equation (1) of the main text (the total counts), and $y_{k,i,1}$ (the disease counts at time $t=1$ as the model is autoregressive)  from our motivating example. We assumed the following true parameter values: $\alpha_{02}=.15$, $\alpha_{12}=.05$, $\alpha_{22}=0$, $\zeta_2=.35$, $\zeta_1=.45$, $\alpha_{62}=.7$, $\alpha_{7}=-.65$, $1/\sigma^2_2=\tau_2=15$, $\alpha_{03}=0$, $\alpha_{13}=-.05$, $\alpha_{23}=-.01$, $\zeta_3=.45$, $\alpha_{63}=.8$, $1/\sigma^2_3=\tau_3=15$, $\eta_{02}=-2$, $\eta_{12}=.15$, $\eta_{22}=.01$, $\eta_{82}=4$, $\rho_2^{AR}=2.5$, $\rho_{32}^{DI}=1$, $\eta_{03}=-4$, $\eta_{13}=.3$, $\eta_{23}=0$, $\eta_{83}=7$, $\rho_{3}^{AR}=4$, $\rho_{23}^{DI}=2$, $\Sigma_{11}^{-1}=\Sigma_{22}^{-1}=2.5$, and $\Sigma_{12}^{-1}=\Sigma_{21}^{-1}=-1.5$. For the multinomial part of the model, we chose true parameter values similar to those estimated in our motivating example. For the Markov chain part, some of the estimates in Table 3 of the main text are extreme, as chikungunya cases emerged at a similar time across all areas (see the discussion in Section 4.2 of the main text). Therefore, we shrunk some of the estimates from Table 3 when choosing the true parameter values for $\bm{\theta}$ (the Markov chain parameters). The true parameter values we assumed still represent strong autoregressive, interaction, and spatial effects, and result in both diseases being absent around 40 percent of the time. Figure \ref{fig3:SM_sim1} shows simulated cases from a single replication of (\ref{eqn:sim_model}) in one area. The simulated counts exhibit long strings of zeroes, similar to the Zika and chikungunya counts analyzed in our motivating example (see Figure 1 of the main text).

\begin{figure}[!t]
 	\centering
 	\includegraphics[width=.85\textwidth]{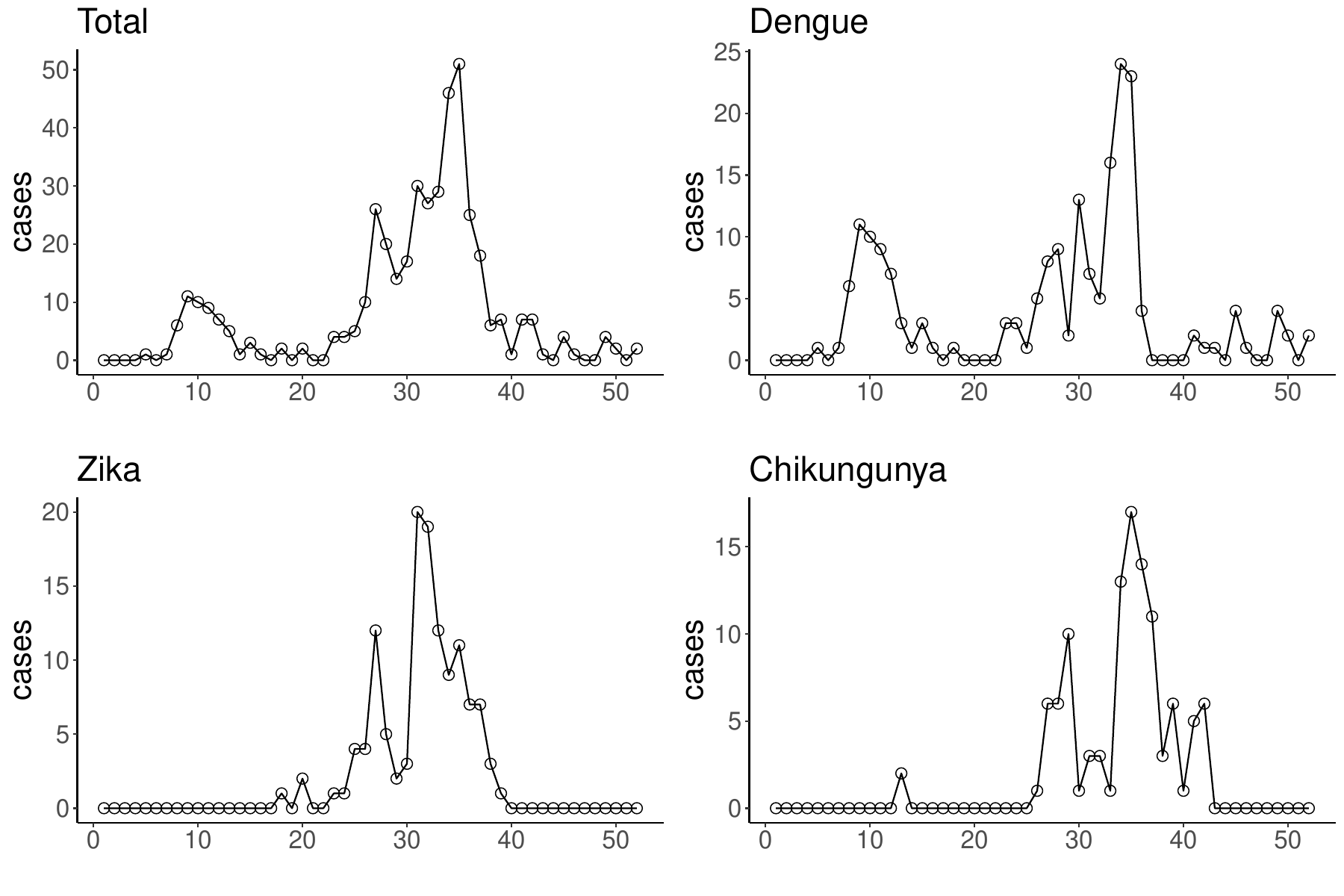}
	\caption{{\color{black} All panels except the top left show simulated counts in one area from a single replication of (\ref{eqn:sim_model}). The top left panel shows the total counts, which were not simulated and taken from our motivating example.}\label{fig3:SM_sim1}} 
\end{figure}

We simulated 200 replications of (\ref{eqn:sim_model}). We then fit a correctly specified MS-ZIARMN model to each replication using our hybrid Gibbs sampling algorithm. To investigate how the estimates would be affected if zero inflation were not accounted for, we also fit an ARMN version of (\ref{eqn:sim_model}) to each replication. That is, (\ref{eqn:sim_model}) fixing $S_{k,i,t}=1$ for all $k$, $i$, and $t$. We checked convergence for each replication using the minimum effective sample size ($>500$) and the maximum Gelman-Rubin statistic ($<1.1$) \citep{plummerCODAConvergenceDiagnosis2006}. Both the MS-ZIARMN and ARMN models converged for all 200 replications.

The sample means and 95\% quantile intervals (0.025 and 0.975 quantiles) of the posterior medians are shown in Figure \ref{fig3:sim_multi} for the multinomial parameters and Figure \ref{fig3:sim_markov} for the Markov chain parameters. When fitting the MS-ZIARMN model, the posteriors are centered close to the true parameter values on average. In contrast, if we ignore zero inflation and fit the ARMN model, the posteriors are often centered quite far from the true parameter values. Ignoring zero inflation tends to lead to an overestimation of the covariate effects and an underestimation of the intercepts. For example, we would likely estimate a smaller inherent favorability of Zika transmission relative to dengue transmission and a larger effect of temperature on the favorability of Zika transmission (compared to the actual values).

\begin{figure}[!htb]
 	\centering
 	\includegraphics[width=\textwidth]{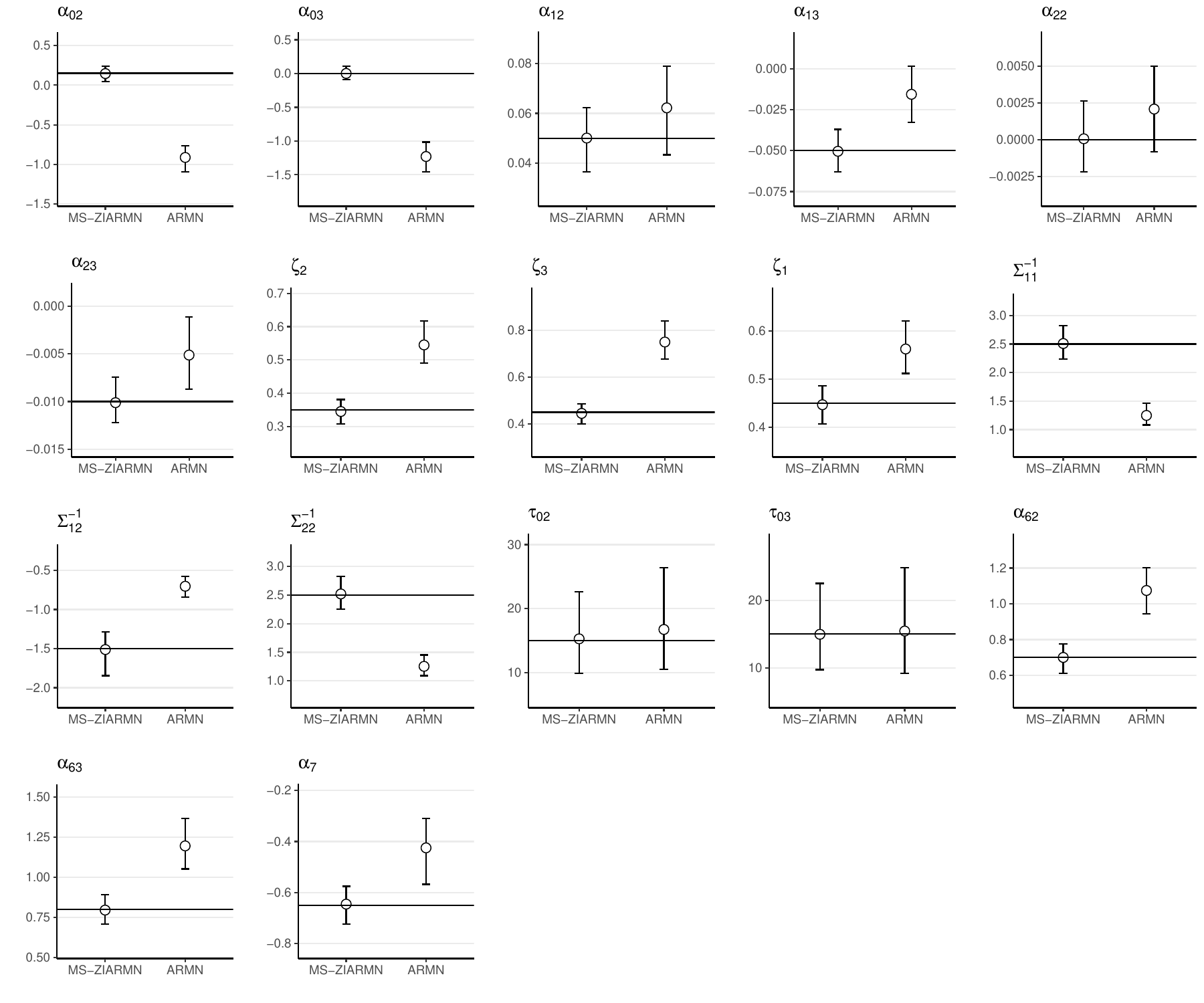}
	\caption{{\color{black} Shows the sample means (circles) and 95\% quantile intervals (caps) of the posterior medians
from fitting the MS-ZIARMN (correctly specified) and ARMN (no zero-inflation) models to 200 replications of (13). The horizontal lines are drawn at the true parameter values. Shows just the parameters from the multinomial part of the model.}\label{fig3:sim_multi}} 
\end{figure}

\begin{figure}[!htb]
 	\centering
 	\includegraphics[width=\textwidth]{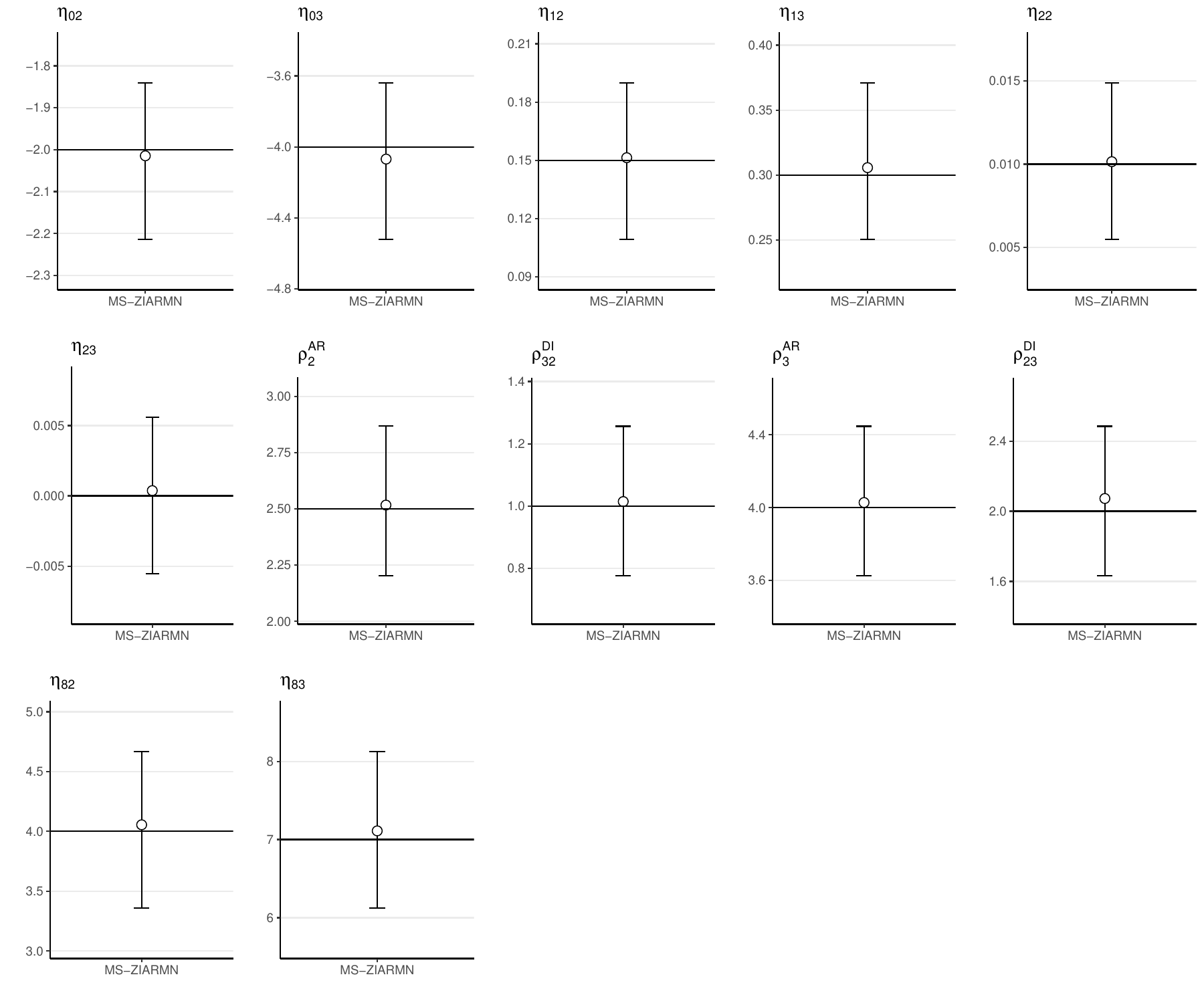}
	\caption{{\color{black} Shows the sample means (circles) and 95\% quantile intervals (caps) of the posterior medians
from fitting the MS-ZIARMN (correctly specified) model to 200 replications of (13). The horizontal lines are drawn at the true parameter values. Shows just the parameters from the Markov chain part of the model.}\label{fig3:sim_markov}} 
\end{figure}

The average coverage of the 95\% credible intervals was 95.4\% (min=92.5\%, max=98.5\%)  for the MS-ZIARMN model and 20\% (min=0\%, max=95\%) for the ARMN model. Solely considering the covariate effects, the average coverage was only 15.1\% (min=0\%, max=65.5\%) for the ARMN model. Therefore, if zero inflation is not accounted for, the 95\% credible intervals will often not cover the true values of important parameters.

In conclusion, our hybrid Gibbs sampling algorithm was able to recover the true parameter values of the MS-ZIARMN model, provided it was correctly specified, and converged consistently. Ignoring zero inflation leads to poor coverage of the 95\% credible intervals and posteriors often centered far from the true parameter values.

}

\clearpage

\section{Widely Applicable Information Criterion (WAIC)} \label{app3:WAIC}

The WAIC for a state space model is more accurate when the latent state indicators are marginalized \citep{auger-methe_guide_2021}. We can use the forward filtering part of the FFBS algorithm to marginalize $S_{it}^*$ from $p(\bm{y}_{it} | \bm{y}_{t-1}, S_{it}^*, \text{total}_{it},\bm{\beta})$,
\begin{align*}
p(\bm{y}_{it}|\bm{y}_{1:(t-1)},\text{total}_{it},\bm{v}) &= \sum_{k=1}^{4} p(\bm{y}_{it} | \bm{y}_{t-1}, S_{it}^*=k, \text{total}_{it},\bm{\beta})P(S_{it}^*=k | \bm{y}_{1:(t-1)},\bm{v}),
\end{align*} see Equation (\ref{eqn:FF}) above. Then, following \cite{gelmanUnderstandingPredictiveInformation2014}, the WAIC can be calculated as,
\begin{align}\label{eqn: WAIC3}
    \begin{split}
   &\text{lpdd} = \sum_{i=1}^{N}\sum_{t=2}^T \log\left(\frac{1}{Q-M}\sum_{m=M+1}^Q p(y_{it}|\bm{y}_{1:(t-1)}, \text{total}_{it},{\bm{v}}^{[m]})\right), \\
   &\text{pwaic} =\sum_{i=1}^{N}\sum_{t=2}^T Var_{m=M+1}^Q \log\left(p(y_{it}|\bm{y}_{1:(t-1)},\text{total}_{it},{\bm{v}}^{[m]})\right),\\
   &\text{WAIC} =-2(\text{lpdd}-\text{pwaic}),
   \end{split}
\end{align} where the superscript $[m]$ denotes a draw from the posterior of the variable and $Var$ denotes the sample variance.

\section{Fitted Values} \label{m3:fittedvalues}

Comparing the fit of a time series model to the observed data is an important part of model checking in time series analysis \citep{knorr-heldHierarchicalModelSpace2003b,liboschikTscountPackageAnalysis2017,chenMarkovSwitchingIntegervalued2019}. We will let $\bm{y}_{it}'$ denote a fitted value. We assume that  $\bm{y}_{it}'$ is a new count from the same MS-ZIARMN model from Equation (8) of the main text that is assumed to have conditionally generated $\bm{y}_{it}$, with the same past counts, parameter values, covariates and present diseases, except we assume a new value for the random effect $\bm{\phi}_{it}'$ corresponding to $\bm{y}_{it}'$. That is, we assume $\bm{y}_{it}'$ comes from the same space and time as $\bm{y}_{it}$ and from the same assumed model (8), except we posit a scenario where the unknown residual space-time factors influencing $\bm{y}_{it}'$ could be different. We assume a new value for the random effect $\phi_{kit}$ as they represent a residual for each observation and, therefore, it would be misleading to add them to the fit of the model.

To be more concrete, let $\bm{v}'$ represent $\bm{v}$ with $\Sigma$ and  $\bm{\phi}_{it}$, for all $i$ and $t$, removed. We assume that ${\bm{y}_{it}'| \bm{y}_{t-1}, S_{it}^*, \text{total}_{it},\bm{v}',\bm{\phi}_{it}'}$ follows the mixture distribution in Equation (8) of the main text with $\lambda_{kit}^*$ replaced by $\lambda_{kit}^{*'}$ where
\begin{align*}
\log(\lambda_{kit}^{*'}) &= \alpha_{0k}+\alpha_{0ki}+\bm{x}_{it}^T \bm{\alpha}_k+\phi_{kit}^{'}+\zeta_k\log(y_{ki(t-1)}+1)-\zeta_1\log(y_{1i(t-1)}+1).
\end{align*} We assume $\bm{\phi}_{it}'$ is a new value for the random effect corresponding to $\bm{y}_{it}'$ representing potentially different residual space-time factors. Therefore, $\bm{\phi}_{it}'$ does not feature in the likelihood and only depends on $\bm{y}$ through $\Sigma$, that is,
\begin{align*}
\bm{\phi}_{it}' | \Sigma, \bm{y}=\bm{\phi}_{it}' | \Sigma \sim  \text{MVN}_{K-1}(\bm{0},\Sigma).
\end{align*} The posterior distribution of $\bm{y}_{it}'$ is then given by,
\begin{align} \label{eqn:postpred}
p(\bm{y}_{it}'|\bm{y}) = \int_{\bm{v}',\Sigma,\bm{\phi}_{it}'} \sum_{S_{it}^*} p(\bm{y}_{it}'| \bm{y}_{t-1}, S_{it}^*, \text{total}_{it},\bm{v}',\bm{\phi}_{it}')p(\bm{\phi}_{it}' | \Sigma)p(\bm{v}',\Sigma,S_{it}^*|\bm{y}) \, d\bm{v}'d\Sigma d\bm{\phi}_{it}'.
\end{align}

Let the superscript $[m]$, for $m=M+1,\dots Q$, where $M$ is the size of the burn-in sample and $Q$ is the total MCMC sample size, denote a draw from the posterior distribution of a variable. From (\ref{eqn:postpred}), to draw $\bm{y}_{it}^{'[m]} \sim p(\bm{y}_{it}'|\bm{y})$ we can first draw  $\bm{\phi}_{it}^{'[m]}$ from $p(\bm{\phi}_{it}' | \Sigma=\Sigma^{[m]})$. Then we can draw $\bm{y}_{it}^{'[m]}$ from $p(\bm{y}_{it}'| \bm{y}_{t-1}, S_{it}^*=S_{it}^{*[m]}, \text{total}_{it},\bm{v}'=\bm{v}^{'[m]},\bm{\phi}_{it}'=\bm{\phi}_{it}^{'[m]})$. To help assess the fit of the model, we then compare the posterior mean and credible interval of $y_{kit}^{'}$ with the observed value $y_{kit}$.

The posterior probability that disease $k$ was present in area $i$ during time $t$ is given by $P(S_{kit}=1|\bm{y})\approx\frac{1}{Q-M}\sum_{m=M+1}^{Q}S_{kit}^{[m]}$. It is important to view plots of $P(S_{kit}=1|\bm{y})$ versus $t$ for various diseases $k$ and areas $i$ for model checking, see Section 4.2 of the main text. We cannot compare $P(S_{kit}=1|\bm{y})$ to the true value of $S_{kit}$ when $y_{kit}=0$, since it is unknown. We usually do not know if the diseases were circulating whenever 0 cases were reported, as they could have been circulating undetected. However, we can still check to make sure the model's estimates of when the diseases were present are reasonable and correspond to our general knowledge of the epidemiological situation. Note, when $y_{kit}>0$, $P(S_{kit}=1|\bm{y})$ is not informative of the model fit as we will always have $P(S_{kit}=1|\bm{y})=1$ if $y_{kit}>0$, see Equation (8) of the main text.

\newpage

\section{Estimates From the Comparison Models} \label{app3:estcompare}

This section provides the estimated coefficients from the multinomial part of the comparison models described in Section 4 of the main text. The tables here mirror Table 2 from the main text. The estimates for the ARMN and Zeng (2022) models are given in Table \ref{tab3:multinomest_zeng} and those for the ZIARMN model in Table \ref{tab3:multinomest_ziarmn}. We combined the ARMN and Zeng (2022) models, as the Zeng (2022) model estimated that the diseases were always present, meaning the multinomial parameter estimates from the two models were nearly identical. The estimates for the sensitivity analysis model described in Section \ref{m3:sensitivity} below are given in Table \ref{tab3:multinomest_sens}.

\renewcommand{\arraystretch}{1.5}
\begin{table}[!hbt]
\centering
\caption{\label{tab3:multinomest_zeng} Posterior means and 95\% posterior credible intervals (in parentheses) for the estimated coefficients from the multinomial part of the fitted Zeng (2022) and ARMN models. The intercept row shows $\lambda_{kit}$ for $k=2$ (Zika) and $k=3$ (chik.) in a typical area at average values of the covariates. Recall, if $\lambda_{kit}>1$ ($\lambda_{kit}<1$) then the share of disease $k$ relative to dengue will tend to grow (shrink) over time. Other rows show the ratio of $\pi_{kit}/\pi_{1it}$ (relative odds ratio) or the ratio of $\lambda_{kit}$ (rate ratio) (both are the same, see Equations (2)-(3) of the main text) corresponding to a unit increase in the covariate. All covariates are standardized. Significant effects are bolded. See Section 4.1 of the main text for an explanation of the covariates.}

\begin{tabular}{lcc} 
 \hline                                    & \multicolumn{2}{c}{\textbf{Relative Odds Ratio or Rate Ratio}}            \\ \hline 
      \textbf{Covariates}                       & \textbf{Zika-dengue} & \textbf{chik.-dengue}   \\  \hline
Intercept  &  .41 (.38, .45)   & .19 (.17, .22)    \\
$\text{verde}_i$ & .98 (.92, 1.05)  & \textbf{.90 (.83, .98)} \\
$\text{SDI}_i$ & 1.03 (.96, 1.10)    & 1 (.92, 1.09)\\
$\text{popdens}_i$ & 1 (.94, 1.09)  & .97 (.88, 1.06)  \\
$\text{favela}_i$ & .96 (.90, 1.03) & .95 (.87, 1.03) \\ 
$\text{temp}_{it}$ & \textbf{1.10 (1.04, 1.16)} & \textbf{.81 (.76, .86)} \\[5pt] 
Neighborhood dengue prevalence & \textbf{.56 (.53, .59)} & \textbf{.56 (.53, .59)} \\[10pt]
Neighborhood Zika prevalence & \textbf{2.56 (2.37, 2.76)} & -- \\[10pt]
Neighborhood chik. prevalence & -- & \textbf{2.13 (2.01, 2.25)} \\[5pt]
Previous Zika cases & -- & \textbf{1.45 (1.36, 1.55)} \\
Previous chik. cases & \textbf{1.30 (1.25, 1.36)} & -- \\
\hline
\end{tabular}
\end{table}

\begin{table}[t]
\centering
\caption{\label{tab3:multinomest_ziarmn} Posterior means and 95\% posterior credible intervals (in parentheses) for the estimated coefficients from the multinomial part of the fitted ZIAMRN model. The intercept row shows $\lambda_{kit}$ for $k=2$ (Zika) and $k=3$ (chik.) in a typical area at average values of the covariates. Recall, if $\lambda_{kit}>1$ ($\lambda_{kit}<1$) then the share of disease $k$ relative to dengue will tend to grow (shrink) over time. Other rows show the ratio of $\pi_{kit}/\pi_{1it}$ (relative odds ratio) or the ratio of $\lambda_{kit}$ (rate ratio) (both are the same, see Equations (2)-(3) of the main text) corresponding to a unit increase in the covariate. All covariates are standardized. Significant effects are bolded. See Section 4.1 of the main text for an explanation of the covariates.}

\begin{tabular}{lcc} 
 \hline                                    & \multicolumn{2}{c}{\textbf{Relative Odds Ratio or Rate Ratio}}            \\ \hline 
      \textbf{Covariates}                       & \textbf{Zika-dengue} & \textbf{chik.-dengue}   \\  \hline
Intercept  &  1.19 (1.07, 1.31)   & .99 (.89, 1.10)    \\
$\text{verde}_i$ & 1 (.94, 1.07)  & \textbf{.91 (.84, .99)} \\
$\text{SDI}_i$ & 1.07 (.99, 1.15)    & 1 (.91, 1.10)\\
$\text{popdens}_i$ & 1.02 (.94, 1.10)  & 1.05 (.96, 1.16)  \\
$\text{favela}_i$ & .97 (.91, 1.04) & .93 (.86, 1.01) \\ 
$\text{temp}_{it}$ & \textbf{1.17 (1.11, 1.24)} & \textbf{.86 (.81, .92)} \\[5pt] 
Neighborhood dengue prevalence & \textbf{.68 (.64, .71)} & \textbf{.68 (.64, .71)} \\[10pt]
Neighborhood Zika prevalence  & \textbf{1.53 (1.44, 1.65)} & -- \\[10pt]
Neighborhood chik. prevalence & -- & \textbf{1.43 (1.36, 1.51)} \\[5pt]
Previous Zika cases& -- & 1.01 (.96, 1.08) \\
Previous chik. cases & 1.03 (.99, 1.07) & -- \\
\hline
\end{tabular}
\end{table}

\clearpage

\section{Map of Temperature in Rio}

\begin{figure}[!htb]
 	\centering
 	\includegraphics[width=\textwidth]{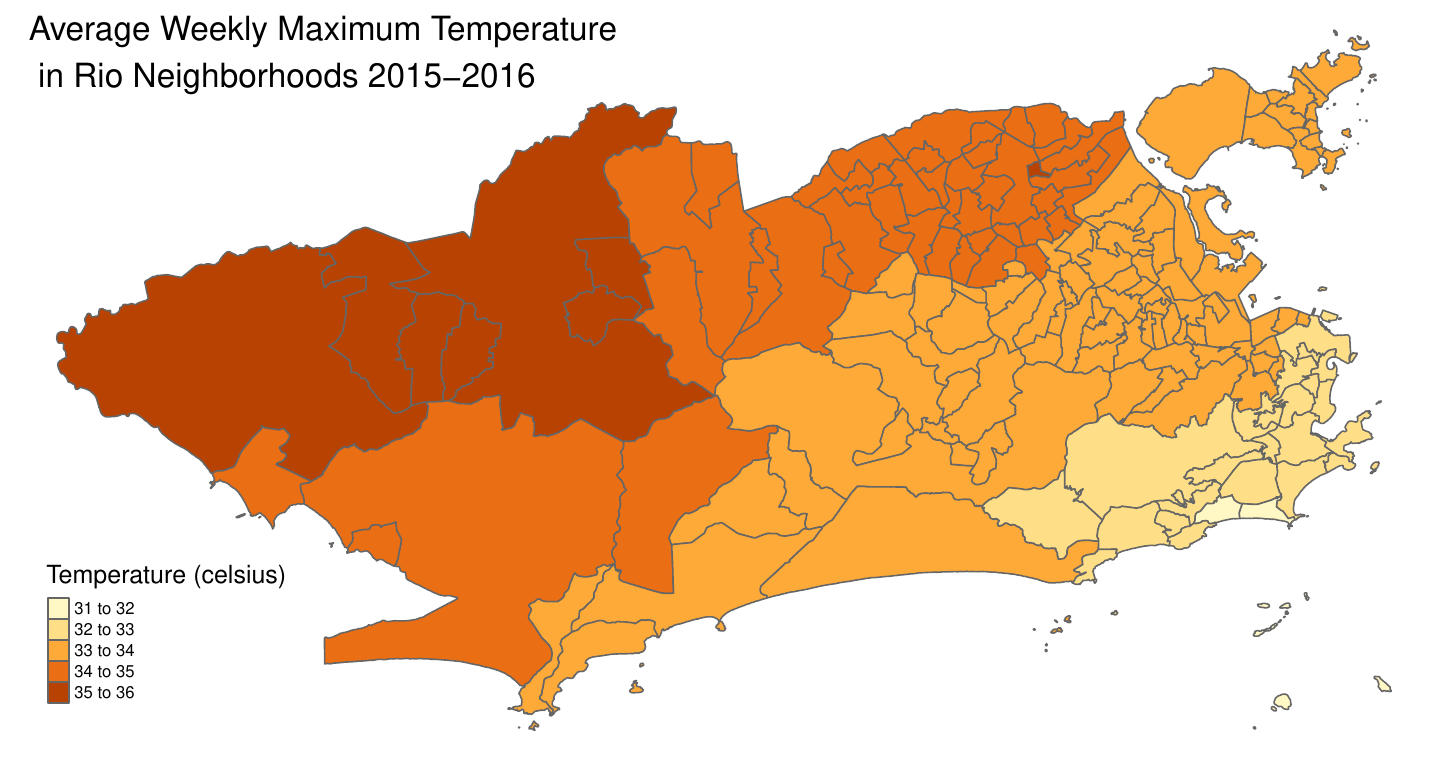}
	\caption{Shows the average weekly maximum temperature across Rio neighborhoods between 2015 and 2016.\label{fig3:SM_temp_map}} 
\end{figure}

Figure \ref{fig3:SM_temp_map} shows a map of the average weekly maximum temperature across Rio between 2015 and 2016. Temperatures tend to be higher in the west of the city and lower downtown.

\newpage

\section{Sensitivity analysis to control for differences in the susceptible populations} 
\label{m3:sensitivity}

As mentioned in Section 2 of the main text, differences in the susceptible populations of the diseases can be an important source of unmeasured confounding. If one of the diseases has a much larger susceptible population, then its transmission will be favored. It is difficult to directly measure susceptible populations of vector-borne diseases (also most diseases) due to underreporting, different strains, cross-immunity, waning immunity and demographic changes \citep{reich_interactions_2013,freitas_spacetime_2019,aogo_effects_2023}. As a proxy for differences in the susceptible populations, we considered the differences in cumulative incidence per person. That is, we refit the MS-ZIARMN model including $\left(\sum_{j=1}^{t-1}y_{kij}-\sum_{j=1}^{t-1}y_{1ij}\right)/\text{pop}_i$ in $\bm{x}_{kit}$ for $k=2$ (Zika) and k=3 (chik.). The multinomial estimates are given in Table \ref{tab3:multinomest_sens} and none are substantially different from those in Table 2 of the main text. The effects of the differences in cumulative incidence were not significant, and the WAIC of the model increased by more than 5, indicating a worse fit. That is, if one of the diseases accumulated many more cases compared to the others, there was likely not a large effect on the favorability of its transmission. One explanation is that the parameters $\zeta_k$ in Equation (1) of the main text are already mostly accounting for this, as the effect of nonhomogeneous mixing on transmission is similar to the effect of the depletion of the susceptible population.

\begin{table}[t]
\centering
\caption{\label{tab3:multinomest_sens} Posterior means and 95\% posterior credible intervals (in parentheses) for the estimated coefficients from the multinomial part of the fitted sensitivity analysis model from Section \ref{m3:sensitivity}. The intercept row shows $\lambda_{kit}$ for $k=2$ (Zika) and $k=3$ (chik.) in a typical area at average values of the covariates. Recall, if $\lambda_{kit}>1$ ($\lambda_{kit}<1$) then the share of disease $k$ relative to dengue will tend to grow (shrink) over time. Other rows show the ratio of $\pi_{kit}/\pi_{1it}$ (relative odds ratio) or the ratio of $\lambda_{kit}$ (rate ratio) (both are the same, see Equations (2)-(3) of the main text) corresponding to a unit increase in the covariate. All covariates are standardized. Significant effects are bolded. See Section \ref{m3:sensitivity} and Section 4.1 of the main text for an explanation of the covariates.}

\begin{tabular}{lcc} 
 \hline                                    & \multicolumn{2}{c}{\textbf{Relative Odds Ratio or Rate Ratio}}            \\ \hline 
      \textbf{Covariates}                       & \textbf{Zika-dengue} & \textbf{chik.-dengue}   \\  \hline
Intercept  &  1.13 (1.02, 1.26)   & 1 (.9, 1.11)    \\
$\text{verde}_i$ & 1.02 (.94, 1.1)  & .92 (.85, 1) \\
$\text{SDI}_i$ & 1.06 (.97, 1.15)    & 1.02 (.92, 1.12)\\
$\text{popdens}_i$ & 1.02 (.93, 1.12)  & 1.06 (.96, 1.17)  \\
$\text{favela}_i$ & .97 (.89, 1.04) & .94 (.86, 1.03) \\ 
$\text{temp}_{it}$ & \textbf{1.14 (1.08, 1.20)} & \textbf{.85 (.80, .90)} \\[5pt] 
Neighborhood dengue prevalence & \textbf{.7 (.67, .74)} & \textbf{.7 (.67, .74)} \\[10pt]
Neighborhood Zika prevalence & \textbf{1.59 (1.48, 1.70)} & -- \\[10pt]
Neighborhood chik. prevalence & -- & \textbf{1.43 (1.36, 1.51)} \\[5pt]
Previous Zika cases & -- & \textbf{.9 (.84, .96)} \\
Previous chik. cases & .98 (.94, 1.02) & -- \\[5pt]
$\left(\sum_{j=1}^{t-1}y_{2ij}-\sum_{j=1}^{t-1}y_{1ij}\right)/\text{pop}_i$ & .94 (.87, 1) & -- \\[5pt]
$\left(\sum_{j=1}^{t-1}y_{3ij}-\sum_{j=1}^{t-1}y_{1ij}\right)/\text{pop}_i$ & -- & .99 (.94, 1.05) \\
\hline
\end{tabular}
\end{table}
\clearpage
{\color{black}
\newpage
\section{\color{black}Traceplots}

This section shows the traceplots of all parameters (except the random effects) of the MS-ZIARMN model from our motivating example in Section 4 of the main text. Note that due to parameter transformations, the traceplots will not necessarily correspond to Tables 2 and 3 from the main text.

\begin{figure}[!htb]
 	\centering
 	\includegraphics[width=\textwidth]{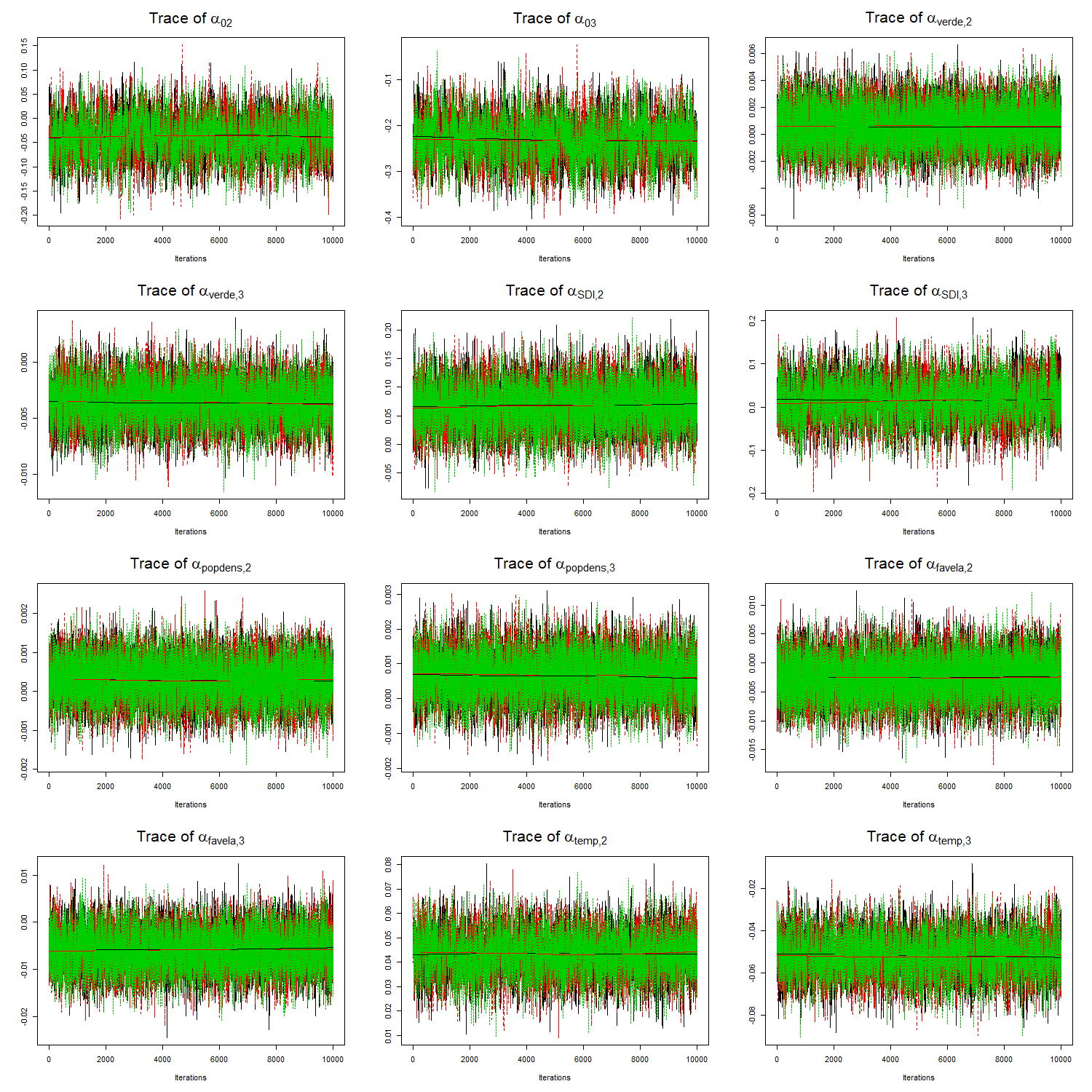}
	\caption{{\color{black}Shows traceplots from the MS-ZIARMN model fit in our motivating example in Section 4 (Part 1).}} 
\end{figure}

\begin{figure}[!htb]
 	\centering
 	\includegraphics[width=\textwidth]{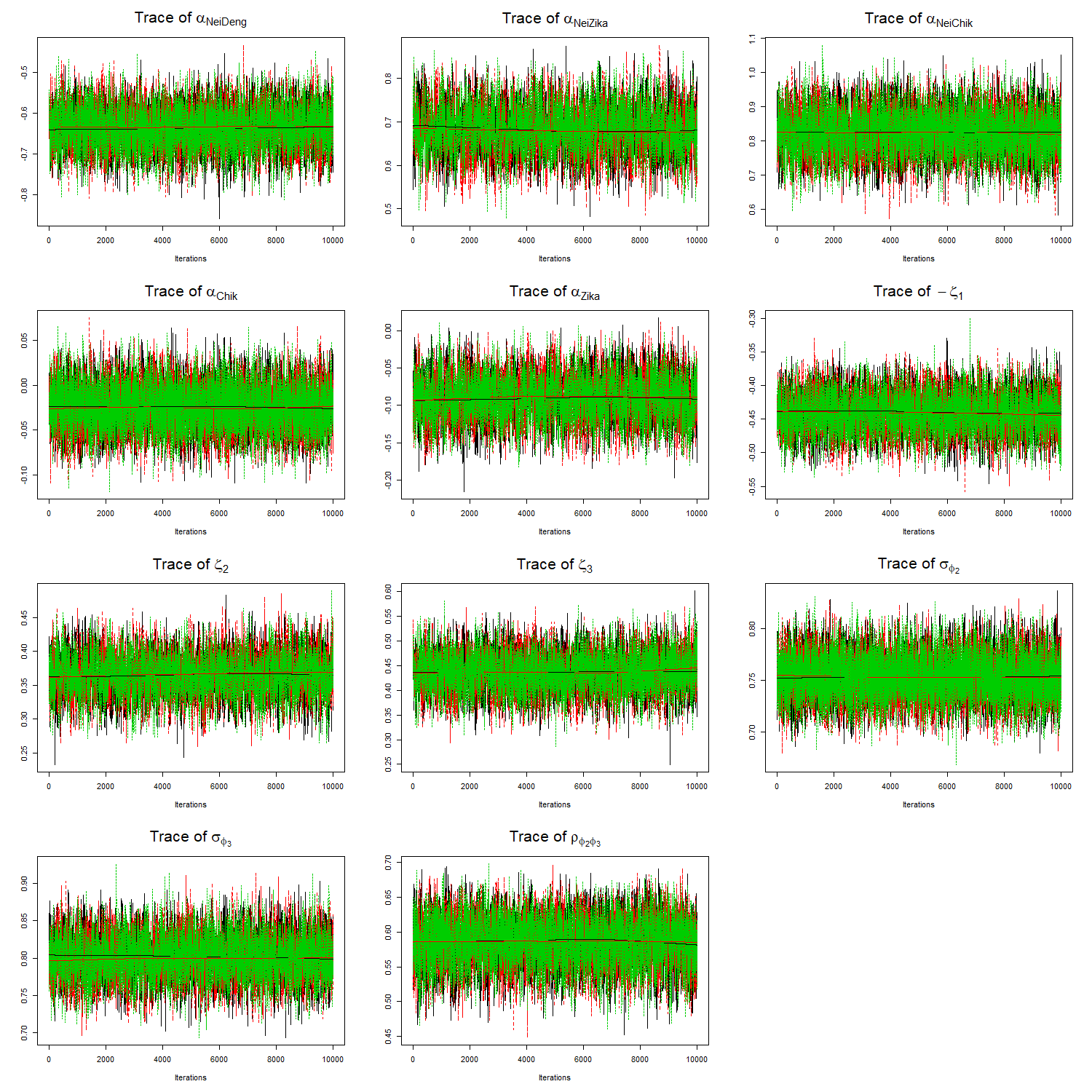}
	\caption{{\color{black}Shows traceplots from the MS-ZIARMN model fit in our motivating example in Section 4 (Part 2).}} 
\end{figure}

\begin{figure}[!htb]
 	\centering
 	\includegraphics[width=\textwidth]{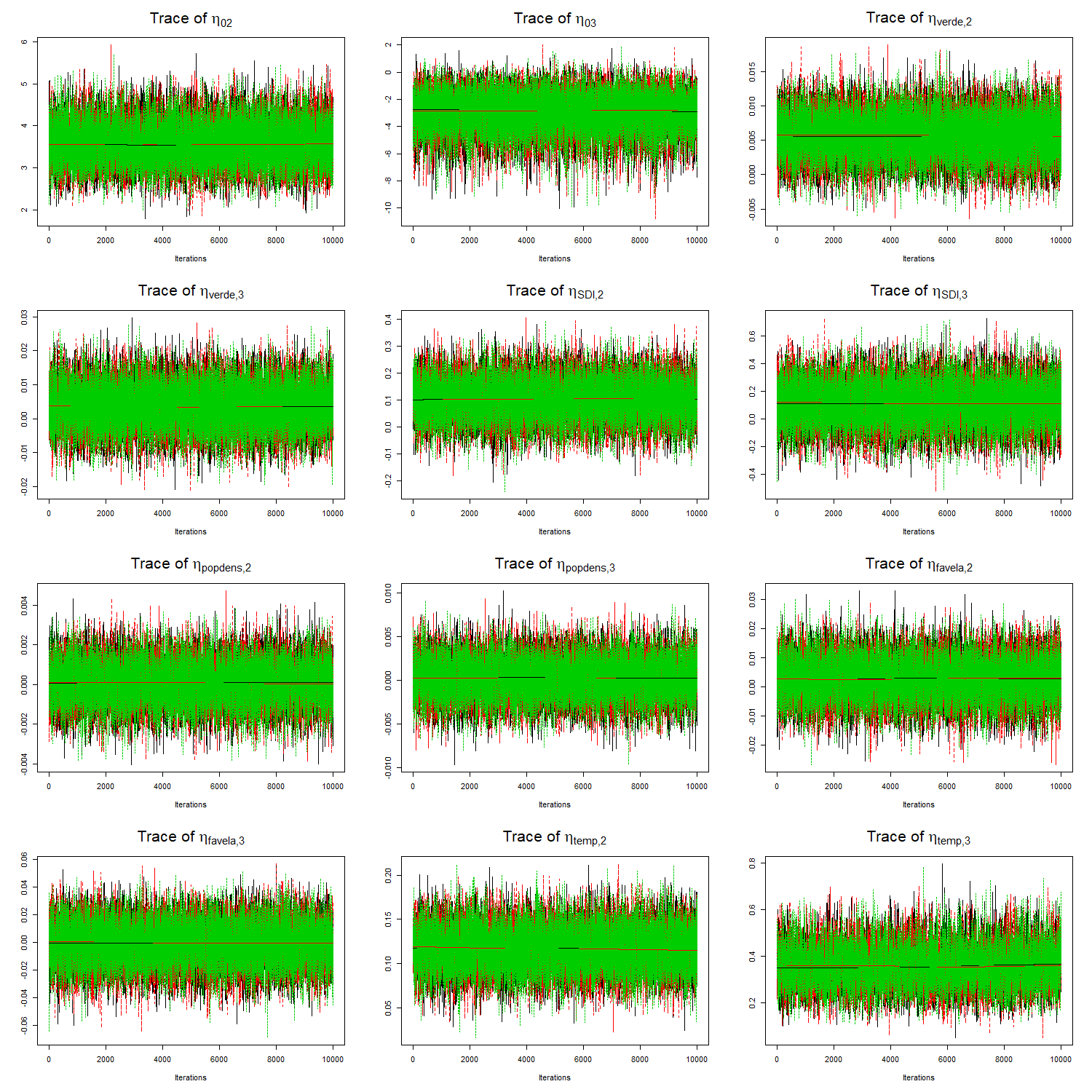}
	\caption{{\color{black}Shows traceplots from the MS-ZIARMN model fit in our motivating example in Section 4 (Part 3).}} 
\end{figure}

\begin{figure}[!htb]
 	\centering
 	\includegraphics[width=\textwidth]{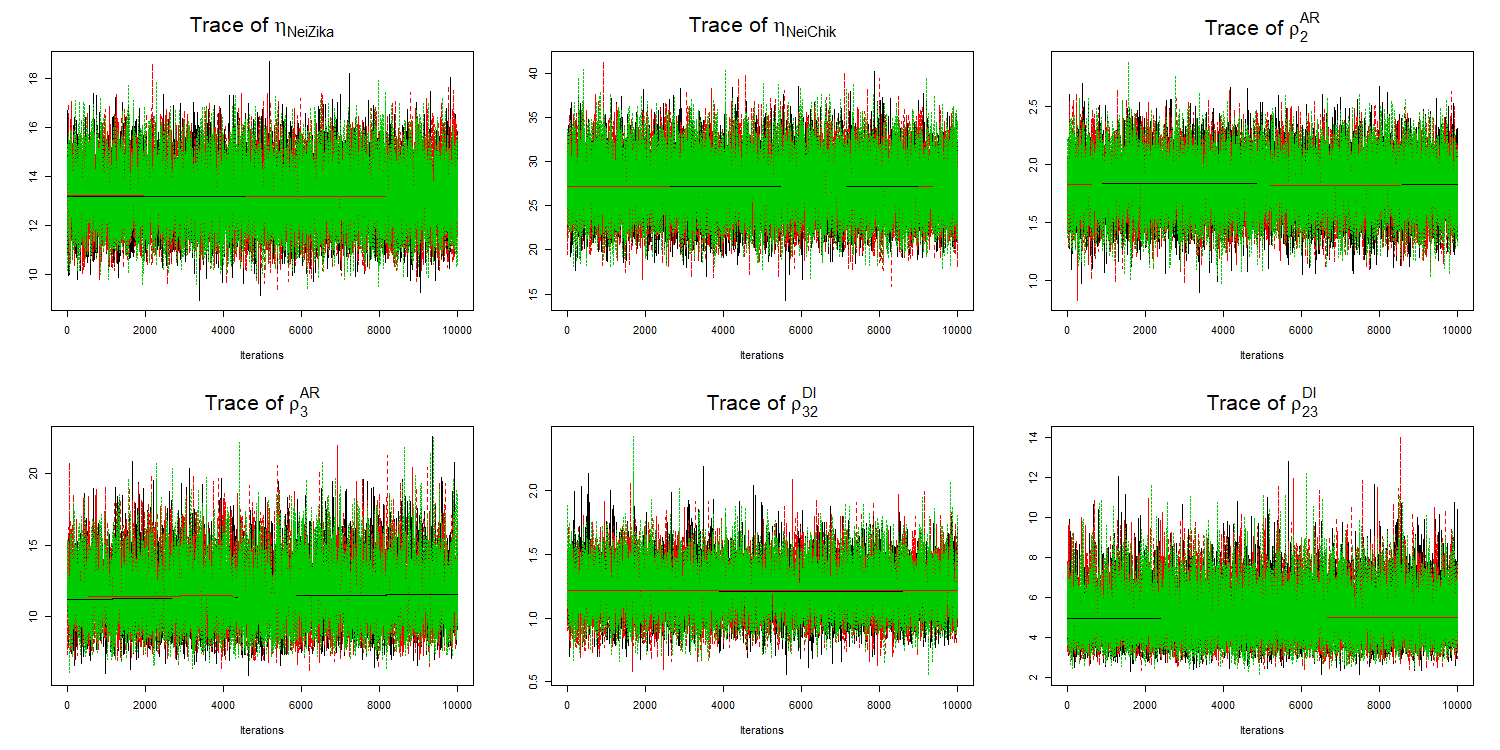}
	\caption{{\color{black}Shows traceplots from the MS-ZIARMN model fit in our motivating example in Section 4 (Part 4).}} 
\end{figure}

}

\clearpage

\section{Results From the ICAR Model}

 Recall, from Equation (3) in the main text, that we allow the log relative odds to depend on past disease cases in neighboring areas. This will induce spatio-temporal correlations between the disease counts and is epidemiologically motivated, as infectious diseases are transmitted from person to person and spread geographically \citep{bauerStratifiedSpaceTime2018}. However, some spatial correlation may remain that is not accounted for by this autoregression. Therefore, following a suggestion from a reviewer, we tried fitting our MS-ZIARMN model from Section 4 of the main text using a conditional auto-regressive (ICAR) model for the intercepts, $\alpha_{0ki} \sim \text{ICAR}(\sigma_{k,\text{ICAR}}^{2})$, in Equation (3) \citep{besag1991bayesian}. That is, we assumed, 
\begin{align*}
\alpha_{0ki} | \bm{\alpha}_{0k(-i)} \sim N\left(\frac{\sum_{j \in NE(i) }\alpha_{0kj}}{|NE(i)|},\frac{\sigma_{k,\text{ICAR}}^{2}}{|NE(i)|}\right),
\end{align*} where $NE(i)$ is the set of neighboring areas of area $i$, $|NE(i)|$ is the number of neighbors and $(-i)$ denotes all areas excluding area $i$. We used gamma(.1,.1) priors for $1/\sigma_{2,\text{ICAR}}^{2}$ and $1/\sigma_{3,\text{ICAR}}^{2}$. The code for fitting the ICAR model is available on GitHub, \newline \url{https://github.com/Dirk-Douwes-Schultz/MS_ZIARMN_code}.

The ICAR model converged and had a WAIC of 22,071 versus 22,091 for the normal random intercept model (the MS-ZIARMN model from Section 4 of the main text), indicating some excess spatial correlation. Table \ref{tab3:multinomest_ICAR} below (which mirrors Table 2 in the main text) shows the multinomial estimates from the ICAR model. There are only very minor differences between Table \ref{tab3:multinomest_ICAR} below and Table 2 in the main text. We compared all other results reported in Section 4, and they were also very similar between the two intercept models. Therefore, using ICAR random intercepts as opposed to normal random intercepts would not lead to any changes in the analysis of the results in Section 4.

Although our MS-ZIARMN model from Section 4 converged with ICAR intercepts, we were unable to obtain convergence for the Zeng or ZIARMN models. As the results of the normal random intercept and ICAR models were very similar (as discussed above), we decided to use normal random intercepts in the main text to ensure that all comparisons in Section 4 were fair.

\renewcommand{\arraystretch}{1.5}
\begin{table}[t]
\centering
\caption{\label{tab3:multinomest_ICAR} Posterior means and 95\% posterior credible intervals (in parentheses) for the estimated coefficients from the multinomial part of the fitted MS-ZIARMN model with ICAR intercepts. Mirrors Table 2 from the main text.}

\begin{tabular}{lcc} 
 \hline                                    & \multicolumn{2}{c}{\textbf{Relative Odds Ratio or Rate Ratio}}            \\ \hline 
      \textbf{Covariates}                       & \textbf{Zika-dengue} & \textbf{chik.-dengue}   \\  \hline
Intercept  &  {1.16 (1.05, 1.27)}   & {1.04 (.94, 1.15)}    \\
$\text{verde}_i$ & {.94 (.87, 1.01)}  & {.96 (.88, 1.05)} \\
$\text{SDI}_i$ & {.98 (.89, 1.06)}    & {.99 (.9, 1.10)}\\
$\text{popdens}_i$ & {.99 (.92, 1.07)}  & {1.04 (.95, 1.14)}  \\
$\text{favela}_i$ & {.97 (.91, 1.03)} & {.93 (.86, 1.00)} \\ 
$\text{temp}_{it}$ & {\textbf{1.16 (1.10, 1.22)}} & {\textbf{.84 (.79, .89)}} \\
Neighborhood dengue prevalence & {\textbf{.73 (.69, .77)}} & {\textbf{.73 (.69, .77)}} \\
 Neighborhood Zika prevalence& {\textbf{1.52 (1.42, 1.63)}} & -- \\
Neighborhood chik. prevalence & -- & {\textbf{1.36 (1.29, 1.43)}} \\
Previous Zika cases & -- & {\textbf{.91 (.86, .97)}} \\
Previous chik. cases & {.98 (.94, 1.02)} & -- \\
\hline
\end{tabular}
\end{table}

\clearpage

\bibliography{supplement}